\documentclass[a4paper]{article}

\usepackage{amsmath, amsfonts, amssymb, theorem, verbatim}

\allowdisplaybreaks[2]

\numberwithin{equation}{section}

\newtheorem{theorem}{Theorem}[section]
\newtheorem{corollary}[theorem]{Corollary}
\newtheorem{lemma}[theorem]{Lemma}
\newtheorem{proposition}[theorem]{Proposition}
{\theorembodyfont{\rmfamily} 
} 
{\theorembodyfont{\rmfamily} 
} 
{\theorembodyfont{\rmfamily} 
}

{\theorembodyfont{\rmfamily} 
}

\makeatletter

\newenvironment{proof}{\noindent\textsc{Proof.}}
{\hspace*{\fill}$\square$\par\bigskip}

\def\section{\@startsection {section}{1}{\z@}{3.5ex plus 1ex minus
    .2ex}{2.3ex plus .2ex}{\large\bf}}
    \def\subsection{\@startsection{subsection}{2}{\z@}{3.25ex plus 1ex minus 
 .2ex}{1.5ex plus .2ex}{\bf}}

\makeatother

\title{Deformation of two body quantum Calogero-Moser-Sutherland models} 
\author{Kenji Taniguchi 
\thanks{Department of Physics and Mathematics, 
Aoyama Gakuin University, 
5-10-1, Fuchinobe, Sagamihara, Kanagawa 229-8558, Japan. 
(taniken@gem.aoyama.ac.jp) }}

\date{\empty}

\begin{document}

\maketitle

\begin{abstract}
The possibility of deformation of two body quantum
Calogero-Moser-Sutherland models is studied. 
Obtained are some necessary conditions for the singular locus of the
potential function. 
Such locus is determined if it consists of two, three or four lines. 
Furthermore, a new deformation of elliptic $B_{2}$ type
Calogero-Moser-Sutherland model is explicitly constructed.

\end{abstract}



\section{Introduction}
\label{section:introduction}
A Schr\"{o}dinger operator 
\[
L := 
-\sum_{i=1}^{n} \frac{\partial^{2}}{\partial x_{i}^{2}} + R(x) 
\]
is called completely integrable if there exist $n$ algebraically
independent differential operators $P_{1} = L, P_{2}, \dots, P_{n}$
which commute each other. 
Let $(\Sigma, W)$ be a pair of a root system and its Weyl group. 
The $n$-body Calogero-Moser-Sutherland (CMS) operator 
\begin{align} 
L &= 
-\sum_{i=1}^{n} \frac{\partial^{2}}{\partial x_{i}^{2}} 
+ \sum_{\alpha \in \Sigma^{+}} 
m_{\alpha} (m_{\alpha} + 1) |\alpha|^{2}
u(\langle \alpha, x \rangle), 
\label{eq:intro-2}
\end{align}
with 
\begin{align*}
u_{\alpha}(t) 
&= 
\begin{cases} 
1/t^{2} \quad \mbox{(rational case)}, 
\\
\omega^{2}/\sin^{2} \omega t, \enskip 
\omega^{2}/\sinh^{2} \omega t 
\quad \mbox{(trigonometric case)}, 
\\
\wp(t) \quad \mbox{(elliptic case)}, 
\end{cases}
\\
m_{w \alpha} &= m_{\alpha} \quad (\alpha \in \Sigma, w \in W), 
\end{align*}
is an example of completely integrable operator. 
Here, $\wp(t)$ is the Weierstrass $\wp$ function. 
The constants $m_{\alpha}$ are called the \textit{coupling constants}. 

Obviously, these potential functions possess inverse square
singularities along the walls of Weyl chambers. 
As a generalisation of CMS operator, 
let us consider a Schr\"{o}dinger operator 
\begin{equation}\label{eq:generalization}
L = 
-\sum_{i=1}^{n} \frac{\partial^{2}}{\partial x_{i}^{2}} + R(x), 
\quad \mbox{with} \quad 
R(x) 
= 
\sum_{\alpha \in \mathcal{H}} 
\frac{C_{\alpha}}
{\langle \alpha, x \rangle^{2}} 
+ 
\widetilde{R}(x). 
\end{equation}
Here, $\mathcal{H}$ is a finite set of mutually non-parallel vectors
in $\boldsymbol{R}^{n}$, $C_{\alpha}$ are non-zero constants and
$\widetilde{R}(x)$ is real analytic at $x = 0$. 
We call $\mathcal{H}$ the {\it singular locus} of $L$ or 
the singular locus of $R(x)$. 
Note that we do not assume the symmetry of either $R(x)$ or $P$, 
nor do we assume $\mathcal{H}$ to be a subset of a root system. 

In \cite{T}, the author investigated what kind of differential
operator $P$ commutes with $L$ in \eqref{eq:generalization}. 
One of the main results of \cite{T} is that, 
if $C_{\alpha} \not\in \{m(m+1) |\alpha|^{2} ; 
m \in \boldsymbol{Z}\}$ for any $\alpha \in \mathcal{H}$, 
then the principal symbol of $P$ is invariant under the action of the
group $W$ generated by reflections $r_{\alpha}$ with respect to the
hyperplanes $\langle \alpha, x \rangle = 0$ 
($\alpha \in \mathcal{H}$). 
Therefore, if $L$ possesses a non-trivial commutant, 
then $W$ must be a finite reflection group and $\mathcal{H}$ must be a
subset of the root system of this reflection group
(\cite[Theorem~4.4]{T}).

On the other hand, if some of the coupling constants are one, i.e. 
$C_{\alpha} = 1 \cdot 2 |\alpha|^{2}$ for some
$\alpha \in \mathcal{H}$, 
it is known that there exist completely integrable Schr\"{o}dinger
operators like \eqref{eq:intro-2}, 
but whose singular loci are not root systems but deformed
ones \cite{CFV, VFC}. 

The final objective of this research is to classify such deformed
completely integrable CMS type operators and to construct such
operators explicitly. 
But, in this paper, we do not consider general cases, 
but restrict our interest to the rank two rational cases. 
Namely, we consider what kind of operator $P$ commutes with 
\begin{align}
L & = 
-\left(
\frac{\partial^{2}}{\partial x_{1}^{2}} 
+ 
\frac{\partial^{2}}{\partial x_{2}^{2}} 
\right) 
+ 
\sum_{\alpha \in \mathcal{H}} 
\frac{C_{\alpha}}{\langle \alpha, x \rangle^{2}}, 
\qquad 
(\mathcal{H} \subset \boldsymbol{R}^{2}, 
C_{\alpha} \not=0). 
\label{eq:rank two rational}
\end{align}
The reason to do so is as follows: 
If the operator $L$ in \eqref{eq:generalization} and a differential
operator $P$ commute, 
then, by ``restricting'' them to a two dimensional subspace, 
we obtain a two body completely integrable CMS type operator $L'$,
whose potential function is a rational function. 
For details, see \S \ref{section:rank two reduction}. 
Therefore, two body rational integrable models are building blocks of
general integrable models, and it is important to classify and
construct them. 

The first result of this paper is the relation between the order of
$P$ and the cardinality of $\mathcal{H}$. 
\begin{theorem}[Corollary~\ref{corollary:order condition}] 
\label{theorem:intro-1}
Assume that the Schr\"{o}dinger operator $L$ in
\eqref{eq:rank two rational} has a non-trivial commutant $P$, 
whose principal symbol is constant with respect to $x$. 
Then the order of $P$ is not less than the cardinality of
$\mathcal{H}$. 
\end{theorem}

Next results are the conditions for the singular locus
$\mathcal{H}$ and the constants $C_{\alpha}$. 
For $\alpha = (\alpha_{1}, \alpha_{2}) \in \mathcal{H}$, 
let $\alpha^{\bot} = (-\alpha_{2}, \alpha_{1})$. 

\begin{theorem}[Theorem~\ref{theorem:linear relation-1},
Theorem~\ref{theorem:linear relation-2}]
\label{theorem:intro-2}
Assume that $L$ and $P$ satisfy the same condition as in
Theorem~\ref{theorem:intro-1}.  
Then, for each $\alpha_{0} \in \mathcal{H}$, 
\begin{equation}
\sum_{\beta \in \mathcal{H}, \beta \not= \alpha_{0}} 
\frac{\langle \alpha_{0}, \beta \rangle}
{\langle \alpha_{0}^{\bot}, \beta \rangle^{3}} 
C_{\beta} = 0 
\qquad \mbox{and} \qquad 
(C_{\alpha_{0}} - 2 |\alpha_{0}|^{2}) 
\sum_{\beta \in \mathcal{H}, \beta \not= \alpha_{0}} 
\frac{\langle \alpha_{0}, \beta \rangle |\beta|^{2}}
{\langle \alpha_{0}^{\bot}, \beta \rangle^{5}} 
C_{\beta} = 0 
\label{eq:main}
\end{equation}
are satisfied. 
\end{theorem}

In \S\S \ref{section:construction of P2,3,4}, 
\ref{section:possibility}, 
we investigate what kind of $\mathcal{H}$ and $C_{\alpha}$ satisfy
\eqref{eq:main}, when $\# \mathcal{H}= 2, 3$ or $4$. 
%
%
Since $\mathcal{H}$ describes the singular locus of the potential
function in \eqref{eq:rank two rational}, 
the norm of each vector in $\mathcal{H}$ is not essential. 
Actually, if you replace $\alpha \in \mathcal{H}$ and $C_{\alpha}$
with $k \alpha$ and $k^{2} C_{\alpha}$ 
($k \in \boldsymbol{R}^{\times}$) respectively, 
the operator $L$ and the conditions \eqref{eq:main} are unchanged. 
Therefore, we consider two singular loci $\mathcal{H}$ and
$\mathcal{H}'$ to be equivalent if each vector in $\mathcal{H}'$ is a
non-zero multiple of a vector in $\mathcal{H}$. 
Moreover, we also consider $\mathcal{H}$ and $\mathcal{H}'$ to be
equivalent if $\mathcal{H}' = \{ g \alpha ; \alpha \in \mathcal{H}\}$
for some $g \in O(2)$.

\begin{theorem}[Corollary~\ref{cor:2-sheets}, Theorem~\ref{theorem:A2},
Theorem~\ref{theorem:4 lines arrangement}]
\label{theorem:intro-3}
\begin{enumerate}
\item
If $\# \mathcal{H} = 2$, 
then the two vectors in $\mathcal{H}$ cross at right angles. 
Therefore, the singular locus is of type 
$A_{1} \times A_{1}$. 
\item
If $\# \mathcal{H} = 3$, 
then $\mathcal{H} = \{e_{1}, \pm a e_{1} + e_{2}\}$ for some 
$a \not= 0$. 
Moreover, if $\mathcal{H}$ is not a positive system of $A_{2}$ type
root system, 
then the coupling constants for $\pm a e_{1} + e_{2}$ must be one and
there is no other completely integrable model than the one constructed
in \cite{CFV}. 
\item
If $\# \mathcal{H} = 4$, 
then $\mathcal{H} = \{e_{1}, e_{2}, \pm a e_{1} + e_{2}\}$ 
for some $a \not= 0$. 
\end{enumerate}
\end{theorem}

As stated above, deformation of CMS operators is known if some of
the coupling constants are one. 
On the other hand, Theorem~\ref{theorem:intro-2} implies that there
may be other deformation of a CMS operator even if no coupling
constant is one. 
In \S \ref{section:new deformation}, 
we present an example of a new deformation of the $B_{2}$ type CMS
operator. 
The result is as follows. 
\begin{theorem}[Theorem~\ref{theorem:new commutative pair}] 
Let $L$ be the Schr\"{o}dinger operator defined by 
\begin{align*}
L = & 
- \left(
\frac{\partial^{2}}{\partial x_{1}^{2}} 
+ \frac{\partial^{2}}{\partial x_{2}^{2}} 
\right) 
+ u_{1}(2 a x_{1}) + u_{2} (2 x_{2}) + u_{+}(a x_{1} + x_{2}) 
+ u_{-}(-a x_{1} + x_{2}), 
\\
& u_{1}(t) = \frac{3}{4} (a^{2} + 1) (3 a^{-2} - 1) \wp(t), 
\qquad 
u_{2}(t) = \frac{3}{4} (a^{2} + 1) (3 a^{2} - 1) \wp(t), 
\\
& u_{+}(t) = u_{-}(t) = 2 \cdot 3 (a^{2} + 1) \wp(t). 
\end{align*}
Then, there exists a sixth order commutant $P$ of $L$, whose principal
symbol is 
\[
a(4 - a^{2}) \xi_{1}^{6} 
+ 5a \xi_{1}^{4} \xi_{2}^{2} 
+ 5 a^{-1} \xi_{1}^{2} \xi_{2}^{4} 
+ a^{-1}(4 - a^{-2}) \xi_{2}^{6}. 
\]
\end{theorem}

\noindent
\textbf{Acknowledgements.} 
This research was supported in part by Grant-in-Aid for Scientific
Research (C)(2) No.\,15540183, Japan Society for the Promotion of
Science.



\section{Rank two reduction}
\label{section:rank two reduction}

To begin, we introduce some notation. 
Let $\{e_{1}, \dots, e_{n}\}$ be the standard basis of
$\boldsymbol{R}^{n}$ and $x = (x_{1}, \dots, x_{n})$ be the
corresponding coordinates. 
For simplicity, denote by $\partial_{x_{i}}$ the partial differential
$\partial/\partial x_{i}$ and define 
$\partial_{x} = (\partial_{x_{1}}, \dots, \partial_{x_{n}})$. 
An $m_{0}$-th order differential operator $P$ is expressed as 
\[
P= 
\sum_{k = 0}^{m_{0}} P_{k}, 
\qquad 
P_{k} = 
\sum_{|p| = m_{0} - k} a_{p}(x) \partial_{x}^{p}, 
\]
where $p = (p_{1}, \dots, p_{n}) \in \boldsymbol{N}^{n}$ is a
multi-index,  
and $|p|$ is the length $\sum_{i} p_{i}$ of $p$. 
Corresponding to this operator, 
we introduce 
\[
\widetilde{P}_{k} 
= 
\sum_{|p| = m_{0} - k} a_{p}(x) \xi^{p} 
\qquad
(\xi 
= (\xi_{1}, \dots, \xi_{n})),  
\]
and call it the \textit{symbol} of $P_{k}$. 
In particular, $\widetilde{P}_{0}$ is called the 
{\it principal symbol} of $P$.

Let $\langle u, v \rangle$ be the standard inner product on
$\boldsymbol{R}^{n}$, and let $|v|$ be the norm of $v$. 
We also use the symbol $\langle \enskip, \enskip \rangle$ to
other couplings. 
For example, 
$\langle \partial_{x}, \partial_{\xi} \rangle 
= 
\sum_{i=1}^{n} \partial_{x_{i}} \partial_{\xi_{i}}$. 
For notational convenience, let 
\begin{align*}
x_{\alpha} &:= \langle \alpha, x \rangle, 
& 
\xi_{\alpha} &:= \langle \alpha, \xi \rangle, 
&
\partial_{x, \alpha} &:= \langle \alpha, \partial_{x} \rangle, 
& 
\partial_{\xi, \alpha} &:= \langle \alpha, \partial_{\xi} \rangle.  
\end{align*}

Assume that the operator $L$ in \eqref{eq:generalization} commutes
with $P$, 
whose principal symbol $\tilde{P}_{0}$ is constant
with respect to $x$. 
Then, by rank one reduction, we have the following results. 

\begin{lemma}{(\cite[Lemma~2.1]{T})}\label{lemma:review-1}
For any $\alpha \in \mathcal{H}$, $P$ is regular singular along the
hyperplane $x_{\alpha} = 0$, 
i.e. $x_{\alpha}^{k} \widetilde{P}_{k}$ is analytic at 
$x_{\alpha} = 0$. 
\end{lemma} 

Put $x' = (x_{1}', x_{2}', x_{3}', \dots, x_{n}') 
= 
(\varepsilon^{-1} x_{1}, \varepsilon^{-1} x_{2}, x_{3}, \dots, x_{n})$
and consider the Laurent expansion of $L$ and $P$ as meromorphic
functions of $\varepsilon$. 
By Lemma~\ref{lemma:review-1} and $\partial_{x} = 
(\varepsilon^{-1} \partial_{x_{1}'}, 
\varepsilon^{-1} \partial_{x_{2}'},
\partial_{x_{3}'}, \dots, \partial_{x_{n}'},)$, 
we have 
\begin{align*}
L &= 
\sum_{i=-2}^{\infty} \varepsilon^{i} L(i),
&
& L(-2) = 
-(\partial_{x_{1}'}^{2} + \partial_{x_{2}'}^{2}) 
+ \sum_{\alpha \in \mathcal{H} \cap 
(\boldsymbol{R} e_{1} + \boldsymbol{R} e_{2})} 
\frac{C_{\alpha}}{x_{\alpha}'{}^{2}} 
\\
P &= 
\sum_{i=-m_{0}}^{\infty} \varepsilon^{i} P(i), 
&
& \mbox{$P(-m_{0})$ is a differential operator 
on $\boldsymbol{R} e_{1} + \boldsymbol{R} e_{2}$}. 
\end{align*}
Here, $L(i)$ and $P(i)$ are differential operators. 
Moreover, by Lemma~\ref{lemma:review-1}, 
the principal symbol of $P(-m_{0})$ is constant with respect to $x$. 
This expansion implies that if $[L, P] = 0$, 
we have $[L(-2), P(-m_{0})] = 0$. 
In other words, we obtain a two body rational CMS type completely
integrable system. 

Therefore, we restrict our interest to this case. 
Namely, let $n = 2$ and define afresh $L$ and $P$ by 
\begin{align*}
L =& -(\partial_{x_{1}}^{2} + \partial_{x_{2}}^{2}) + R(x), 
\quad 
R(x) = 
\sum_{\alpha \in \mathcal{H}} u_{\alpha}(x_{\alpha}), 
\quad 
u_{\alpha}(t) := \frac{C_{\alpha}}{t^{2}}, 
\\
P =& \sum_{k=0}^{m_{0}} P_{k}, 
\quad 
P_{k} = \sum_{p \in \boldsymbol{N}^{2}, |p| = m_{0} - k} 
a_{p}(x) \partial_{x}^{p},  
\end{align*}
where $\mathcal{H} \subset \boldsymbol{R}^{2}$, $C_{\alpha} \not= 0$
for any $\alpha \in \mathcal{H}$, 
and $a_{p}(x)$ is a homogeneous rational function of degree $-k$ 
if $|p| = k$. 
Hereafter, we will seek conditions for $\mathcal{H}$,
$C_{\alpha}$ or $P$ so that $L$ and $P$ commute. 
For notational convenience, we will abbreviate
$u_{\alpha}^{(k)}(x_{\alpha})$ to $u_{\alpha}^{(k)}$.



\section{Order condition for $P$}
\label{section:conditions for P} 


By Leibniz rule, we have 
\[
[P_{k}, R(x)]\tilde{\ }
= 
\sum_{j=1}^{m_{0}-k} 
\frac{1}{j!} 
\sum_{\alpha \in \mathcal{H}} 
u_{\alpha}^{(j)} 
\partial_{\xi, \alpha}^{j} 
\widetilde{P}_{k}. 
\]
Therefore, we have the following lemma. 
\begin{lemma}\label{lemma:order-1}
The condition $[L, P] = 0$ is equivalent to 
\[
2 \langle \xi,  \partial_{x} \rangle \widetilde{P}_{k} 
+ 
\Delta \widetilde{P}_{k-1} 
+ 
\sum_{j=1}^{k-1} 
\frac{1}{j!} 
\sum_{\alpha \in \mathcal{H}} 
u_{\alpha}^{(j)} 
\partial_{\xi, \alpha}^{j} 
\widetilde{P}_{k-j-1}
= 0
\]
for any $k = 0, \dots, m_{0}$. 
Here, we set $\widetilde{P}_{-1} = \widetilde{P}_{m_{0}+1} = 0$, 
and we defined 
$\langle \xi,  \partial_{x} \rangle 
:= 
\xi_{1} \partial_{x_{1}} + \xi_{2} \partial_{x_{2}}$, 
$\Delta := \partial_{x_{1}}^{2} + \partial_{x_{2}}^{2}$. 
\end{lemma}

By Lemma~2.2 and Lemma~3.2 in \cite{T}, 
we can easily show the following proposition. 
\begin{proposition}\label{proposition:review-2} 
Choose $\alpha \in \mathcal{H}$ and express $\widetilde{P}_{0}$ as a
polynomial in $\xi_{\alpha}$, $\xi_{\alpha^{\bot}}$; 
\[\widetilde{P}_{0} 
= \sum_{k=0}^{m_{0}} c_{k} \xi_{\alpha}^{k} 
\xi_{\alpha^{\bot}}^{m_{0}-k}. 
\] 
\begin{enumerate}
\item
If $C_{\alpha} \not= m(m+1)|\alpha|^{2}$ for 
any $m \in \boldsymbol{Z}$, 
then $c_{k} = 0$ for all odd $k$. 
\item
If $C_{\alpha} = m(m+1)|\alpha|^{2}$ for 
some $m \in \boldsymbol{Z}_{> 0}$, 
then $c_{1} = c_{3} = \dots = c_{2m-1} = 0$. 
\end{enumerate}
Especially, 
$\partial_{\xi, \alpha} 
\widetilde{P}_{0}|_{\xi_{\alpha} \to 0} = 0$ for any
$\alpha \in \mathcal{H}$ since $C_{\alpha} \not= 0$. 

\end{proposition}

\begin{proposition}\label{proposition:rotation}
Let $D_{\theta}$ be the differential operator 
$\xi_{2} \partial_{\xi_{1}} - \xi_{1} \partial_{\xi_{2}}$. 
If $c_{1}$ in the above proposition is $0$, 
$D_{\theta} \widetilde{P}_{0}$ is divisible by 
$\xi_{\alpha}$. 
\end{proposition}
\begin{proof} 
For any $v = (v_{1}, v_{2})$, 
$w = (w_{1}, w_{2}) \in \boldsymbol{R}^{2}$, 
we have 
\begin{align}
\xi_{w} \partial_{\xi, v} 
- 
\xi_{v} \partial_{\xi, w} 
&= 
(w_{1} \xi_{1} + w_{2} \xi_{2}) 
(v_{1} \partial_{\xi_{1}} + v_{2} \partial_{\xi_{2}}) 
- 
(v_{1} \xi_{1} + v_{2} \xi_{2}) 
(w_{1} \partial_{\xi_{1}} + w_{2} \partial_{\xi_{2}})
\notag\\
&= (w_{2} v_{1} - w_{1} v_{2}) 
(\xi_{2} \partial_{\xi_{1}} - \xi_{1} \partial_{\xi_{2}}) 
\notag\\
&= \langle v^{\bot}, w \rangle D_{\theta}. 
\label{eq:D_theta} 
\end{align}
Therefore, 
\begin{align*}
D_{\theta} \widetilde{P}_{0} 
&= 
\frac{1}{|\alpha|^{2}} 
(\xi_{\alpha^{\bot}} \partial_{\xi, \alpha} 
- 
\xi_{\alpha} \partial_{\xi, \alpha^{\bot}}) 
\left(
\sum_{k=0}^{m_{0}} c_{k} \xi_{\alpha}^{k} 
\xi_{\alpha^{\bot}}^{m_{0}-k}
\right)
\\
&= \sum_{k=2}^{m_{0}} k c_{k} 
\xi_{\alpha}^{k-1} 
\xi_{\alpha^{\bot}}^{m_{0}-k+1}
-
\sum_{k=0}^{m_{0}} (m_{0} - k) c_{k} 
\xi_{\alpha}^{k+1} 
\xi_{\alpha^{\bot}}^{m_{0}-k-1},
\end{align*}
since $c_{1} = 0$. 
The right hand side is divisible by $\xi_{\alpha}$. 
\end{proof}

\begin{corollary}\label{corollary:order condition} 
If $L$ and $P$ commute, then $D_{\theta} \widetilde{P}_{0}$ is
divisible by $\prod_{\alpha \in \mathcal{H}} \xi_{\alpha}$. 
Therefore, if $\widetilde{P}_{0}$ is not a polynomial 
in $\xi_{1}^{2} + \xi_{2}^{2}$, 
the order of $P$ is not less than the cardinality of $\mathcal{H}$. 
\end{corollary}
\begin{proof}
The first part is a direct consequence of
Proposition~\ref{proposition:review-2}, \ref{proposition:rotation}. 
Since $D_{\theta} \widetilde{P}_{0} = 0$ is equivalent to  
$\widetilde{P}_{0} \in \boldsymbol{C} [\xi_{1}^{2} + \xi_{2}^{2}]$, 
the second assertion follows from the first one. 
\end{proof}



\section{Construction of $\widetilde{P}_{2}$, $\widetilde{P}_{3}$ and
$\widetilde{P}_{4}$} 
\label{section:construction of P2,3,4}

For a differential operator 
$Q = \sum_{p} a_{p}(x) \partial_{x}^{p}$, 
let 
${}^{t} Q$ be the formal adjoint operator 
$\sum_{p}(-\partial_{x})^{p} \circ a_{p}(x)$ of $Q$. 
Since $L$ is formally self-adjoint, 
if $P$ commutes with $L$, so does ${}^{t}P$. 
Therefore, we may assume that $P$ is formally (skew-)self-adjoint,
that is, ${}^{t}P = (-1)^{\mathrm{ord} P} P$. 
\begin{lemma}\label{lemma:order-2}
If $P$ is formally (skew-)self-adjoint, then 
\[
\widetilde{P}_{2k+1} 
= 
\frac{1}{2} 
\sum_{j=1}^{2k+1} 
\frac{(-1)^{j+1}}{j!} 
\langle \partial_{x}, \partial_{\xi} \rangle^{j} 
\widetilde{P}_{2k+1-j}. 
\]
\end{lemma}
\begin{proof}
By the Leibniz rule, we have 
\[
(-1)^{\mathrm{ord} P} \times {}^{t} P 
= 
\sum_{l=0}^{\mathrm{ord} P} 
(-1)^{k} 
\sum_{j=0}^{k} 
\frac{(-1)^{j}}{j!} 
\langle \partial_{x}, \partial_{\xi} \rangle^{j} 
\widetilde{P}_{l-j}. 
\]
The lemma is easily deduced from this equation. 
\end{proof} 

Since $\widetilde{P}_{0}$ is constant with respect to $x$, 
this lemma implies $\widetilde{P}_{1} = 0$. 
By Lemma~\ref{lemma:order-1}, 
$\widetilde{P}_{2}$ satisfies 
\[
2 \langle \xi, \partial_{x} \rangle \widetilde{P}_{2} 
+ 
\sum_{\alpha \in \mathcal{H}} 
u_{\alpha}' 
\partial_{\xi, \alpha} \widetilde{P}_{0}
= 0 
\quad \Leftrightarrow \quad 
\langle \xi, \partial_{x} \rangle 
\left(
\widetilde{P}_{2} 
+ \frac{1}{2} \sum_{\alpha \in \mathcal{H}} 
u_{\alpha} \frac{\partial_{\xi, \alpha} \widetilde{P}_{0}}
{\xi_{\alpha}} 
\right)
= 0. 
\]
Note that Proposition~\ref{proposition:review-2} implies that 
$\partial_{\xi, \alpha} \widetilde{P}_{0} / \xi_{\alpha}$ is 
a polynomial. 

Let $\widetilde{Q} = 
\widetilde{P}_{2} 
+ (1/2) \sum_{\alpha \in \mathcal{H}} 
u_{\alpha} \partial_{\xi, \alpha} \widetilde{P}_{0} / \xi_{\alpha}$.  
It is a polynomial in $\xi$ of degree $m_{0} - 2$ and 
its coefficients are homogeneous rational functions of degree $-2$. 
On the other hand, since $\widetilde{Q}$ satisfies 
$\langle \xi, \partial_{x} \rangle \widetilde{Q} = 0$, 
it is a function in $\xi_{1}, \xi_{2}$ 
and $x_{2} \xi_{1} - x_{1} \xi_{2}$. 
From these conditions, we can conclude $\widetilde{Q} = 0$. 
Therefore, 
\begin{equation}\label{eq:P_2}
\widetilde{P}_{2} 
= 
- \frac{1}{2} 
\sum_{\alpha \in \mathcal{H}} 
u_{\alpha} \widetilde{P}_{2}^{\alpha}, 
\quad 
\widetilde{P}_{2}^{\alpha} 
:= 
\frac{\partial_{\xi, \alpha} \widetilde{P}_{0}}
{\xi_{\alpha}} 
\quad \mbox{and} \quad 
\widetilde{P}_{3} 
= \frac{1}{2} 
\langle \partial_{x}, \partial_{\xi} \rangle \widetilde{P}_{2} 
= 
- \frac{1}{4} 
\sum_{\alpha \in \mathcal{H}} 
u_{\alpha}' \partial_{\xi, \alpha} 
\widetilde{P}_{2}^{\alpha}. 
\end{equation}
By these formulae and Lemma~\ref{lemma:order-1}, 
$\widetilde{P}_{4}$ satisfies the following equation: 
\begin{align}
\langle \xi, \partial_{x} \rangle 
\widetilde{P}_{4} 
&= 
\frac{1}{8} 
\sum_{\alpha \in \mathcal{H}} 
|\alpha|^{2} u_{\alpha}^{(3)} \partial_{\xi, \alpha}
\widetilde{P}_{2}^{\alpha} 
+ \frac{1}{4} 
\sum_{\alpha, \beta \in \mathcal{H}} 
u_{\alpha}' u_{\beta}  
\partial_{\xi, \alpha} \widetilde{P}_{2}^{\beta} 
-\frac{1}{12} 
\sum_{\alpha \in \mathcal{H}} 
u_{\alpha}^{(3)} \partial_{\xi, \alpha}^{3} 
\widetilde{P}_{0}
\notag\\
& = 
\langle \xi, \partial_{x} \rangle 
\sum_{\alpha \in \mathcal{H}} 
\left(
(u_{\alpha}^{2} + |\alpha|^{2} u_{\alpha}'') 
\frac{\partial_{\xi, \alpha} \widetilde{P}_{2}^{\alpha}}
{8 \xi_{\alpha}} 
- 
u_{\alpha}'' 
\frac{\partial_{\xi, \alpha}^{3} \widetilde{P}_{0}}
{12 \xi_{\alpha}}
\right) 
+ \frac{1}{4} 
\sum_{\alpha, \beta \in \mathcal{H} \atop \alpha \not= \beta} 
u_{\alpha}' u_{\beta} \partial_{\xi, \alpha} 
\widetilde{P}_{2}^{\beta}. 
\label{eq:P4 cross}
\end{align}
Now, since $u_{\alpha}(t) = C_{\alpha}/t^{2}$, 
we have $u_{\alpha}(t)^{2} = C_{\alpha} u_{\alpha}''(t)/6$ and 
\begin{align}
\sum_{\alpha \in \mathcal{H}} 
& \left(
(u_{\alpha}^{2} + |\alpha|^{2} u_{\alpha}'') 
\frac{\partial_{\xi, \alpha} \widetilde{P}_{2}^{\alpha}}
{8 \xi_{\alpha}} 
- 
u_{\alpha}'' 
\frac{\partial_{\xi, \alpha}^{3} \widetilde{P}_{0}}
{12 \xi_{\alpha}}
\right) 
= 
\frac{1}{48} 
\sum_{\alpha \in \mathcal{H}} 
u_{\alpha}'' 
\widetilde{P}_{4}^{\alpha} 
\notag
\\
& \mbox{where } 
\widetilde{P}_{4}^{\alpha} 
= \frac{(C_{\alpha} + 6 |\alpha|^{2}) 
\partial_{\xi, \alpha} \widetilde{P}_{2}^{\alpha} 
- 4 \partial_{\xi, \alpha}^{3} \widetilde{P}_{0}}
{\xi_{\alpha}}. 
\notag
\end{align}
By Proposition~\ref{proposition:review-2}, 
the coefficients $c_{1}, c_{3}$ in the expression 
$\widetilde{P}_{0} 
= \sum_{k=0}^{m_{0}} 
c_{k} \xi_{\alpha}^{k} 
\xi_{\alpha^{\bot}}^{m_{0}-k}$ 
satisfy $c_{1} = c_{3} (C_{\alpha} - 2 |\alpha|^{2}) = 0$. 
Therefore, $\widetilde{P}_{4}^{\alpha}$ is a polynomial in $\xi$,
since 
\begin{align*}
(C_{\alpha} + 6 |\alpha|^{2}) 
\partial_{\xi, \alpha} \widetilde{P}_{2}^{\alpha} 
- 4 \partial_{\xi, \alpha}^{3} \widetilde{P}_{0} 
&= 
|\alpha|^{4} 
\sum_{k=0}^{m_{0}} 
k(k-2) 
\{C_{\alpha} + (10 - 4k) |\alpha|^{2}\} c_{k} \xi_{\alpha}^{k-3}
\xi_{\alpha^{\bot}}^{m_{0}-k} 
\\
&= 
|\alpha|^{4} 
\sum_{k=4}^{m_{0}} 
k(k-2) 
\{C_{\alpha} + (10 - 4k) |\alpha|^{2}\} c_{k} \xi_{\alpha}^{k-3}
\xi_{\alpha^{\bot}}^{m_{0}-k}, 
\end{align*}
is divisible by $\xi_{\alpha}$. 
Moreover, the last term in \eqref{eq:P4 cross} is expressed as 
\begin{equation}\label{eq:compatibility for P2}
\langle \xi, \partial_{x} \rangle 
F(x, \xi) 
= 
\frac{1}{4} 
\sum_{\alpha, \beta \in \mathcal{H} \atop \alpha \not= \beta} 
u_{\alpha}' u_{\beta} \partial_{\xi, \alpha} 
\widetilde{P}_{2}^{\beta}, 
\quad 
F(x, \xi) 
:= \widetilde{P}_{4} 
- \frac{1}{48} 
\sum_{\alpha \in \mathcal{H}} 
u_{\alpha}'' \widetilde{P}_{4}^{\alpha}.  
\end{equation}
Here, $F(x, \xi)$ is a polynomial in $\xi$ and a meromorphic function
in $x$ with poles along $x_{\alpha} = 0$ of order at most two 
for each $\alpha \in \mathcal{H}$. 
Therefore, we have 
\begin{align}
& \lim_{x_{\alpha_{0}} \to 0} 
(x_{\alpha_{0}} \langle \xi, \partial_{x} \rangle 
+ 2 \xi_{\alpha_{0}}) 
(x_{\alpha_{0}} \langle \xi, \partial_{x} \rangle 
+ \xi_{\alpha_{0}}) 
(x_{\alpha} \langle \xi, \partial_{x} \rangle) 
F(x, \xi)
= 0 
\notag \\
& \quad \Leftrightarrow \quad 
\lim_{x_{\alpha_{0}} \to 0} 
\langle \xi, \partial_{x} \rangle^{2} 
x_{\alpha_{0}}^{3} 
\sum_{\alpha, \beta \in \mathcal{H} \atop \alpha \not= \beta} 
u_{\alpha}' u_{\beta} \partial_{\xi, \alpha} 
\widetilde{P}_{2}^{\beta} 
= 0 
\label{eq:pole condition for P4}
\end{align}
for each $\alpha_{0} \in \mathcal{H}$. 
Here, $\lim_{x_{\alpha_{0}} \to 0} G(x)$ is the limit 
$\lim_{x_{\alpha_{0}} \to 0} 
\hat{G}(x_{\alpha_{0}}, x_{\alpha_{0}^{\bot}})$, 
where $\hat{G}(x_{\alpha_{0}}, x_{\alpha_{0}^{\bot}})$ is the
expression of $G(x)$ as a function of  $x_{\alpha_{0}}$,
$x_{\alpha_{0}^{\bot}}$. 

\begin{theorem}\label{theorem:linear relation-1} 
If $\widetilde{P}_{0}$ is not a polynomial 
in $\xi_{1}^{2} + \xi_{2}^{2}$, 
then, for each $\alpha_{0} \in \mathcal{H}$, 
\begin{equation}\label{eq:first relation}
\sum_{\beta \in \mathcal{H} \atop \beta \not= \alpha_{0}}
\frac{\langle \alpha_{0}, \beta \rangle }
{\langle \alpha_{0}^{\bot}, \beta \rangle^{3}} 
C_{\beta}
= 0 
\end{equation}
holds. 
\end{theorem}
\begin{proof}
Since $u_{\alpha}(t) = C_{\alpha}/t^{2}$ and 
\begin{align*}
\lim_{x_{\alpha_{0}} \to 0} 
u_{\beta}'' 
&= 
\lim_{x_{\alpha_{0}} \to 0} 
(-2)(-3) C_{\beta} 
\left(
\frac{\langle \alpha_{0}, \beta \rangle x_{\alpha_{0}} 
+ 
\langle \alpha_{0}^{\bot}, \beta \rangle x_{\alpha_{0}^{\bot}}}
{|\alpha_{0}|^{2}}
\right)^{-4} 
= 
\frac{6 C_{\beta} |\alpha_{0}|^{8}}
{\langle \alpha_{0}^{\bot}, \beta \rangle^{4} 
x_{\alpha_{0}^{\bot}}^{4}},
\end{align*}
we have 
\begin{align*}
& 
\lim_{x_{\alpha_{0}} \to 0} 
\langle \xi, \partial_{x} \rangle^{2} 
x_{\alpha_{0}}^{3} 
\sum_{\alpha, \beta \in \mathcal{H} \atop \alpha \not= \beta} 
u_{\alpha}' u_{\beta} \partial_{\xi, \alpha} 
\widetilde{P}_{2}^{\beta} 
= 0 
\\
& \quad \Leftrightarrow \quad 
\lim_{x_{\alpha_{0}} \to 0} 
\langle \xi, \partial_{x} \rangle^{2} 
x_{\alpha_{0}}^{3} 
\sum_{\beta \in \mathcal{H} \atop \beta \not= \alpha_{0}} 
C_{\alpha_{0}} 
\left(-2 x_{\alpha_{0}}^{-3} u_{\beta} 
\partial_{\xi, \alpha_{0}} \widetilde{P}_{2}^{\beta} 
+ x_{\alpha_{0}}^{-2} u_{\beta}' 
\partial_{\xi, \beta} \widetilde{P}_{2}^{\alpha_{0}} 
\right)
= 0
\\
& \quad \Leftrightarrow \quad 
\lim_{x_{\alpha_{0}} \to 0} 
C_{\alpha_{0}} 
\sum_{\beta \in \mathcal{H} \atop \beta \not= \alpha_{0}} 
\left(
-2 u_{\beta}'' \xi_{\beta}^{2} 
\partial_{\xi, \alpha_{0}} \widetilde{P}_{2}^{\beta} 
+ 2 u_{\beta}'' \xi_{\alpha_{0}} \xi_{\beta} 
\partial_{\xi, \beta} \widetilde{P}_{2}^{\alpha_{0}} 
\right)
= 0
\\
& \quad \Leftrightarrow \quad 
C_{\alpha_{0}} 
\sum_{\beta \in \mathcal{H} \atop \beta \not= \alpha_{0}} 
\frac{C_{\beta}}{\langle \alpha_{0}^{\bot}, \beta \rangle^{4}} 
\xi_{\beta} 
(\xi_{\beta} \partial_{\xi, \alpha_{0}} \widetilde{P}_{2}^{\beta} 
- 
\xi_{\alpha_{0}} \partial_{\xi, \beta} \widetilde{P}_{2}^{\alpha_{0}}) 
= 0 
\\ 
& \quad \Leftrightarrow \quad 
C_{\alpha_{0}} 
\sum_{\beta \in \mathcal{H} \atop \beta \not= \alpha_{0}} 
\frac{\langle \alpha_{0}, \beta \rangle C_{\beta}}
{\langle \alpha_{0}^{\bot}, \beta \rangle^{4}} 
\xi_{\beta} 
(\widetilde{P}_{2}^{\alpha_{0}} - \widetilde{P}_{2}^{\beta}) 
= 0 
\\ 
& \quad \Leftrightarrow \quad 
C_{\alpha_{0}} 
\sum_{\beta \in \mathcal{H} \atop \beta \not= \alpha_{0}} 
\frac{\langle \alpha_{0}, \beta \rangle C_{\beta}}
{\langle \alpha_{0}^{\bot}, \beta \rangle^{4}} 
\frac{\xi_{\beta} \partial_{\xi, \alpha_{0}} \widetilde{P}_{0}  
- 
\xi_{\alpha_{0}} \partial_{\xi, \beta} \widetilde{P}_{0}}
{\xi_{\alpha_{0}}} 
= 0 
\\ 
& \quad \Leftrightarrow \quad 
C_{\alpha_{0}} 
\sum_{\beta \in \mathcal{H} \atop \beta \not= \alpha_{0}} 
\frac{\langle \alpha_{0}, \beta \rangle C_{\beta}}
{\langle \alpha_{0}^{\bot}, \beta \rangle^{3}} 
\frac{D_{\theta} \widetilde{P}_{0}}
{\xi_{\alpha_{0}}} 
= 0. 
\end{align*}
Here, we used 
$\xi_{\beta} \partial_{\xi, \alpha} \widetilde{P}_{2}^{\beta} 
= 
\xi_{\beta} 
\partial_{\xi, \alpha} 
(\partial_{\xi, \beta} \widetilde{P}_{0} / \xi_{\beta})  
= 
\partial_{\xi, \alpha} \partial_{\xi, \beta} \widetilde{P}_{0}
- \langle \alpha, \beta \rangle \widetilde{P}_{2}^{\beta}$
and \eqref{eq:D_theta}. 
\end{proof}

\begin{corollary}\label{cor:2-sheets} 
Under the assumption of Theorem~\ref{theorem:linear relation-1}, 
if $\mathcal{H}$ consists of two vectors, 
then they cross at right angles and the singular locus is of
type $A_{1} \times A_{1}$. 
\end{corollary}
\begin{proof}
Let $\mathcal{H} = \{\alpha, \beta\}$. 
Then by Theorem~\ref{theorem:linear relation-1}, 
we have 
$\langle \alpha, \beta \rangle C_{\beta} / 
\langle \alpha^{\bot}, \beta \rangle^{3} 
= 0$. 
Since $C_{\beta} \not= 0$, 
this implies $\langle \alpha, \beta \rangle = 0$. 
\end{proof}

Next, let us consider the last term in \eqref{eq:P4 cross}. 
Let $N := \# \mathcal{H}$ and 
\begin{align*}
d_{\alpha, \beta, \gamma} 
& := 
\frac{\langle \beta, \gamma \rangle}
{\langle \beta^{\bot}, \gamma \rangle^{3} C_{\alpha}} 
+ 
\frac{\langle \gamma, \alpha\rangle}
{\langle \gamma^{\bot}, \alpha \rangle^{3} C_{\beta}} 
+ 
\frac{\langle \alpha, \beta \rangle}
{\langle \alpha^{\bot}, \beta \rangle^{3} C_{\gamma}}.  
\end{align*}
\begin{lemma}\label{lemma:d_abc} 
The constant $d_{\alpha, \beta, \gamma}$ is skew-symmetric with
respect to $\alpha$, $\beta$, $\gamma$ and satisfies 
\begin{equation}\label{eq:d_abc} 
\sum_{\gamma \in \mathcal{H} \atop \gamma \not= \alpha, \beta} 
C_{\gamma} d_{\alpha, \beta, \gamma} 
= N 
\frac{\langle \alpha, \beta \rangle}{
\langle \alpha^{\bot}, \beta \rangle^{3}}.
\end{equation}
\end{lemma}
\begin{proof}
The first statement follows from 
$\langle \alpha^{\bot}, \beta \rangle 
= - \langle \beta^{\bot}, \alpha \rangle$. 

The second statement is a consequence of \eqref{eq:first relation}: 
\begin{align*}
\sum_{\gamma \in \mathcal{H} \atop \gamma \not= \alpha, \beta} 
C_{\gamma} d_{\alpha, \beta, \gamma} 
&= 
\sum_{\gamma \in \mathcal{H} \atop \gamma \not= \alpha, \beta} 
\left(
\frac{1}{C_{\alpha}} 
\frac{\langle \beta, \gamma \rangle C_{\gamma}}
{\langle \beta^{\bot}, \gamma \rangle^{3}} 
+ 
\frac{1}{C_{\beta}} 
\frac{\langle \gamma, \alpha\rangle C_{\gamma}}
{\langle \gamma^{\bot}, \alpha \rangle^{3}} 
+ 
\frac{\langle \alpha, \beta \rangle}
{\langle \alpha^{\bot}, \beta \rangle^{3}} 
\right)
\\&= 
-\frac{1}{C_{\alpha}} 
\frac{\langle \beta, \alpha \rangle C_{\alpha}}
{\langle \beta^{\bot}, \alpha \rangle^{3}} 
- 
\frac{1}{C_{\beta}} 
\frac{\langle \beta, \alpha \rangle C_{\beta}}
{\langle \beta^{\bot}, \alpha \rangle^{3}} 
+ (N-2)  
\frac{\langle \alpha, \beta \rangle}
{\langle \alpha^{\bot}, \beta \rangle^{3}} 
\\
&= 
N
\frac{\langle \alpha, \beta \rangle}
{\langle \alpha^{\bot}, \beta \rangle^{3}}. 
\end{align*}
\end{proof}

Since 
$u_{\alpha}(t) = C_{\alpha}/t^{2}$ and  
\begin{equation}
\label{eq:triangle}
\langle \beta^{\bot}, \gamma \rangle \alpha 
+ 
\langle \gamma^{\bot}, \alpha \rangle \beta 
+ 
\langle \alpha^{\bot}, \beta \rangle \gamma 
= 0
\end{equation}
for any $\alpha$, $\beta$, $\gamma \in \mathcal{H}$, 
we have 
\begin{equation}
\label{eq:pe-relation-1}
\begin{vmatrix}
\langle \beta^{\bot}, \gamma \rangle^{3} C_{\alpha} 
& 
\langle \gamma^{\bot}, \alpha \rangle^{3} C_{\beta}
& 
\langle \alpha^{\bot}, \beta \rangle^{3} C_{\gamma} 
\\
\langle \beta^{\bot}, \gamma \rangle u_{\alpha} 
& 
\langle \gamma^{\bot}, \alpha \rangle u_{\beta}
& 
\langle \alpha^{\bot}, \beta \rangle u_{\gamma} 
\\
u_{\alpha}' 
& 
u_{\beta}'
& 
u_{\gamma}' 
\end{vmatrix} 
= 0. 
\end{equation}
By Corollary~\ref{corollary:order condition}, 
$D_{\theta} \widetilde{P}_{0} / \xi_{\alpha} \xi_{\beta} \xi_{\gamma}$ 
is a polynomial in $\xi$ and it is symmetric with respect to $\alpha,
\beta, \gamma$. 
Then by \eqref{eq:pe-relation-1}, we have 
\begin{equation}
\label{eq:pe-relation-2} 
\sum_{\alpha, \beta, \gamma \in \mathcal{H} \atop 
\alpha \not= \beta \not= \gamma \not= \alpha} 
\langle \alpha^{\bot}, \beta \rangle^{3} 
\langle \gamma^{\bot}, \alpha \rangle 
C_{\gamma} d_{\alpha, \beta, \gamma} 
\frac{D_{\theta} \widetilde{P}_{0}}
{\xi_{\alpha} \xi_{\beta} \xi_{\gamma}}
u_{\alpha}' u_{\beta} 
= 0. 
\end{equation}

\begin{lemma}\label{lemma:P4ab another} 
For $\alpha, \beta \in \mathcal{H}$, let 
\begin{align}
\widetilde{P}_{4}^{\alpha, \beta} 
&= 
-\frac{
\langle \gamma^{\bot}, \alpha \rangle 
\partial_{\xi, \beta} \widetilde{P}_{2}^{\alpha}
+
\langle \beta^{\bot}, \gamma \rangle 
\partial_{\xi, \alpha} \widetilde{P}_{2}^{\beta}}
{\langle \alpha^{\bot}, \beta \rangle \xi_{\gamma}} 
+ 
\frac{\langle \alpha^{\bot}, \beta \rangle^{3}}{N} 
\sum_{\delta \in \mathcal{H} \atop \delta \not= \alpha, \beta, \gamma} 
\langle \delta^{\bot}, \gamma \rangle C_{\delta} 
d_{\alpha, \beta, \delta} 
\frac{D_{\theta} \widetilde{P}_{0}}
{\xi_{\alpha} \xi_{\beta} \xi_{\gamma} \xi_{\delta}}, 
\label{eq:P4ab} 
\end{align}
where $\gamma$ is any vector in $\mathcal{H}$ 
other than $\alpha, \beta$. 
Then, $\widetilde{P}_{4}^{\alpha, \beta}$ is a polynomial 
in $\xi$ and satisfies 
\begin{equation}\label{eq:P4ab another}
\xi_{\alpha} \widetilde{P}_{4}^{\alpha, \beta} 
= 
\partial_{\xi, \alpha} \widetilde{P}_{2}^{\beta} 
+ 
\frac{\langle \alpha^{\bot}, \beta \rangle^{3}}{N} 
\sum_{\delta \in \mathcal{H} \atop \delta \not= \alpha, \beta} 
\langle \delta^{\bot}, \alpha \rangle C_{\delta} 
d_{\alpha, \beta, \delta} 
\frac{D_{\theta} \widetilde{P}_{0}}
{\xi_{\alpha} \xi_{\beta} \xi_{\delta}}. 
\end{equation}
\end{lemma}
\begin{proof}
By Corollary~\ref{corollary:order condition}, 
$D_{\theta} \widetilde{P}_{0}/
\xi_{\alpha} \xi_{\beta} \xi_{\gamma} \xi_{\delta}$ is a polynomial. 
Therefore, to prove $\widetilde{P}_{4}^{\alpha, \beta}$ being a
polynomial, we have only to show that 
$\langle \gamma^{\bot}, \alpha \rangle 
\partial_{\xi, \beta} \widetilde{P}_{2}^{\alpha}
+
\langle \beta^{\bot}, \gamma \rangle 
\partial_{\xi, \alpha} \widetilde{P}_{2}^{\beta}$ is divisible by
$\xi_{\gamma}$. 
Since 
\begin{align*}
& 
\xi_{\alpha} \xi_{\beta} 
(\langle \gamma^{\bot}, \alpha \rangle 
\partial_{\xi, \beta} \widetilde{P}_{2}^{\alpha}
+
\langle \beta^{\bot}, \gamma \rangle 
\partial_{\xi, \alpha} \widetilde{P}_{2}^{\beta}) 
+ \langle \alpha^{\bot}, \beta \rangle \xi_{\gamma} 
\partial_{\xi, \alpha} \partial_{\xi, \beta} \widetilde{P}_{0} 
\\
&= 
\langle \gamma^{\bot}, \alpha \rangle \xi_{\beta} 
(\partial_{\xi, \beta} \partial_{\xi, \alpha} \widetilde{P}_{0} 
- 
\langle \alpha, \beta \rangle \widetilde{P}_{2}^{\alpha}) 
+ 
\langle \beta^{\bot}, \gamma \rangle \xi_{\alpha} 
(\partial_{\xi, \alpha} \partial_{\xi, \beta} \widetilde{P}_{0} 
- 
\langle \alpha, \beta \rangle \widetilde{P}_{2}^{\beta}) 
+ \langle \alpha^{\bot}, \beta \rangle \xi_{\gamma} 
\partial_{\xi, \alpha} \partial_{\xi, \beta} \widetilde{P}_{0} 
\\
&= 
(\langle \gamma^{\bot}, \alpha \rangle \xi_{\beta} 
+ \langle \beta^{\bot}, \gamma \rangle \xi_{\alpha} 
+ \langle \alpha^{\bot}, \beta \rangle \xi_{\gamma}) 
\partial_{\xi, \alpha} \partial_{\xi, \beta} \widetilde{P}_{0} 
- \langle \alpha, \beta \rangle 
(\langle \gamma^{\bot}, \alpha \rangle \xi_{\beta} 
\widetilde{P}_{2}^{\alpha} 
+ 
\langle \beta^{\bot}, \gamma \rangle \xi_{\alpha} 
\widetilde{P}_{2}^{\beta}) 
\\
&= 
\langle \alpha, \beta \rangle 
\{
(\langle \alpha^{\bot}, \beta \rangle \xi_{\gamma} 
+ \langle \beta^{\bot}, \gamma \rangle \xi_{\alpha}) 
\widetilde{P}_{2}^{\alpha} 
+ 
(\langle \gamma^{\bot}, \alpha \rangle \xi_{\beta} 
+ \langle \alpha^{\bot}, \beta \rangle \xi_{\gamma}) 
\widetilde{P}_{2}^{\beta} 
\}
\\
&= 
\langle \alpha, \beta \rangle 
\langle \alpha^{\bot}, \beta \rangle \xi_{\gamma} 
(\widetilde{P}_{2}^{\alpha} + \widetilde{P}_{2}^{\beta}) 
+ 
\langle \alpha, \beta \rangle 
(\langle \beta^{\bot}, \gamma \rangle \partial_{\xi, \alpha} 
+ \langle \gamma^{\bot}, \alpha \rangle \partial_{\xi, \beta}) 
\widetilde{P}_{0} 
\\
&= 
\langle \alpha, \beta \rangle 
\langle \alpha^{\bot}, \beta \rangle \xi_{\gamma} 
(\widetilde{P}_{2}^{\alpha} + \widetilde{P}_{2}^{\beta} 
- \widetilde{P}_{2}^{\gamma}), 
\end{align*}
we have 
\[
\langle \gamma^{\bot}, \alpha \rangle 
\partial_{\xi, \beta} \widetilde{P}_{2}^{\alpha}
+
\langle \beta^{\bot}, \gamma \rangle 
\partial_{\xi, \alpha} \widetilde{P}_{2}^{\beta}
= \frac{\langle \alpha^{\bot}, \beta \rangle \xi_{\gamma}}
{\xi_{\alpha} \xi_{\beta}} 
\{-\partial_{\xi, \alpha} \partial_{\xi, \beta} \widetilde{P}_{0} 
+ \langle \alpha, \beta \rangle 
(\widetilde{P}_{2}^{\alpha} + \widetilde{P}_{2}^{\beta} 
- \widetilde{P}_{2}^{\gamma})\}.
\]
By the uniqueness of factorization, 
this is divisible by $\xi_{\gamma}$. 

Let us prove \eqref{eq:P4ab another}. 
Firstly, we have  
\begin{align*}
- \xi_{\alpha} 
& 
\frac{
\langle \gamma^{\bot}, \alpha \rangle 
\partial_{\xi, \beta} \widetilde{P}_{2}^{\alpha}
+
\langle \beta^{\bot}, \gamma \rangle 
\partial_{\xi, \alpha} \widetilde{P}_{2}^{\beta}}
{\langle \alpha^{\bot}, \beta \rangle \xi_{\gamma}} 
- \partial_{\xi, \alpha} \widetilde{P}_{2}^{\beta} 
\\
&= 
-\frac{
\langle \gamma^{\bot}, \alpha \rangle \xi_{\alpha} 
\partial_{\xi, \beta} \widetilde{P}_{2}^{\alpha}
+
(\langle \beta^{\bot}, \gamma \rangle \xi_{\alpha} 
+ 
\langle \alpha^{\bot}, \beta \rangle \xi_{\gamma}) 
\partial_{\xi, \alpha} \widetilde{P}_{2}^{\beta}}
{\langle \alpha^{\bot}, \beta \rangle \xi_{\gamma}} 
\\
&= 
-\frac{
\langle \gamma^{\bot}, \alpha \rangle 
(\xi_{\alpha} 
\partial_{\xi, \beta} \widetilde{P}_{2}^{\alpha}
- \xi_{\beta} 
\partial_{\xi, \alpha} \widetilde{P}_{2}^{\beta})}
{\langle \alpha^{\bot}, \beta \rangle \xi_{\gamma}} 
\\
&= 
\langle \gamma^{\bot}, \alpha \rangle 
\langle \alpha, \beta \rangle 
\frac{D_{\theta} \widetilde{P}_{0}}
{\xi_{\alpha} \xi_{\beta} \xi_{\gamma}}. 
\end{align*}
Here, we used \eqref{eq:triangle} and calculated as in the proof of
Theorem~\ref{theorem:linear relation-1}. 

Secondly, 
since 
\[
\frac{\langle \delta^{\bot}, \gamma \rangle}
{\xi_{\beta} \xi_{\gamma} \xi_{\delta}} 
- 
\frac{\langle \delta^{\bot}, \alpha \rangle}
{\xi_{\alpha} \xi_{\beta} \xi_{\delta}} 
= 
\frac{\langle \delta^{\bot}, \gamma \rangle \xi_{\alpha} 
+ 
\langle \alpha^{\bot}, \delta \rangle \xi_{\gamma}}
{\xi_{\alpha} \xi_{\beta} \xi_{\gamma} \xi_{\delta}} 
= 
- \frac{\langle \gamma^{\bot}, \alpha \rangle}
{\xi_{\alpha} \xi_{\beta} \xi_{\gamma}}, 
\]
we have 
\begin{align*}
& 
\frac{\langle \alpha^{\bot}, \beta \rangle^{3}}{N} 
\left(
\xi_{\alpha} 
\sum_{\delta \in \mathcal{H} \atop \delta \not= \alpha, \beta, \gamma} 
\langle \delta^{\bot}, \gamma \rangle C_{\delta} 
d_{\alpha, \beta, \delta} 
\frac{D_{\theta} \widetilde{P}_{0}}
{\xi_{\alpha} \xi_{\beta} \xi_{\gamma} \xi_{\delta}} 
- 
\sum_{\delta \in \mathcal{H} \atop \delta \not= \alpha, \beta} 
\langle \delta^{\bot}, \alpha \rangle C_{\delta} 
d_{\alpha, \beta, \delta} 
\frac{D_{\theta} \widetilde{P}_{0}}
{\xi_{\alpha} \xi_{\beta} \xi_{\delta}} 
\right) 
\\
&= 
\frac{\langle \alpha^{\bot}, \beta \rangle^{3}}{N} 
\left(
- \frac{\langle \gamma^{\bot}, \alpha \rangle 
D_{\theta} \widetilde{P}_{0}}
{\xi_{\alpha} \xi_{\beta} \xi_{\gamma}} 
\sum_{\delta \in \mathcal{H} \atop \delta \not= \alpha, \beta, \gamma} 
C_{\delta} d_{\alpha, \beta, \delta} 
- 
\langle \gamma^{\bot}, \alpha \rangle C_{\gamma} 
d_{\alpha, \beta, \gamma} 
\frac{D_{\theta} \widetilde{P}_{0}}
{\xi_{\alpha} \xi_{\beta} \xi_{\gamma}} 
\right) 
\\
&= 
\frac{\langle \alpha^{\bot}, \beta \rangle^{3}}{N} 
\left(
- \frac{\langle \gamma^{\bot}, \alpha \rangle 
D_{\theta} \widetilde{P}_{0}}
{\xi_{\alpha} \xi_{\beta} \xi_{\gamma}} 
\right)
N 
\frac{\langle \alpha, \beta \rangle}{
\langle \alpha^{\bot}, \beta \rangle^{3}} 
\\
&= 
-\langle \gamma^{\bot}, \alpha \rangle 
\langle \alpha, \beta \rangle 
\frac{D_{\theta} \widetilde{P}_{0}}
{\xi_{\alpha} \xi_{\beta} \xi_{\gamma}}, 
\end{align*}
by \eqref{eq:d_abc}. 
Therefore, \eqref{eq:P4ab another} is proved. 
\end{proof}

\begin{proposition}\label{proposition:P4ab} 
$\widetilde{P}_{4}^{\alpha, \beta}$
satisfies 
\begin{equation}\label{eq:determination of P4ab} 
\sum_{\alpha, \beta \in \mathcal{H} \atop \alpha \not= \beta} 
u_{\alpha}' u_{\beta} \partial_{\xi, \alpha} 
\widetilde{P}_{2}^{\beta} 
= 
\langle \xi, \partial_{x} \rangle 
\left(
\sum_{\{\alpha, \beta\} \subset \mathcal{H} \atop \alpha \not= \beta} 
u_{\alpha} u_{\beta} \widetilde{P}_{4}^{\alpha, \beta} 
\right). 
\end{equation}
\end{proposition}
\begin{proof} 
By Lemma~\ref{lemma:P4ab another} and \eqref{eq:pe-relation-2}, 
we have 
\begin{align*}
\langle \xi, \partial_{x} \rangle 
& 
\left(
\sum_{\{\alpha, \beta\} \subset \mathcal{H} 
\atop \alpha \not= \beta} 
u_{\alpha} u_{\beta} \widetilde{P}_{4}^{\alpha, \beta} 
\right)
- 
\sum_{\alpha, \beta \in \mathcal{H} \atop \alpha \not= \beta} 
u_{\alpha}' u_{\beta} \partial_{\xi, \alpha} 
\widetilde{P}_{2}^{\beta} 
\\
&= 
\sum_{\alpha, \beta \in \mathcal{H} \atop \alpha \not= \beta} 
u_{\alpha}' u_{\beta} 
\left(\xi_{\alpha} \widetilde{P}_{4}^{\alpha, \beta} 
- \partial_{\xi, \alpha} \widetilde{P}_{2}^{\beta} 
\right)
\\
&= 
\sum_{\alpha, \beta, \delta \in \mathcal{H} 
\atop \alpha \not= \beta \not= \delta \not= \alpha} 
u_{\alpha}' u_{\beta} 
\frac{\langle \alpha^{\bot}, \beta \rangle^{3}}{N} 
\langle \delta^{\bot}, \alpha \rangle C_{\delta} 
d_{\alpha, \beta, \delta} 
\frac{D_{\theta} \widetilde{P}_{0}}
{\xi_{\alpha} \xi_{\beta} \xi_{\delta}}
\\
&= 0. 
\end{align*}
\end{proof}

Putting together these results, 
we obtain the explicit expression of $\widetilde{P}_{4}$. 
\begin{proposition}\label{proposition:explicit form of P4} 
$\widetilde{P}_{4}$ is expressed as 
\[
\widetilde{P}_{4} 
= 
\frac{1}{48} 
\sum_{\alpha \in \mathcal{H}} 
u_{\alpha}'' \widetilde{P}_{4}^{\alpha} 
+ 
\frac{1}{4} 
\sum_{\{\alpha, \beta \} \subset \mathcal{H} 
\atop \alpha \not= \beta} 
u_{\alpha} u_{\beta} \widetilde{P}_{4}^{\alpha, \beta}. 
\]
\end{proposition}
Moreover, we can write down $\widetilde{P}_{5}$ explicitly by using
Lemma~\ref{lemma:order-2}.



\section{Second condition for $\mathcal{H}$ and $C_{\alpha}$} 
\label{section:second condition} 

In the previous section, 
we obtained a condition \eqref{eq:first relation} for $\mathcal{H}$
and $C_{\alpha}$ by investigating the pole of $\widetilde{P}_{4}$
at $x_{\alpha_{0}} = 0$. 
In this section, we investigate the pole of $\widetilde{P}_{6}$ at
$x_{\alpha_{0}} = 0$ and obtain another condition for $\mathcal{H}$
and $C_{\alpha}$.

\begin{lemma}\label{lemma:basic equation for P6} 
Let 
\[
\widetilde{P}_{6}^{\alpha} 
:= 
\frac{1}{\xi_{\alpha}} 
\{
24 \partial_{\xi, \alpha}^{5} \widetilde{P}_{0} 
+ (C_{\alpha} - 90 |\alpha|^{2}) 
\partial_{\xi, \alpha}^{3} \widetilde{P}_{2}^{\alpha} 
+ C_{\alpha} \partial_{\xi, \alpha}
\widetilde{P}_{4}^{\alpha} 
\}.
\]
Then, $\widetilde{P}_{6}^{\alpha}$ is a polynomial in $\xi$
and $\widetilde{P}_{6}$ satisfies the following equation: 
\begin{align}
\langle \xi, \partial_{x} \rangle \widetilde{P}_{6} 
&= 
\langle \xi, \partial_{x} \rangle 
\left(
- \frac{1}{8} \langle \partial_{\xi}, \partial_{x} \rangle^{2} 
\widetilde{P}_{4} 
+ \frac{1}{2} \langle \partial_{\xi}, \partial_{x} \rangle
\widetilde{P}_{5} 
- \frac{1}{5760} 
\sum_{\alpha \in \mathcal{H}} u_{\alpha}^{(4)} 
\widetilde{P}_{6}^{\alpha} 
\right) 
\label{eq:compatibility for P4} 
\\
& \quad 
+ \frac{1}{96} 
\sum_{\alpha, \beta \in \mathcal{H} \atop \alpha \not= \beta} 
\{
u_{\alpha}^{(3)} u_{\beta} 
(\partial_{\xi, \alpha}^{3} \widetilde{P}_{2}^{\beta} 
- C_{\alpha} \partial_{\xi, \alpha} 
\widetilde{P}_{4}^{\alpha, \beta}) 
- 
u_{\alpha}'' u_{\beta}' 
(3 \partial_{\xi, \alpha}^{2} \partial_{\xi, \beta} 
\widetilde{P}_{2}^{\alpha} 
+
\partial_{\xi, \beta} \widetilde{P}_{4}^{\alpha}) 
\}
\notag\\
& \quad 
-\frac{1}{8}  
\sum_{\alpha \in \mathcal{H}} 
\sum_{\{\beta, \gamma \} \subset \mathcal{H} \atop 
\alpha \not= \beta \not= \gamma \not= \alpha} 
u_{\alpha}' u_{\beta} u_{\gamma} 
\partial_{\xi, \alpha} \widetilde{P}_{4}^{\beta, \gamma}. 
\notag
\end{align}
\end{lemma} 
\begin{proof}
Let us consider the expression 
$\widetilde{P}_{0} 
= \sum_{k=0}^{m_{0}} 
c_{k} \xi_{\alpha}^{k} 
\xi_{\alpha^{\bot}}^{m_{0}-k}$. 
For this expression, we have 
\[
\widetilde{P}_{6}^{\alpha} 
= 
15 |\alpha|^{6} 
c_{5} (C_{\alpha} - 2 |\alpha|^{2}) (C_{\alpha} - 6 |\alpha|^{2}) 
\frac{\xi_{\alpha^{\bot}}^{m_{0} - 5}}{\xi_{\alpha}} 
+ \mbox{(a polynomial in $\xi$)}. 
\]
Since $c_{5}$ satisfies $c_{5} (C_{\alpha} - 2 |\alpha|^{2}) 
(C_{\alpha} - 6 |\alpha|^{2}) = 0$ 
by Proposition~\ref{proposition:review-2}, 
$\widetilde{P}_{6}^{\alpha}$ is a polynomial in $\xi$. 
The second assertion is a consequence of Lemma~\ref{lemma:order-1},
\eqref{eq:P_2}, Proposition~\ref{proposition:P4ab}, 
$\Delta = 
[
\langle \partial_{x}, \partial_{\xi} \rangle, 
\langle \xi, \partial_{x} \rangle]$ 
and 
$\langle \partial_{x}, \partial_{\xi} \rangle \Delta 
= 
(1/2)
[
\langle \partial_{x}, \partial_{\xi} \rangle^{2}, 
\langle \xi, \partial_{x} \rangle]$. 
\end{proof}

We obtained \eqref{eq:pole condition for P4} by investigating 
the pole of $\widetilde{P}_{4}$ at $x_{\alpha_{0}} = 0$. 
In the same way, we obtain 
\begin{align*}
\lim_{x_{\alpha_{0}} \to 0} 
&
\langle \xi, \partial_{x} \rangle^{4} x_{\alpha_{0}}^{5} 
\left(
\sum_{\alpha, \beta \in \mathcal{H} \atop \alpha \not= \beta} 
\left\{
u_{\alpha}^{(3)} u_{\beta} 
(\partial_{\xi, \alpha}^{3} \widetilde{P}_{2}^{\beta} 
- C_{\alpha} \partial_{\xi, \alpha} 
\widetilde{P}_{4}^{\alpha, \beta}) 
- 
u_{\alpha}'' u_{\beta}' 
(3 \partial_{\xi, \alpha}^{2} \partial_{\xi, \beta} 
\widetilde{P}_{2}^{\alpha} 
+
\partial_{\xi, \beta} \widetilde{P}_{4}^{\alpha})
\right\}
\right.
\\
& \hspace{40mm}
\left. 
-12 
\sum_{\alpha \in \mathcal{H}} 
\sum_{\{\beta, \gamma \} \subset \mathcal{H} \atop 
\alpha \not= \beta \not= \gamma \not= \alpha} 
u_{\alpha}' u_{\beta} u_{\gamma} 
\partial_{\xi, \alpha} \widetilde{P}_{4}^{\beta, \gamma} 
\right) 
\\
= 0& 
\end{align*}
for each $\alpha_{0} \in \mathcal{H}$. 
Since $u_{\alpha} = C_{\alpha}/x_{\alpha}^{2}$, 
this is equivalent to 
\[
S_{1} + S_{2} + S_{3} = 0, 
\]
where 
\begin{align*} 
& S_{1} := 
- 
\sum_{\beta \in \mathcal{H} \atop \beta \not= \alpha_{0}} 
\frac{C_{\beta}}{\langle \alpha_{0}^{\bot}, \beta \rangle^{6}} 
\xi_{\beta}^{3} 
\left\{
\xi_{\beta} 
(\partial_{\xi, \alpha_{0}}^{3} \widetilde{P}_{2}^{\beta} 
- C_{\alpha_{0}} 
\partial_{\xi, \alpha_{0}} \widetilde{P}_{4}^{\alpha_{0}, \beta}) 
+ \xi_{\alpha_{0}} 
(3 \partial_{\xi, \alpha_{0}}^{2} \partial_{\xi, \beta}
\widetilde{P}_{2}^{\alpha_{0}} 
+ 
\partial_{\xi, \beta} \widetilde{P}_{4}^{\alpha_{0}}) 
\right\}, 
\\
& S_{2} :=  
\sum_{\beta \in \mathcal{H} \atop \beta \not= \alpha_{0}} 
\frac{C_{\beta}}{\langle \alpha_{0}^{\bot}, \beta \rangle^{6}} 
\xi_{\alpha_{0}}^{2} \xi_{\beta} 
\left\{ 
\xi_{\alpha_{0}} 
(\partial_{\xi, \beta}^{3} \widetilde{P}_{2}^{\alpha_{0}} 
- C_{\beta} 
\partial_{\xi, \beta} \widetilde{P}_{4}^{\alpha_{0}, \beta}) 
+ \xi_{\beta} 
(3 \partial_{\xi, \beta}^{2} \partial_{\xi, \alpha_{0}}
\widetilde{P}_{2}^{\beta} 
+ 
\partial_{\xi, \alpha_{0}} \widetilde{P}_{4}^{\beta}) 
\right\} 
\\ \intertext{and} 
& S_{3} := 
 \frac{1}{5} 
\sum_{\beta, \gamma \in \mathcal{H} \atop 
\beta \not= \gamma \not= \alpha_{0} \not= \beta} 
\frac{C_{\beta} C_{\gamma}}
{\langle \alpha_{0}^{\bot}, \beta \rangle^{4} 
\langle \alpha_{0}^{\bot}, \gamma \rangle^{3}} 
\xi_{\alpha_{0}}^{2} 
(3\langle \alpha_{0}^{\bot}, \gamma \rangle \xi_{\beta} 
+ 2 \langle \alpha_{0}^{\bot}, \beta \rangle \xi_{\gamma}) 
(\xi_{\beta} \partial_{\xi, \alpha_{0}} 
\widetilde{P}_{4}^{\beta, \gamma} 
- 
\xi_{\alpha_{0}} \partial_{\xi, \beta} 
\widetilde{P}_{4}^{\alpha_{0}, \gamma}). 
\end{align*}

\begin{theorem}\label{theorem:linear relation-2} 
If $\widetilde{P}_{0}$ is not a polynomial 
in $\xi_{1}^{2} + \xi_{2}^{2}$, 
then, for each $\alpha_{0} \in \mathcal{H}$, 
\begin{equation}\label{eq:second relation}
(C_{\alpha_{0}} - 2 |\alpha_{0}|^{2}) 
\sum_{\beta \in \mathcal{H} \atop \beta \not= \alpha_{0}}
\frac{\langle \alpha_{0}, \beta \rangle |\beta|^{2}}
{\langle \alpha_{0}^{\bot}, \beta \rangle^{5}} 
C_{\beta}
= 0 
\end{equation}
holds. 
\end{theorem}

\noindent\textsc{Proof.} 
The proof is divided into many lemmas. 
Let 
\[\widetilde{Q}_{\alpha, \beta, \gamma} 
:= D_{\theta} \widetilde{P}_{0}/\xi_{\alpha} \xi_{\beta} \xi_{\gamma} 
\quad \mbox{and} \quad  
\widetilde{Q}_{4}^{\alpha, \beta} 
:= \sum_{\gamma \in \mathcal{H} \atop \gamma \not= \alpha, \beta} 
\langle \gamma^{\bot}, \alpha \rangle 
C_{\gamma} d_{\alpha, \beta, \gamma} 
\widetilde{Q}_{\alpha, \beta, \gamma}. 
\]
By \eqref{eq:P4ab another}, we have  
\begin{align*}
& 
\xi_{\beta} \partial_{\xi, \alpha_{0}} 
\widetilde{P}_{4}^{\beta, \gamma} 
- 
\xi_{\alpha_{0}} \partial_{\xi, \beta} 
\widetilde{P}_{4}^{\alpha_{0}, \gamma}
\\
&= 
\partial_{\xi, \alpha_{0}} 
(\xi_{\beta} 
\widetilde{P}_{4}^{\beta, \gamma}) 
- 
\partial_{\xi, \beta} 
(\xi_{\alpha_{0}} 
\widetilde{P}_{4}^{\alpha_{0}, \gamma}) 
- \langle \alpha_{0}, \beta \rangle 
(\widetilde{P}_{4}^{\beta, \gamma} 
- \widetilde{P}_{4}^{\alpha_{0}, \gamma}) 
\\
&= 
\frac{\langle \beta^{\bot}, \gamma \rangle^{3}}{N} 
\sum_{\delta \in \mathcal{H} \atop \delta \not= \beta, \gamma} 
\langle \delta^{\bot}, \beta \rangle 
C_{\delta} d_{\beta, \gamma, \delta} 
\partial_{\xi, \alpha_{0}} 
\widetilde{Q}_{\beta, \gamma, \delta} 
- 
\frac{\langle \alpha_{0}^{\bot}, \gamma \rangle^{3}}{N} 
\partial_{\xi, \beta} 
\widetilde{Q}_{4}^{\alpha_{0}, \gamma} 
+ \langle \alpha_{0}, \beta \rangle 
(\widetilde{P}_{4}^{\alpha_{0}, \gamma} 
- \widetilde{P}_{4}^{\beta, \gamma}). 
\end{align*}
Therefore, we put 
\[ 
S_{3} = S_{4} + S_{5} + S_{6} + S_{7} + S_{8}, 
\]
where 
\begin{align*}
S_{4} :&= 
\frac{3}{5} 
\sum_{\beta, \gamma \in \mathcal{H} \atop 
\beta \not= \gamma \not= \alpha_{0} \not= \beta} 
\frac{\langle \alpha_{0}, \beta \rangle C_{\beta} C_{\gamma}}
{\langle \alpha_{0}^{\bot}, \beta \rangle^{4} 
\langle \alpha_{0}^{\bot}, \gamma \rangle^{2}} 
\xi_{\alpha_{0}}^{2} \xi_{\beta} 
(\widetilde{P}_{4}^{\alpha_{0}, \gamma} 
- \widetilde{P}_{4}^{\beta, \gamma}), 
\\
S_{5} :&= 
\frac{2}{5} 
\sum_{\beta, \gamma \in \mathcal{H} \atop 
\beta \not= \gamma \not= \alpha_{0} \not= \beta} 
\frac{\langle \alpha_{0}, \beta \rangle C_{\beta} C_{\gamma}}
{\langle \alpha_{0}^{\bot}, \beta \rangle^{3} 
\langle \alpha_{0}^{\bot}, \gamma \rangle^{3}} 
\xi_{\alpha_{0}}^{2} \xi_{\gamma} 
(\widetilde{P}_{4}^{\alpha_{0}, \gamma} 
- \widetilde{P}_{4}^{\beta, \gamma}) 
\\
&= 
\frac{2}{5} 
\sum_{\beta, \gamma \in \mathcal{H} \atop 
\beta \not= \gamma \not= \alpha_{0} \not= \beta} 
\frac{\langle \alpha_{0}, \gamma \rangle C_{\beta} C_{\gamma}}
{\langle \alpha_{0}^{\bot}, \beta \rangle^{3} 
\langle \alpha_{0}^{\bot}, \gamma \rangle^{3}} 
\xi_{\alpha_{0}}^{2} \xi_{\beta} 
(\widetilde{P}_{4}^{\alpha_{0}, \beta} 
- \widetilde{P}_{4}^{\beta, \gamma}), 
\\
S_{6} :&= 
\frac{1}{5N} 
\sum_{
\mbox{\tiny 
$\begin{matrix} 
\beta, \gamma, \delta \in \mathcal{H} 
\\
\alpha_{0}, \beta, \gamma, \delta 
\\
\mbox{ are all different}
\end{matrix}$
}}
\frac{\langle \beta^{\bot}, \gamma \rangle^{3} 
\langle \delta^{\bot}, \beta \rangle 
C_{\beta} C_{\gamma} C_{\delta} d_{\beta, \gamma, \delta}}
{\langle \alpha_{0}^{\bot}, \beta \rangle^{4} 
\langle \alpha_{0}^{\bot}, \gamma \rangle^{3}} 
\xi_{\alpha_{0}}^{2} 
(3\langle \alpha_{0}^{\bot}, \gamma \rangle \xi_{\beta} 
+ 2 \langle \alpha_{0}^{\bot}, \beta \rangle \xi_{\gamma}) 
\partial_{\xi, \alpha_{0}} 
\widetilde{Q}_{\beta, \gamma, \delta} 
\\
S_{7} :&= 
\frac{1}{5N} 
\sum_{\beta, \gamma \in \mathcal{H} \atop 
\beta \not= \gamma \not= \alpha_{0} \not= \beta} 
\frac{\langle \beta^{\bot}, \gamma \rangle^{3} 
\langle \alpha_{0}^{\bot}, \beta \rangle 
C_{\beta} C_{\gamma} C_{\alpha_{0}} d_{\beta, \gamma, \alpha_{0}}}
{\langle \alpha_{0}^{\bot}, \beta \rangle^{4} 
\langle \alpha_{0}^{\bot}, \gamma \rangle^{3}} 
\xi_{\alpha_{0}}^{2} 
(3\langle \alpha_{0}^{\bot}, \gamma \rangle \xi_{\beta} 
+ 2 \langle \alpha_{0}^{\bot}, \beta \rangle \xi_{\gamma}) 
\partial_{\xi, \alpha_{0}} 
\widetilde{Q}_{\beta, \gamma, \alpha_{0}} 
\\
&= 
\frac{1}{N} 
\sum_{\beta, \gamma \in \mathcal{H} \atop 
\beta \not= \gamma \not= \alpha_{0} \not= \beta} 
\frac{\langle \beta^{\bot}, \gamma \rangle^{3} 
C_{\alpha_{0}} C_{\beta} C_{\gamma} d_{\alpha_{0}, \beta, \gamma}}
{\langle \alpha_{0}^{\bot}, \beta \rangle^{3} 
\langle \alpha_{0}^{\bot}, \gamma \rangle^{2}} 
\xi_{\alpha_{0}}^{2} \xi_{\beta} 
\partial_{\xi, \alpha_{0}} 
\widetilde{Q}_{\alpha_{0}, \beta, \gamma}, 
\\
S_{8} :&= 
-\frac{1}{5N} 
\sum_{\beta, \gamma \in \mathcal{H} \atop 
\beta \not= \gamma \not= \alpha_{0} \not= \beta} 
\frac{C_{\beta} C_{\gamma}}
{\langle \alpha_{0}^{\bot}, \beta \rangle^{4}}
\xi_{\alpha_{0}}^{2} 
(3\langle \alpha_{0}^{\bot}, \gamma \rangle \xi_{\beta} 
+ 2 \langle \alpha_{0}^{\bot}, \beta \rangle \xi_{\gamma}) 
\partial_{\xi, \beta} \widetilde{Q}_{4}^{\alpha_{0}, \gamma}. 
\end{align*}

\begin{lemma}\label{lemma:S6} 
\quad $S_{6} = 0$. 
\end{lemma} 
\begin{proof} 
For $\eta = (\eta_{1}, \eta_{2})$, let 
\[
\bar{S}_{6}(\eta) 
= 
\sum_{
\mbox{\tiny 
$\begin{matrix} 
\beta, \gamma, \delta \in \mathcal{H} 
\\
\alpha_{0}, \beta, \gamma, \delta 
\\
\mbox{ are all different}
\end{matrix}$
}}
\frac{\langle \beta^{\bot}, \gamma \rangle^{3} 
\langle \delta^{\bot}, \beta \rangle 
C_{\beta} C_{\gamma} C_{\delta} d_{\beta, \gamma, \delta}}
{\langle \alpha_{0}^{\bot}, \beta \rangle^{4} 
\langle \alpha_{0}^{\bot}, \gamma \rangle^{3}} 
\xi_{\alpha_{0}}^{2} 
(3\langle \alpha_{0}^{\bot}, \gamma \rangle \eta_{\beta} 
+ 2 \langle \alpha_{0}^{\bot}, \beta \rangle \eta_{\gamma}) 
\partial_{\xi, \alpha_{0}} 
\widetilde{Q}_{\beta, \gamma, \delta}. 
\]
We prove $\bar{S}_{6}(\eta) = 0$. 
If so, we have  $S_{6} = 0$, 
since $S_{6} = \bar{S}_{6}(\xi)/5N$. 

For an ordered triple $\{\beta, \gamma, \delta\} \subset 
\mathcal{H} \setminus \{\alpha_{0}\}$, 
let 
\begin{align*}
A_{\beta} 
&:= 
\langle \gamma^{\bot}, \delta \rangle 
\langle \alpha_{0}^{\bot}, \beta \rangle, 
&
A_{\gamma} 
&:= 
\langle \delta^{\bot}, \beta \rangle 
\langle \alpha_{0}^{\bot}, \gamma \rangle, 
&
A_{\delta} 
&:= 
\langle \beta^{\bot}, \gamma \rangle 
\langle \alpha_{0}^{\bot}, \delta \rangle. 
\end{align*}
Note that they satisfy 
$\sigma(A_{\varepsilon}) 
= (\mathrm{sgn} \sigma) A_{\sigma(\varepsilon)}$ 
($\varepsilon \in \{\beta, \gamma, \delta\}$) 
for a permutation $\sigma$ of $\{\beta, \gamma, \delta\}$, 
and 
\[
A_{\beta} + A_{\gamma} + A_{\delta} = 0 
\]
by \eqref{eq:triangle}. 
Hereafter, we denote by 
${\sum_{\beta, \gamma, \delta}}' F(b, c, d)$ the sum 
$F(\beta, \gamma, \delta) + F(\gamma, \delta, \beta) 
+ F(\delta, \beta, \gamma)$. 

Firstly, we have 
\begin{align*}
\bar{S}_{6}(\eta) 
& = 
\sum_{
\mbox{\tiny 
$\begin{matrix} 
\beta, \gamma, \delta \in \mathcal{H} 
\\
\alpha_{0}, \beta, \gamma, \delta 
\\
\mbox{ are all different}
\end{matrix}$
}}
\frac{\langle \beta^{\bot}, \gamma \rangle^{3} 
\langle \delta^{\bot}, \beta \rangle 
\langle \alpha_{0}^{\bot}, \gamma \rangle
\langle \alpha_{0}^{\bot}, \delta \rangle^{4}
C_{\beta} C_{\gamma} C_{\delta} d_{\beta, \gamma, \delta}}
{\langle \alpha_{0}^{\bot}, \beta \rangle^{4} 
\langle \alpha_{0}^{\bot}, \gamma \rangle^{4}
\langle \alpha_{0}^{\bot}, \delta \rangle^{4}
} 
\xi_{\alpha_{0}}^{2} 
\\
& \hspace{40mm} 
\times 
(3\langle \alpha_{0}^{\bot}, \gamma \rangle \eta_{\beta} 
+ 2 \langle \alpha_{0}^{\bot}, \beta \rangle \eta_{\gamma}) 
\partial_{\xi, \alpha_{0}} 
\widetilde{Q}_{\beta, \gamma, \delta}
\\
& = 
\sum_{
\mbox{\tiny 
$\begin{matrix} 
\beta, \gamma, \delta \in \mathcal{H} 
\\
\alpha_{0}, \beta, \gamma, \delta 
\\
\mbox{ are all different}
\end{matrix}$
}}
\frac{\langle \alpha_{0}^{\bot}, \delta \rangle 
A_{\gamma} A_{\delta}^{3} \xi_{\alpha_{0}}^{2} 
C_{\beta} C_{\gamma} C_{\delta} d_{\beta, \gamma, \delta}}
{\langle \alpha_{0}^{\bot}, \beta \rangle^{4} 
\langle \alpha_{0}^{\bot}, \gamma \rangle^{4}
\langle \alpha_{0}^{\bot}, \delta \rangle^{4}
} 
(3\langle \alpha_{0}^{\bot}, \gamma \rangle \eta_{\beta} 
+ 2 \langle \alpha_{0}^{\bot}, \beta \rangle \eta_{\gamma}) 
\partial_{\xi, \alpha_{0}} 
\widetilde{Q}_{\beta, \gamma, \delta}
\\
& = 
\sum_{
\mbox{\tiny 
$\begin{matrix} 
\{\beta, \gamma, \delta\} \subset \mathcal{H} 
\\
\alpha_{0}, \beta, \gamma, \delta 
\\
\mbox{ are all different}
\end{matrix}$
}}
\frac{C_{\beta} C_{\gamma} C_{\delta} d_{\beta, \gamma, \delta}}
{\langle \alpha_{0}^{\bot}, \beta \rangle^{4} 
\langle \alpha_{0}^{\bot}, \gamma \rangle^{4} 
\langle \alpha_{0}^{\bot}, \delta \rangle^{4}} 
(3 S_{9} + 2 S_{10})
\xi_{\alpha_{0}}^{2} 
\partial_{\xi, \alpha_{0}} 
\widetilde{Q}_{\beta, \gamma, \delta}, 
\end{align*}
where 
\begin{align*}
S_{9} 
:= & 
{\sum_{\beta, \gamma, \delta}}' 
\langle \alpha_{0}^{\bot}, d \rangle A_{d}^{3} 
(\langle \alpha_{0}^{\bot}, c \rangle A_{c} \eta_{b} 
- \langle \alpha_{0}^{\bot}, b \rangle A_{b} \eta_{c}), 
\\
S_{10} 
:= & 
{\sum_{\beta, \gamma, \delta}}' 
\langle \alpha_{0}^{\bot}, d \rangle A_{d}^{3} 
(\langle \alpha_{0}^{\bot}, b \rangle A_{c} \eta_{c} 
- \langle \alpha_{0}^{\bot}, c \rangle A_{b} \eta_{b}). 
\end{align*}
Since 
\begin{align*}
S_{9} 
= & 
{\sum_{\beta, \gamma, \delta}}' 
\langle \alpha_{0}^{\bot}, b \rangle 
\langle \alpha_{0}^{\bot}, c \rangle 
A_{b} A_{c} \eta_{d} (A_{c}^{2} - A_{b}^{2}) 
&
& 
=
A_{\beta} A_{\gamma} A_{\delta} 
{\sum_{\beta, \gamma, \delta}}' 
(A_{b} - A_{c}) 
\langle \alpha_{0}^{\bot}, b \rangle 
\langle \alpha_{0}^{\bot}, c \rangle 
\eta_{d} 
\\
= & 
A_{\beta} A_{\gamma} A_{\delta} 
{\sum_{\beta, \gamma, \delta}}' 
\langle \alpha_{0}^{\bot}, b \rangle A_{b} 
(\langle \alpha_{0}^{\bot}, c \rangle \eta_{d} 
- 
\langle \alpha_{0}^{\bot}, d \rangle \eta_{c}) 
&
&=
- A_{\beta} A_{\gamma} A_{\delta} 
\eta_{\alpha_{0}} 
{\sum_{\beta, \gamma, \delta}}' 
\langle \alpha_{0}^{\bot}, b \rangle 
\langle c^{\bot}, d \rangle 
A_{b} 
\\
=& 
- A_{\beta} A_{\gamma} A_{\delta} 
(A_{\beta}^{2} + A_{\gamma}^{2} + A_{\delta}^{2}) \eta_{\alpha_{0}} 
&
& 
=
2 A_{\beta} A_{\gamma} A_{\delta} 
(A_{\beta} A_{\gamma} + A_{\gamma} A_{\delta} + A_{\delta} A_{\beta}) 
\eta_{\alpha_{0}} 
\end{align*}
and 
\begin{align*}
S_{10} =& 
{\sum_{\beta, \gamma, \delta}}' 
\langle \alpha_{0}^{\bot}, d \rangle A_{d}^{3} 
\{
\langle \alpha_{0}^{\bot}, b \rangle 
(A_{c} - A_{b}) \eta_{c} 
+ 
A_{b} 
(\langle \alpha_{0}^{\bot}, b \rangle \eta_{c} 
- 
\langle \alpha_{0}^{\bot}, c \rangle \eta_{b})
\} 
\\
=& 
{\sum_{\beta, \gamma, \delta}}' 
\{
\langle \alpha_{0}^{\bot}, d \rangle 
\langle \alpha_{0}^{\bot}, b \rangle 
A_{d}^{2} (A_{b}^{2} - A_{c}^{2}) \eta_{c} 
- 
\langle \alpha_{0}^{\bot}, d \rangle 
\langle b^{\bot}, c \rangle A_{d}^{3} A_{b} \eta_{\alpha_{0}} 
\}
\\
=& 
{\sum_{\beta, \gamma, \delta}}' 
\{
A_{b}^{2} A_{c}^{2} 
\langle \alpha_{0}^{\bot}, c \rangle 
(\langle \alpha_{0}^{\bot}, b \rangle \eta_{d} 
- 
\langle \alpha_{0}^{\bot}, d \rangle \eta_{b}) 
- A_{d}^{4} A_{b} \eta_{\alpha_{0}}
\}
\\
=& 
\eta_{\alpha_{0}} 
{\sum_{\beta, \gamma, \delta}}' 
(A_{b}^{2} A_{c}^{3} - A_{b}^{4} A_{c}) 
\\
=& 
\eta_{\alpha_{0}} 
{\sum_{\beta, \gamma, \delta}}' 
A_{b}^{2} A_{c} (A_{c}^{2} - A_{b}^{2}) 
\\
=& 
\eta_{\alpha_{0}} 
{\sum_{\beta, \gamma, \delta}}' 
A_{b}^{2} A_{c} A_{d} (A_{b} - A_{c}) 
\\
=& 
A_{\beta} A_{\gamma} A_{\delta} \eta_{\alpha_{0}} 
{\sum_{\beta, \gamma, \delta}}' 
(A_{b}^{2} - A_{b} A_{c}) 
\\
=& 
-3 A_{\beta} A_{\gamma} A_{\delta} 
(A_{\beta} A_{\gamma} + A_{\gamma} A_{\delta} + A_{\delta} A_{\beta}) 
\eta_{\alpha_{0}}, 
\end{align*}
we have $3 S_{9} + 2 S_{10} = 0$ and $\bar{S}_{6}(\eta) = 0$. 
\end{proof}

\begin{lemma}\label{lemma:S4} 
\quad $\displaystyle S_{4} = -\frac{3}{5} (S_{11} + S_{12})$, where 
\begin{align*}
S_{11} 
&:= 
\sum_{\beta \in \mathcal{H} \atop \beta \not= \alpha_{0}} 
\frac{\langle \alpha_{0}, \beta \rangle C_{\beta}^{2}}
{\langle \alpha_{0}^{\bot}, \beta \rangle^{6}} 
\xi_{\alpha_{0}}^{2} 
(\xi_{\beta} \widetilde{P}_{4}^{\alpha_{0}, \beta} 
- \partial_{\xi, \beta} \widetilde{P}_{2}^{\beta}), 
&
S_{12} &:= 
\frac{1}{N} 
\sum_{\beta, \gamma \in \mathcal{H} \atop 
\beta \not= \gamma \not= \alpha_{0} \not= \beta} 
\frac{\langle \beta^{\bot}, \gamma \rangle^{3} 
\langle \alpha_{0}, \beta \rangle
C_{\beta} C_{\gamma}}
{\langle \alpha_{0}^{\bot}, \beta \rangle^{4} 
\langle \alpha_{0}^{\bot}, \gamma \rangle^{2}} 
\xi_{\alpha_{0}}^{2} \widetilde{Q}_{4}^{\beta, \gamma}. 
\end{align*}
\end{lemma}
\begin{proof}
We use \eqref{eq:D_theta}, 
\eqref{eq:first relation} and \eqref{eq:P4ab another}. 
Firstly, 
\begin{align*}
\frac{5}{3} 
S_{4} &= 
\sum_{\beta, \gamma \in \mathcal{H} \atop 
\beta \not= \gamma \not= \alpha_{0} \not= \beta} 
\frac{\langle \alpha_{0}, \beta \rangle C_{\beta} C_{\gamma}}
{\langle \alpha_{0}^{\bot}, \beta \rangle^{4} 
\langle \alpha_{0}^{\bot}, \gamma \rangle^{2}} 
\xi_{\alpha_{0}} 
\\
& \hspace{20mm} 
\times 
\left\{ 
\xi_{\beta} 
\left(
\partial_{\xi, \alpha_{0}} 
\widetilde{P}_{2}^{\gamma} 
+ 
\frac{\langle \alpha_{0}^{\bot}, \gamma \rangle^{3}}{N} 
\widetilde{Q}_{4}^{\alpha_{0}, \gamma}
\right) 
-
\xi_{\alpha_{0}} 
\left(
\partial_{\xi, \beta} 
\widetilde{P}_{2}^{\gamma} 
+ 
\frac{\langle \beta^{\bot}, \gamma \rangle^{3}}{N} 
\widetilde{Q}_{4}^{\beta, \gamma}
\right) 
\right\}
\\
&= 
\sum_{\beta, \gamma \in \mathcal{H} \atop 
\beta \not= \gamma \not= \alpha_{0} \not= \beta} 
\frac{\langle \alpha_{0}, \beta \rangle 
C_{\beta} C_{\gamma}}
{\langle \alpha_{0}^{\bot}, \beta \rangle^{3} 
\langle \alpha_{0}^{\bot}, \gamma \rangle^{2}} 
\xi_{\alpha_{0}} D_{\theta} \widetilde{P}_{2}^{\gamma} 
+ 
\frac{1}{N} 
\sum_{\beta, \gamma \in \mathcal{H} \atop 
\beta \not= \gamma \not= \alpha_{0} \not= \beta} 
\frac{\langle \alpha_{0}, \beta \rangle 
\langle \alpha_{0}^{\bot}, \gamma \rangle 
C_{\beta} C_{\gamma}}
{\langle \alpha_{0}^{\bot}, \beta \rangle^{4}} 
\xi_{\alpha_{0}} \xi_{\beta} 
\widetilde{Q}_{4}^{\alpha_{0}, \gamma}
- S_{12}.
\end{align*}
The lemma is a consequence of the following calculations: 
\begin{align*}
\mbox{(i)} \quad & 
\sum_{\beta, \gamma \in \mathcal{H} \atop 
\beta \not= \gamma \not= \alpha_{0} \not= \beta} 
\frac{\langle \alpha_{0}, \beta \rangle 
C_{\beta} C_{\gamma}}
{\langle \alpha_{0}^{\bot}, \beta \rangle^{3} 
\langle \alpha_{0}^{\bot}, \gamma \rangle^{2}} 
\xi_{\alpha_{0}} D_{\theta} \widetilde{P}_{2}^{\gamma} 
\\
& 
= 
\sum_{\gamma \in \mathcal{H} \atop \gamma \not= \alpha_{0}} 
\frac{C_{\gamma}}{\langle \alpha_{0}^{\bot}, \gamma \rangle^{2}} 
\xi_{\alpha_{0}} D_{\theta} \widetilde{P}_{2}^{\gamma} 
\sum_{\beta \in \mathcal{H} \atop 
\beta \not= \alpha_{0}, \gamma} 
\frac{\langle \alpha_{0}, \beta \rangle C_{\beta}}
{\langle \alpha_{0}^{\bot}, \beta \rangle^{3}} 
= 
- 
\sum_{\gamma \in \mathcal{H} \atop \gamma \not \alpha_{0}} 
\frac{\langle \alpha_{0}, \gamma \rangle C_{\gamma}^{2}}
{\langle \alpha_{0}^{\bot}, \gamma \rangle^{5}} 
\xi_{\alpha_{0}} D_{\theta} \widetilde{P}_{2}^{\gamma} 
\\
&= 
- 
\sum_{\gamma \in \mathcal{H} \atop \beta \not \alpha_{0}} 
\frac{\langle \alpha_{0}, \beta \rangle C_{\beta}^{2}}
{\langle \alpha_{0}^{\bot}, \beta \rangle^{5}} 
\xi_{\alpha_{0}} D_{\theta} \widetilde{P}_{2}^{\beta}. 
\\
\mbox{(ii)} \quad & 
\sum_{\beta, \gamma \in \mathcal{H} \atop 
\beta \not= \gamma \not= \alpha_{0} \not= \beta} 
\frac{\langle \alpha_{0}, \beta \rangle 
\langle \alpha_{0}^{\bot}, \gamma \rangle 
C_{\beta} C_{\gamma}}
{\langle \alpha_{0}^{\bot}, \beta \rangle^{4}} 
\xi_{\alpha_{0}} \xi_{\beta} 
\widetilde{Q}_{4}^{\alpha_{0}, \gamma}
\\
&= 
\sum_{\beta \in \mathcal{H} \atop \beta \not= \alpha_{0}} 
\frac{\langle \alpha_{0}, \beta \rangle C_{\beta}} 
{\langle \alpha_{0}^{\bot}, \beta \rangle^{4}} 
\xi_{\alpha_{0}} \xi_{\beta} 
\sum_{\gamma \in \mathcal{H} \atop \gamma \not= \alpha_{0}, \beta} 
\sum_{\delta \in \mathcal{H} \atop \delta \not= \alpha_{0}, \gamma} 
\langle \alpha_{0}^{\bot}, \gamma \rangle 
\langle \delta^{\bot}, \alpha_{0} \rangle 
C_{\gamma} C_{\delta} 
d_{\alpha_{0}, \gamma, \delta} 
\widetilde{Q}_{\alpha_{0}, \gamma, \delta}
\\
&= 
\sum_{\beta \in \mathcal{H} \atop \beta \not= \alpha_{0}} 
\frac{\langle \alpha_{0}, \beta \rangle C_{\beta}} 
{\langle \alpha_{0}^{\bot}, \beta \rangle^{4}} 
\xi_{\alpha_{0}} \xi_{\beta} 
\sum_{\gamma \in \mathcal{H} \atop \gamma \not= \alpha_{0}, \beta} 
\langle \alpha_{0}^{\bot}, \gamma \rangle 
\langle \beta^{\bot}, \alpha_{0} \rangle 
C_{\gamma} C_{\beta} 
d_{\alpha_{0}, \gamma, \beta} 
\widetilde{Q}_{\alpha_{0}, \gamma, \beta}
\\
&= 
-
\sum_{\beta \in \mathcal{H} \atop \beta \not= \alpha_{0}} 
\frac{\langle \alpha_{0}, \beta \rangle C_{\beta}^{2}} 
{\langle \alpha_{0}^{\bot}, \beta \rangle^{3}} 
\xi_{\alpha_{0}} \xi_{\beta} 
\sum_{\gamma \in \mathcal{H} \atop \gamma \not= \alpha_{0}, \beta} 
\langle \gamma^{\bot}, \alpha_{0} \rangle 
C_{\gamma} 
d_{\alpha_{0}, \beta, \gamma} 
\widetilde{Q}_{\alpha_{0}, \beta, \gamma} 
\\
&= 
-
\sum_{\beta \in \mathcal{H} \atop \beta \not= \alpha_{0}} 
\frac{\langle \alpha_{0}, \beta \rangle C_{\beta}^{2}} 
{\langle \alpha_{0}^{\bot}, \beta \rangle^{3}} 
\xi_{\alpha_{0}} \xi_{\beta} 
\widetilde{Q}_{4}^{\alpha_{0}, \beta}. 
\\
\mbox{(iii)} \quad & 
\langle \alpha_{0}^{\bot}, \beta \rangle 
\xi_{\alpha_{0}} D_{\theta} \widetilde{P}_{2}^{\beta} 
+ \frac{\langle \alpha_{0}^{\bot}, \beta \rangle^{3}}{N} 
\xi_{\alpha_{0}} \xi_{\beta} \widetilde{Q}_{4}^{\alpha_{0}, \beta} 
\\
& = 
\xi_{\alpha_{0}} \xi_{\beta} 
\left( \partial_{\alpha_{0}} \widetilde{P}_{2}^{\beta} 
+ \frac{\langle \alpha_{0}^{\bot}, \beta \rangle^{3}}{N} 
\widetilde{Q}_{4}^{\alpha_{0}, \beta} 
\right) 
- \xi_{\alpha_{0}}^{2} \partial_{\beta} \widetilde{P}_{2}^{\beta} 
= 
\xi_{\alpha_{0}}^{2} 
(\xi_{\beta} \widetilde{P}_{4}^{\alpha_{0}, \beta} 
- \partial_{\beta} \widetilde{P}_{2}^{\beta}). 
\end{align*}
\end{proof}

Let us rewrite $S_{5}$ analogously. 
By \eqref{eq:first relation}, there exists a constant $K_{\alpha_{0}}$
such that 
\[
\sum_{\beta \in \mathcal{H} \atop \beta \not= \alpha_{0}} 
\frac{C_{\beta}}{\langle \alpha_{0}^{\bot}, \beta \rangle^{3}} 
\partial_{\xi, \beta} 
= K_{\alpha_{0}} \partial_{\xi, \alpha_{0}^{\bot}}. 
\]
\begin{lemma}\label{lemma:S5} 
\quad 
$\displaystyle 
S_{5} = - \frac{2}{5} (S_{11} + S_{12}) - S_{13} + S_{14}$, 
where 
\begin{align*}
S_{13} 
=& 
\frac{2 K_{\alpha_{0}}}{5} 
\sum_{\beta \in \mathcal{H} \atop \beta \not= \alpha_{0}} 
\frac{\langle \alpha_{0}, \beta \rangle C_{\beta}}
{\langle \alpha_{0}^{\bot}, \beta \rangle^{3}} 
\xi_{\alpha_{0}}^{2} 
\partial_{\xi, \alpha_{0}^{\bot}} \widetilde{P}_{2}^{\beta}, 
\\
S_{14} 
=& 
\frac{1}{N} 
\sum_{
\mbox{\tiny 
$\begin{matrix} 
\beta, \gamma, \delta \in \mathcal{H} 
\\
\alpha_{0}, \beta, \gamma, \delta 
\\
\mbox{ are all different}
\end{matrix}$
}}
\frac{\langle \beta^{\bot}, \gamma \rangle^{3} 
\langle \delta^{\bot}, \beta \rangle
\langle \alpha_{0}, \beta \rangle 
C_{\beta} C_{\gamma} C_{\delta} 
d_{\beta, \gamma, \delta}}
{\langle \alpha_{0}^{\bot}, \beta \rangle^{4} 
\langle \alpha_{0}^{\bot}, \gamma \rangle^{2}} 
\xi_{\alpha_{0}}^{2} 
\widetilde{Q}_{\beta, \gamma, \delta}. 
\end{align*}
\end{lemma}
\begin{proof} 
Let us calculate $S_{5} + 2 S_{12}/5 - S_{14}$: 
\begin{align}
S_{5} &+ \frac{2}{5} S_{12} - S_{14} 
\notag 
\\
&= 
\frac{2}{5} 
\sum_{\beta \in \mathcal{H} \atop \beta \not= \alpha_{0}} 
\frac{C_{\beta}}{\langle \alpha_{0}^{\bot}, \beta \rangle^{3}}
\xi_{\alpha_{0}}^{2} \xi_{\beta} 
\widetilde{P}_{4}^{\alpha_{0}, \beta} 
\sum_{\gamma \in \mathcal{H} \atop \gamma \not= \alpha_{0}, \beta} 
\frac{\langle \alpha_{0}, \gamma \rangle C_{\gamma}}
{\langle \alpha_{0}^{\bot}, \gamma \rangle^{3}}
\notag 
\\
& \qquad 
-
\frac{2}{5} 
\sum_{\beta, \gamma \in \mathcal{H} \atop 
\beta \not= \gamma \not= \alpha_{0} \not= \beta} 
\frac{\langle \alpha_{0}, \gamma \rangle C_{\beta} C_{\gamma}}
{\langle \alpha_{0}^{\bot}, \beta \rangle^{3} 
\langle \alpha_{0}^{\bot}, \gamma \rangle^{3}} 
\xi_{\alpha_{0}}^{2} 
\left(\partial_{\xi, \beta} \widetilde{P}_{2}^{\gamma} 
+ 
\frac{\langle \beta^{\bot}, \gamma \rangle^{3}}{N} 
\widetilde{Q}_{4}^{\beta, \gamma} 
\right)
+ \frac{2}{5} S_{12} - S_{14} 
\notag 
\\
&= 
-\frac{2}{5}
\sum_{\beta \in \mathcal{H} \atop \beta \not= \alpha_{0}} 
\frac{\langle \alpha_{0}, \beta \rangle C_{\beta}^{2}}
{\langle \alpha_{0}^{\bot}, \beta \rangle^{6}}
\xi_{\alpha_{0}}^{2} \xi_{\beta} 
\widetilde{P}_{4}^{\alpha_{0}, \beta} 
-\frac{2}{5} 
\sum_{\gamma \in \mathcal{H} \atop \gamma \not= \alpha_{0}} 
\frac{\langle \alpha_{0}, \gamma \rangle C_{\gamma}}
{\langle \alpha_{0}^{\bot}, \gamma \rangle^{3}} 
\xi_{\alpha_{0}}^{2} 
\left(
\sum_{\beta \in \mathcal{H} \atop \beta \not= \alpha_{0}, \gamma} 
\frac{C_{\beta}}
{\langle \alpha_{0}^{\bot}, \beta \rangle^{3}} 
\partial_{\xi, \beta} 
\right) 
\widetilde{P}_{2}^{\gamma} 
\notag 
\\
& \qquad 
- \frac{2}{5N} 
\sum_{\beta, \gamma \in \mathcal{H} \atop 
\beta \not= \gamma \not= \alpha_{0} \not= \beta} 
\frac{\langle \beta^{\bot}, \gamma \rangle^{3} 
\langle \alpha_{0}, \gamma \rangle 
C_{\beta} C_{\gamma}}
{\langle \alpha_{0}^{\bot}, \beta \rangle^{3} 
\langle \alpha_{0}^{\bot}, \gamma \rangle^{3}} 
\xi_{\alpha_{0}}^{2} 
\widetilde{Q}_{4}^{\beta, \gamma}
+ \frac{2}{5} S_{12} - S_{14} 
\notag 
\\
&= 
-\frac{2}{5}
\sum_{\beta \in \mathcal{H} \atop \beta \not= \alpha_{0}} 
\frac{\langle \alpha_{0}, \beta \rangle C_{\beta}^{2}}
{\langle \alpha_{0}^{\bot}, \beta \rangle^{6}}
\xi_{\alpha_{0}}^{2} \xi_{\beta} 
\widetilde{P}_{4}^{\alpha_{0}, \beta} 
-\frac{2}{5} 
\sum_{\gamma \in \mathcal{H} \atop \gamma \not= \alpha_{0}} 
\frac{\langle \alpha_{0}, \gamma \rangle C_{\gamma}}
{\langle \alpha_{0}^{\bot}, \gamma \rangle^{3}} 
\xi_{\alpha_{0}}^{2} 
\left(
K_{\alpha_{0}} \partial_{\xi, \alpha_{0}^{\bot}} 
- 
\frac{C_{\gamma}}
{\langle \alpha_{0}^{\bot}, \gamma \rangle^{3}} 
\partial_{\xi, \gamma} 
\right) 
\widetilde{P}_{2}^{\gamma} 
\label{eq:S_5-1}
\\
& \qquad 
-\frac{1}{5 N} 
\sum_{\mbox{\tiny 
$\begin{matrix} 
\beta, \gamma, \delta \in \mathcal{H} 
\\
\alpha_{0}, \beta, \gamma, \delta 
\\
\mbox{ are all different}
\end{matrix}$
}}
\frac{\langle \beta^{\bot}, \gamma \rangle^{3} 
\langle \delta^{\bot}, \beta \rangle 
C_{\beta} C_{\gamma} C_{\delta} d_{\beta, \gamma, \delta}}
{\langle \alpha_{0}^{\bot}, \beta \rangle^{4} 
\langle \alpha_{0}^{\bot}, \gamma \rangle^{3}} 
\xi_{\alpha_{0}}^{2} 
\notag
\\
& \hspace{40mm} 
\times
(3 \langle \alpha_{0}^{\bot}, \gamma \rangle 
\langle \alpha_{0}, \beta \rangle 
+ 2 \langle \alpha_{0}^{\bot}, \beta \rangle 
\langle \alpha_{0}, \gamma \rangle) 
\widetilde{Q}_{\beta, \gamma, \delta}
\label{eq:S_5-2}
\\
& \qquad 
- \frac{2}{5N} 
\sum_{\beta, \gamma \in \mathcal{H} \atop 
\beta \not= \gamma \not= \alpha_{0} \not= \beta} 
\frac{\langle \beta^{\bot}, \gamma \rangle^{3} 
C_{\beta} C_{\gamma} d_{\alpha_{0}, \beta, \gamma}}
{\langle \alpha_{0}^{\bot}, \beta \rangle^{3} 
\langle \alpha_{0}^{\bot}, \gamma \rangle^{3}} 
(\langle \alpha_{0}^{\bot}, \gamma \rangle 
\langle \alpha_{0}, \beta \rangle 
- \langle \alpha_{0}^{\bot}, \beta \rangle 
\langle \alpha_{0}, \gamma \rangle) 
\xi_{\alpha_{0}}^{2} 
\widetilde{Q}_{\alpha_{0}, \beta, \gamma}.
\label{eq:S_5-3}
\end{align}
The terms in \eqref{eq:S_5-1} are equal to $-2 S_{11}/5 - S_{13}$. 
The remaining terms vanish, 
since \eqref{eq:S_5-2} is $-\bar{S}_{6}(\alpha_{0})/5N$ 
and the summand in \eqref{eq:S_5-3} is 
skew-symmetric with respect to $\beta, \gamma$. 
\end{proof}

\begin{lemma}\label{lemma:S8} 
\quad 
$\displaystyle 
S_{8} 
= S_{13} + S_{15}$, where 
$\displaystyle 
S_{15} := 
\frac{1}{N} 
\sum_{\beta \in \mathcal{H} \atop \beta \not= \alpha_{0}} 
\frac{C_{\beta}^{2}}
{\langle \alpha_{0}^{\bot}, \beta \rangle^{3}} 
\xi_{\alpha_{0}}^{2} \xi_{\beta} 
\partial_{\xi, \beta} 
\widetilde{Q}_{4}^{\alpha_{0}, \beta}$.  
\end{lemma}

\begin{proof}
This lemma is a consequence of the following calculations: 
\begin{align*}
\mbox{(i)} \quad & 
\sum_{\beta, \gamma \in \mathcal{H} \atop 
\beta \not= \gamma \not= \alpha_{0} \not= \beta} 
\frac{
\langle \alpha_{0}^{\bot}, \gamma \rangle 
C_{\beta} C_{\gamma}}
{\langle \alpha_{0}^{\bot}, \beta \rangle^{4}} 
\xi_{\alpha_{0}}^{2} \xi_{\beta} 
\partial_{\xi, \beta} 
\widetilde{Q}_{4}^{\alpha_{0}, \gamma} 
\\
& \quad = 
\sum_{\beta \in \mathcal{H} \atop \beta \not= \alpha_{0}} 
\frac{C_{\beta}}{\langle \alpha_{0}^{\bot}, \beta \rangle^{4}} 
\xi_{\alpha_{0}}^{2} \xi_{\beta} 
\sum_{
{\mbox{\tiny 
$\begin{matrix} 
\gamma, \delta \in \mathcal{H} 
\\
\alpha_{0}, \beta, \gamma, \delta 
\\
\mbox{ are all different}
\end{matrix}$
}}}
\langle \alpha_{0}^{\bot}, \gamma \rangle 
\langle \delta^{\bot}, \alpha_{0} \rangle 
C_{\gamma} C_{\delta} 
d_{\alpha_{0}, \gamma, \delta} 
\partial_{\xi, \beta} \widetilde{Q}_{\alpha_{0}, \gamma, \delta} 
\\
& \qquad \qquad 
+ 
\sum_{\beta \in \mathcal{H} \atop \beta \not= \alpha_{0}} 
\frac{C_{\beta}}{\langle \alpha_{0}^{\bot}, \beta \rangle^{4}} 
\xi_{\alpha_{0}}^{2} \xi_{\beta} 
\sum_{\gamma \in \mathcal{H} \atop \gamma \not= \alpha_{0}, \beta} 
\langle \alpha_{0}^{\bot}, \gamma \rangle 
\langle \beta^{\bot}, \alpha_{0} \rangle 
C_{\gamma} C_{\beta} 
d_{\alpha_{0}, \gamma, \beta} 
\partial_{\xi, \beta} \widetilde{Q}_{\alpha_{0}, \gamma, \beta} 
\\
& \quad = 
- 
\sum_{\beta \in \mathcal{H} \atop \beta \not= \alpha_{0}} 
\frac{C_{\beta}^{2}}{\langle \alpha_{0}^{\bot}, \beta \rangle^{3}} 
\xi_{\alpha_{0}}^{2} \xi_{\beta} 
\partial_{\xi, \beta} 
\left(
\sum_{\gamma \in \mathcal{H} \atop \gamma \not= \alpha_{0}, \beta} 
\langle \gamma^{\bot}, \alpha_{0} \rangle 
C_{\gamma} 
d_{\alpha_{0}, \beta, \gamma} 
\widetilde{Q}_{\alpha_{0}, \beta, \gamma} 
\right)
\\
& \quad = 
- 
\sum_{\beta \in \mathcal{H} \atop \beta \not= \alpha_{0}} 
\frac{C_{\beta}^{2}}{\langle \alpha_{0}^{\bot}, \beta \rangle^{3}} 
\xi_{\alpha_{0}}^{2} \xi_{\beta} 
\partial_{\xi, \beta} 
\widetilde{Q}_{4}^{\alpha_{0}, \beta}
\\
& \quad = - N S_{15}.  
\\
\mbox{(ii)} \quad & 
\sum_{\beta, \gamma \in \mathcal{H} \atop 
\beta \not= \gamma \not= \alpha_{0} \not= \beta} 
\frac{C_{\beta} C_{\gamma}}
{\langle \alpha_{0}^{\bot}, \beta \rangle^{3}} 
\xi_{\alpha_{0}}^{2} \xi_{\gamma} 
\partial_{\xi, \beta} 
\widetilde{Q}_{4}^{\alpha_{0}, \gamma} 
\\
& \quad 
= 
\sum_{\gamma \in \mathcal{H} \atop \gamma \not= \alpha_{0}} 
C_{\gamma} \xi_{\alpha_{0}}^{2} \xi_{\gamma} 
\left(
\sum_{\beta \in \mathcal{H} \atop \beta \not= \alpha_{0}, \gamma} 
\frac{C_{\beta}}{\langle \alpha_{0}^{\bot}, \beta \rangle^{3}} 
\partial_{\xi, \beta} 
\right) 
\widetilde{Q}_{4}^{\alpha_{0}, \gamma} 
\\
& \quad 
= 
\sum_{\gamma \in \mathcal{H} \atop \gamma \not= \alpha_{0}} 
C_{\gamma} \xi_{\alpha_{0}}^{2} \xi_{\gamma} 
\left(
K_{\alpha_{0}} \partial_{\xi, \alpha_{0}^{\bot}} 
- 
\frac{C_{\gamma}}{\langle \alpha_{0}^{\bot}, \gamma \rangle^{3}} 
\partial_{\xi, \gamma} 
\right) 
\widetilde{Q}_{4}^{\alpha_{0}, \gamma} 
\\
& \quad 
= 
K_{\alpha_{0}} 
\sum_{\beta \in \mathcal{H} \atop \beta \not= \alpha_{0}} 
C_{\beta} \xi_{\alpha_{0}}^{2} \xi_{\beta} 
\partial_{\xi, \alpha_{0}^{\bot}} 
\widetilde{Q}_{4}^{\alpha_{0}, \beta} 
- N S_{15}. 
\\
\mbox{(iii)} \quad & 
\sum_{\beta \in \mathcal{H} \atop \beta \not= \alpha_{0}} 
C_{\beta} \xi_{\alpha_{0}}^{2} \xi_{\beta} 
\partial_{\xi, \alpha_{0}^{\bot}} 
\widetilde{Q}_{4}^{\alpha_{0}, \beta} 
\\
& \quad = 
\xi_{\alpha_{0}}^{2} 
\sum_{\beta \in \mathcal{H} \atop \beta \not= \alpha_{0}} 
C_{\beta} \xi_{\beta} 
\partial_{\xi, \alpha_{0}^{\bot}} 
\left(
\sum_{\gamma \in \mathcal{H} \atop \gamma \not= \alpha_{0}, \beta} 
\langle \gamma^{\bot}, \alpha_{0} \rangle 
C_{\gamma} 
d_{\alpha_{0}, \beta, \gamma} 
\frac{D_{\theta} \widetilde{P}_{0}}
{\xi_{\alpha_{0}} \xi_{\beta} \xi_{\gamma}}
\right)
\\
& \quad = 
\xi_{\alpha_{0}}^{2} 
\sum_{\beta, \gamma \in \mathcal{H} \atop 
\beta \not= \gamma \not= \alpha_{0} \not= \beta} 
\langle \gamma^{\bot}, \alpha_{0} \rangle 
C_{\beta} C_{\gamma} 
d_{\alpha_{0}, \beta, \gamma} 
\left\{
\partial_{\xi, \alpha_{0}^{\bot}} 
\left(
\frac{D_{\theta} \widetilde{P}_{0}}
{\xi_{\alpha_{0}} \xi_{\gamma}}
\right) 
- 
\langle \alpha_{0}^{\bot}, \beta \rangle 
\frac{D_{\theta} \widetilde{P}_{0}}
{\xi_{\alpha_{0}} \xi_{\beta} \xi_{\gamma}}
\right\}
\\
& \quad = 
\xi_{\alpha_{0}}^{2} 
\sum_{\gamma \in \mathcal{H} \atop \gamma \not= \alpha_{0}} 
\langle \gamma^{\bot}, \alpha_{0} \rangle 
C_{\gamma} 
\partial_{\xi, \alpha_{0}^{\bot}} 
\left(
\frac{D_{\theta} \widetilde{P}_{0}}
{\xi_{\alpha_{0}} \xi_{\gamma}}
\right) 
\sum_{\beta \in \mathcal{H} \atop \beta \not= \alpha_{0}, \gamma} 
C_{\beta} 
d_{\alpha_{0}, \beta, \gamma} 
\\
& \quad = 
-\xi_{\alpha_{0}}^{2} 
\sum_{\gamma \in \mathcal{H} \atop \gamma \not= \alpha_{0}} 
\langle \gamma^{\bot}, \alpha_{0} \rangle 
C_{\gamma} 
\partial_{\xi, \alpha_{0}^{\bot}} 
\left(
\frac{D_{\theta} \widetilde{P}_{0}}
{\xi_{\alpha_{0}} \xi_{\gamma}}
\right) 
\times N 
\frac{\langle \alpha_{0}, \gamma \rangle}
{\langle \alpha_{0}^{\bot}, \gamma \rangle^{3}} 
\\
& \quad = 
- N \xi_{\alpha_{0}}^{2} 
\sum_{\beta \in \mathcal{H} \atop \beta \not= \alpha_{0}}
\frac{\langle \alpha_{0}, \beta \rangle C_{\beta}}
{\langle \alpha_{0}^{\bot}, \beta \rangle^{3}} 
\partial_{\xi, \alpha_{0}^{\bot}} 
(\widetilde{P}_{2}^{\beta} - \widetilde{P}_{2}^{\alpha_{0}}) 
\\
& \quad = 
- N \xi_{\alpha_{0}}^{2} 
\sum_{\beta \in \mathcal{H} \atop \beta \not= \alpha_{0}}
\frac{\langle \alpha_{0}, \beta \rangle C_{\beta}}
{\langle \alpha_{0}^{\bot}, \beta \rangle^{3}} 
\partial_{\xi, \alpha_{0}^{\bot}} 
\widetilde{P}_{2}^{\beta} 
+
N \xi_{\alpha_{0}}^{2} 
\partial_{\xi, \alpha_{0}^{\bot}} 
\widetilde{P}_{2}^{\alpha_{0}} 
\sum_{\beta \in \mathcal{H} \atop \beta \not= \alpha_{0}}
\frac{\langle \alpha_{0}, \beta \rangle C_{\beta}}
{\langle \alpha_{0}^{\bot}, \beta \rangle^{3}} 
\\
& \quad = 
- N \xi_{\alpha_{0}}^{2} 
\sum_{\beta \in \mathcal{H} \atop \beta \not= \alpha_{0}}
\frac{\langle \alpha_{0}, \beta \rangle C_{\beta}}
{\langle \alpha_{0}^{\bot}, \beta \rangle^{3}} 
\partial_{\xi, \alpha_{0}^{\bot}} 
\widetilde{P}_{2}^{\beta}
\\
& \quad = -\frac{5 N S_{13}}{2 K_{\alpha_{0}}}. 
\end{align*}
\end{proof}

\begin{lemma}\label{lemma:S3 second}
\quad 
$- S_{12} + S_{14} = S_{16} + S_{17}$, 
where 
\begin{align*}
S_{16} :=& 
\frac{1}{N} 
\sum_{\beta, \gamma \in \mathcal{H} 
\atop \beta \not= \gamma \not= \alpha_{0} \not= \beta} 
\frac{C_{\alpha_{0}} C_{\beta} C_{\gamma} 
d_{\alpha_{0}, \beta, \gamma}}
{\langle \alpha_{0}^{\bot}, \beta \rangle^{3}} 
(\langle \gamma^{\bot}, \alpha_{0} \rangle 
\langle \alpha_{0}, \beta \rangle 
- \langle \alpha_{0}^{\bot}, \beta \rangle 
\langle \alpha_{0}, \gamma \rangle) 
\xi_{\beta}^{3} 
\frac{\widetilde{Q}_{\alpha_{0}, \beta, \gamma}}{\xi_{\alpha_{0}}}, 
\\
S_{17} :=&
\frac{3}{N} 
\sum_{\beta, \gamma \in \mathcal{H} 
\atop \beta \not= \gamma \not= \alpha_{0} \not= \beta} 
\frac{\langle \beta^{\bot}, \gamma \rangle 
\langle \alpha_{0}, \beta \rangle
C_{\alpha_{0}} C_{\beta} C_{\gamma} 
d_{\alpha_{0}, \beta, \gamma}}
{\langle \alpha_{0}^{\bot}, \beta \rangle^{2} 
\langle \alpha_{0}^{\bot}, \gamma \rangle} 
\xi_{\beta} \xi_{\gamma} 
\widetilde{Q}_{\alpha_{0}, \beta, \gamma}. 
\end{align*}
\end{lemma}

\begin{proof}
\begin{align*}
N(-S_{12} + S_{14}) 
=& 
-\sum_{\beta, \gamma \in \mathcal{H} \atop 
\beta \not= \gamma \not= \alpha_{0} \not= \beta} 
\frac{\langle \beta^{\bot}, \gamma \rangle^{3} 
\langle \alpha_{0}, \beta \rangle 
C_{\beta} C_{\gamma}}
{\langle \alpha_{0}^{\bot}, \beta \rangle^{4} 
\langle \alpha_{0}^{\bot}, \gamma \rangle^{2}} 
\xi_{\alpha_{0}}^{2} 
\sum_{\delta \in \mathcal{H} 
\atop 
\delta \not= \beta, \gamma} 
\langle \delta^{\bot}, \beta \rangle C_{\delta} 
d_{\beta, \gamma, \delta} 
\widetilde{Q}_{\beta, \gamma, \delta} 
\\
& \qquad \qquad + 
\sum_{
\mbox{\tiny 
$\begin{matrix} 
\beta, \gamma, \delta \in \mathcal{H} 
\\
\alpha_{0}, \beta, \gamma, \delta 
\\
\mbox{ are all different}
\end{matrix}$
}}
\frac{\langle \beta^{\bot}, \gamma \rangle^{3} 
\langle \delta^{\bot}, \beta \rangle
\langle \alpha_{0}, \beta \rangle 
C_{\beta} C_{\gamma} C_{\delta} 
d_{\beta, \gamma, \delta}}
{\langle \alpha_{0}^{\bot}, \beta \rangle^{4} 
\langle \alpha_{0}^{\bot}, \gamma \rangle^{2}} 
\xi_{\alpha_{0}}^{2} 
\widetilde{Q}_{\beta, \gamma, \delta} 
\\
=& 
-\sum_{\beta, \gamma \in \mathcal{H} \atop 
\beta \not= \gamma \not= \alpha_{0} \not= \beta} 
\frac{\langle \alpha_{0}, \beta \rangle 
C_{\alpha_{0}} C_{\beta} C_{\gamma} 
d_{\beta, \gamma, \alpha_{0}}}
{\langle \alpha_{0}^{\bot}, \beta \rangle^{3} 
\langle \alpha_{0}^{\bot}, \gamma \rangle^{2}} 
(\langle \beta^{\bot}, \gamma \rangle 
\xi_{\alpha_{0}})^{3} 
\frac{\widetilde{Q}_{\beta, \gamma, \alpha_{0}}}
{\xi_{\alpha}}
\\
=& 
\sum_{\beta, \gamma \in \mathcal{H} \atop 
\beta \not= \gamma \not= \alpha_{0} \not= \beta} 
\langle \alpha_{0}, \beta \rangle 
C_{\alpha_{0}} C_{\beta} C_{\gamma} 
d_{\alpha_{0}, \beta, \gamma} 
\frac{\widetilde{Q}_{\alpha_{0}, \beta, \gamma}} 
{\xi_{\alpha}} 
\\
& \qquad 
\times 
\left(
\frac{\langle \gamma^{\bot}, \alpha_{0} \rangle \xi_{\beta}^{3}}
{\langle \alpha_{0}^{\bot}, \beta \rangle^{3}} 
+ 3 
\frac{\langle \beta^{\bot}, \gamma \rangle 
\xi_{\alpha_{0}} \xi_{\beta} \xi_{\gamma}^{2}}
{\langle \alpha_{0}^{\bot}, \beta \rangle^{2} 
\langle \gamma^{\bot}, \alpha_{0} \rangle} 
+ 
\frac{\xi_{\gamma}^{3}}
{\langle \gamma^{\bot}, \alpha_{0} \rangle^{2}} 
\right)
\\
=& N (S_{16} + S_{17}). 
\end{align*}
\end{proof}

We summarise the above calculations once.
%
\begin{corollary}\label{corollary:S3}
\quad
$S_{3} = S_{7} - S_{11} + S_{15} + S_{16} + S_{17}$. 
\end{corollary}

Next, let us calculate the terms $S_{1}$, $S_{2}$. 
\begin{lemma}\label{lemma:S1+S2} 
\quad 
$S_{1} + S_{2} 
= 
S_{11} - S_{15} + S_{18} + S_{19} + S_{20}$, 
where 
\begin{align*}
S_{18} 
&= 
- 
\sum_{\beta \in \mathcal{H} \atop \beta \not= \alpha_{0}} 
\frac{\langle \alpha_{0}, \beta \rangle C_{\beta}}
{\langle \alpha_{0}^{\bot}, \beta \rangle^{6}} 
\frac{1}{\xi_{\alpha_{0}}} 
(\xi_{\beta}^{3} \partial_{\xi, \alpha_{0}}^{3} \widetilde{P}_{0} 
- 3 \xi_{\alpha_{0}} \xi_{\beta}^{3} 
\partial_{\xi, \alpha_{0}}^{2} \widetilde{P}_{2}^{\beta} 
+ 3 \xi_{\alpha_{0}}^{3} \xi_{\beta} \xi_{\beta} 
\partial_{\xi, \beta}^{2} \widetilde{P}_{2}^{\alpha_{0}} 
- \xi_{\alpha_{0}}^{3} 
\partial_{\xi, \beta}^{3} \widetilde{P}_{0}), 
\\
S_{19} 
&= 
- C_{\alpha_{0}} 
\sum_{\beta \in \mathcal{H} \atop \beta \not= \alpha_{0}} 
\frac{\langle \alpha_{0}, \beta \rangle C_{\beta}}
{\langle \alpha_{0}^{\bot}, \beta \rangle^{6}} 
\frac{\xi_{\beta}^{3}}{\xi_{\alpha_{0}}} 
(\xi_{\alpha_{0}} \widetilde{P}_{4}^{\alpha_{0}, \beta} 
- 
\partial_{\xi, \alpha_{0}} \widetilde{P}_{2}^{\alpha_{0}}), 
\\
S_{20} 
&= 
- \frac{C_{\alpha_{0}}}{N} 
\sum_{\beta \in \mathcal{H} \atop \beta \not= \alpha_{0}} 
\frac{C_{\beta}}
{\langle \alpha_{0}^{\bot}, \beta \rangle^{3}} 
\xi_{\beta}^{3} 
\partial_{\xi, \alpha_{0}} \widetilde{Q}_{4}^{\beta, \alpha_{0}}. 
\end{align*}
\end{lemma}

\begin{proof}
Since 
\begin{align*}
\xi_{\beta} & 
(\partial_{\xi, \alpha}^{3} \widetilde{P}_{2}^{\beta} 
- C_{\alpha} \partial_{\xi, \alpha} 
\widetilde{P}_{4}^{\alpha, \beta}) 
+ 
\xi_{\alpha}  
(3 \partial_{\xi, \alpha}^{2} \partial_{\xi, \beta} 
\widetilde{P}_{2}^{\alpha} 
+ 
\partial_{\xi, \beta} 
\widetilde{P}_{4}^{\alpha}) 
\\
&= 
\langle \alpha, \beta \rangle 
(-3 \partial_{\xi, \alpha}^{2} \widetilde{P}_{2}^{\beta} 
+ C_{\alpha} \widetilde{P}_{4}^{\alpha, \beta} 
-3 \partial_{\xi, \alpha}^{2} \widetilde{P}_{2}^{\alpha} 
- \widetilde{P}_{4}^{\alpha}) 
- 6 |\alpha|^{2} \partial_{\xi, \alpha} \partial_{\xi, \beta} 
\widetilde{P}_{2}^{\alpha} 
+ \partial_{\xi, \alpha}^{3} \partial_{\xi, \beta} 
\widetilde{P}_{0} 
\\
& \qquad 
- C_{\alpha} \partial_{\xi, \alpha} 
\left( 
\partial_{\xi, \beta} \widetilde{P}_{2}^{\alpha} 
+ \frac{\langle \beta^{\bot}, \alpha \rangle^{3}}{N} 
\widetilde{Q}_{4}^{\beta, \alpha} 
\right) 
+ 3 \partial_{\xi, \alpha}^{3} \partial_{\xi, \beta} 
\widetilde{P}_{0} 
+ \partial_{\xi, \beta} 
\{
(C_{\alpha} + 6 |\alpha|^{2}) \partial_{\xi, \alpha} 
\widetilde{P}_{2}^{\alpha} 
- 4 \partial_{\xi, \alpha}^{3} \widetilde{P}_{0} 
\}
\\
&= 
\frac{\langle \alpha, \beta \rangle}{\xi_{\alpha}} 
\{
-3 \xi_{\alpha} \partial_{\xi, \alpha}^{2} \widetilde{P}_{2}^{\beta} 
-3 \xi_{\alpha} \partial_{\xi, \alpha}^{2} \widetilde{P}_{2}^{\alpha} 
- 6 |\alpha|^{2} \partial_{\xi, \alpha} \widetilde{P}_{2}^{\alpha} 
+ 4 \partial_{\xi, \alpha}^{3} \widetilde{P}_{0} 
+ C_{\alpha} 
(\xi_{\alpha} \widetilde{P}_{4}^{\alpha, \beta} 
- \partial_{\xi, \alpha} \widetilde{P}_{2}^{\alpha}) 
\} 
\\
& \qquad 
+ 
\frac{1}{N} C_{\alpha} \langle \alpha^{\bot}, \beta \rangle^{3} 
\partial_{\xi, \alpha} \widetilde{Q}_{4}^{\beta, \alpha} 
\\
& = 
\frac{\langle \alpha, \beta \rangle}{\xi_{\alpha}} 
\left\{
-3 \xi_{\alpha} \partial_{\xi, \alpha}^{2} \widetilde{P}_{2}^{\beta} 
+ \partial_{\xi, \alpha}^{3} \widetilde{P}_{0} 
+ C_{\alpha} 
(\xi_{\alpha} \widetilde{P}_{4}^{\alpha, \beta} 
- \partial_{\xi, \alpha} \widetilde{P}_{2}^{\alpha}) 
\right\} 
+ 
\frac{1}{N} C_{\alpha} \langle \alpha^{\bot}, \beta \rangle^{3} 
\partial_{\xi, \alpha} \widetilde{Q}_{4}^{\beta, \alpha}, 
\end{align*}
we have 
\begin{align*}
S_{1} &+ S_{2} 
\\
=& 
- 
\sum_{\beta \in \mathcal{H} \atop \beta \not= \alpha_{0}} 
\frac{\langle \alpha_{0}, \beta \rangle C_{\beta}}
{\langle \alpha_{0}^{\bot}, \beta \rangle^{6}} 
\frac{1}{\xi_{\alpha_{0}}} 
(\xi_{\beta}^{3} \partial_{\xi, \alpha_{0}}^{3} \widetilde{P}_{0} 
- 3 \xi_{\alpha_{0}} \xi_{\beta}^{3} 
\partial_{\xi, \alpha_{0}}^{2} \widetilde{P}_{2}^{\beta} 
+ 3 \xi_{\alpha_{0}}^{3} \xi_{\beta} \xi_{\beta} 
\partial_{\xi, \beta}^{2} \widetilde{P}_{2}^{\alpha_{0}} 
- \xi_{\alpha_{0}}^{3} 
\partial_{\xi, \beta}^{3} \widetilde{P}_{0}) 
\\
& 
- C_{\alpha_{0}} 
\sum_{\beta \in \mathcal{H} \atop \beta \not= \alpha_{0}} 
\frac{\langle \alpha_{0}, \beta \rangle C_{\beta}}
{\langle \alpha_{0}^{\bot}, \beta \rangle^{6}} 
\frac{\xi_{\beta}^{3}}{\xi_{\alpha_{0}}} 
(
\xi_{\alpha_{0}} \widetilde{P}_{4}^{\alpha_{0}, \beta} 
- 
\partial_{\xi, \alpha_{0}} \widetilde{P}_{2}^{\alpha_{0}} 
)
+ \sum_{\beta \in \mathcal{H} \atop \beta \not= \alpha_{0}} 
\frac{\langle \alpha_{0}, \beta \rangle C_{\beta}^{2}}
{\langle \alpha_{0}^{\bot}, \beta \rangle^{6}} 
\xi_{\alpha_{0}}^{2} 
(\xi_{\beta} \widetilde{P}_{4}^{\alpha_{0}, \beta} 
- \partial_{\xi, \beta} \widetilde{P}_{2}^{\beta})
\\
& 
- \frac{C_{\alpha_{0}}}{N} 
\sum_{\beta \in \mathcal{H} \atop \beta \not= \alpha_{0}} 
\frac{C_{\beta}}
{\langle \alpha_{0}^{\bot}, \beta \rangle^{3}} 
\xi_{\beta}^{3} 
\partial_{\xi, \alpha_{0}} \widetilde{Q}_{4}^{\beta, \alpha_{0}}
- 
\frac{1}{N} 
\sum_{\beta \in \mathcal{H} \atop \beta \not= \alpha_{0}} 
\frac{C_{\beta}^{2}}
{\langle \alpha_{0}^{\bot}, \beta \rangle^{3}} 
\xi_{\alpha_{0}}^{2} \xi_{\beta} 
\partial_{\xi, \beta} 
\widetilde{Q}_{4}^{\alpha_{0}, \beta}
\\
=& S_{18} + S_{19} + S_{11} + S_{20} - S_{15}. 
\end{align*} 
\end{proof}

\begin{lemma}\label{lemma:S7+S11+S17+S18} 
\quad
$S_{7} + S_{16} + S_{19} + S_{20} = 0$. 
\end{lemma} 
\begin{proof} 
\begin{align*}
S_{7}& + S_{20} 
\\
=& 
\frac{1}{N} 
\sum_{\beta, \gamma \in \mathcal{H} \atop 
\beta \not= \gamma \not= \alpha_{0} \not= \beta} 
\frac{\langle \beta^{\bot}, \gamma \rangle^{3} 
C_{\alpha_{0}} C_{\beta} C_{\gamma} d_{\alpha_{0}, \beta, \gamma}}
{\langle \alpha_{0}^{\bot}, \beta \rangle^{3} 
\langle \alpha_{0}^{\bot}, \gamma \rangle^{2}} 
\xi_{\alpha_{0}}^{2} \xi_{\beta} 
\partial_{\xi, \alpha_{0}} 
\widetilde{Q}_{\alpha_{0}, \beta, \gamma} 
- \frac{C_{\alpha_{0}}}{N} 
\sum_{\beta \in \mathcal{H} \atop \beta \not= \alpha_{0}} 
\frac{C_{\beta}}
{\langle \alpha_{0}^{\bot}, \beta \rangle^{3}} 
\xi_{\beta}^{3} 
\partial_{\xi, \alpha_{0}} \widetilde{Q}_{4}^{\beta, \alpha_{0}} 
\\
=& 
\frac{C_{\alpha_{0}}}{N} 
\sum_{\beta, \gamma \in \mathcal{H} \atop 
\beta \not= \gamma \not= \alpha_{0} \not= \beta} 
\frac{C_{\beta} C_{\gamma} d_{\alpha_{0}, \beta, \gamma}}
{\langle \alpha_{0}^{\bot}, \beta \rangle^{3}} 
\frac{\xi_{\beta}}{\xi_{\alpha_{0}}} 
\left(
\frac{(\langle \beta^{\bot}, \gamma \rangle \xi_{\alpha_{0}})^{3}}
{\langle \alpha_{0}^{\bot}, \gamma \rangle^{2}} 
- 
\langle \beta^{\bot}, \gamma \rangle \xi_{\alpha_{0}} \xi_{\beta}^{2} 
\right) 
\partial_{\xi, \alpha_{0}} 
\widetilde{Q}_{\alpha_{0}, \beta, \gamma} 
\\
=& 
\frac{C_{\alpha_{0}}}{N} 
\sum_{\beta, \gamma \in \mathcal{H} \atop 
\beta \not= \gamma \not= \alpha_{0} \not= \beta} 
\frac{C_{\beta} C_{\gamma} d_{\alpha_{0}, \beta, \gamma}}
{\langle \alpha_{0}^{\bot}, \beta \rangle^{3}} 
\frac{\xi_{\beta}}{\xi_{\alpha_{0}}} 
\partial_{\xi, \alpha_{0}} 
\widetilde{Q}_{\alpha_{0}, \beta, \gamma} 
\\
& \hspace{8mm} 
\times
\left(
\langle \alpha_{0}^{\bot}, \gamma \rangle \xi_{\beta}^{3} 
- 3 \langle \alpha_{0}^{\bot}, \beta \rangle 
\xi_{\beta}^{2} \xi_{\gamma} 
+ 3
\frac{\langle \alpha_{0}^{\bot}, \beta \rangle^{2}}
{\langle \alpha_{0}^{\bot}, \gamma \rangle} 
\xi_{\beta} \xi_{\gamma}^{2} 
- 
\frac{\langle \alpha_{0}^{\bot}, \beta \rangle^{3}}
{\langle \alpha_{0}^{\bot}, \gamma \rangle^{2}} 
\xi_{\gamma}^{3} 
- \langle \alpha_{0}^{\bot}, \gamma \rangle \xi_{\beta}^{3} 
+ \langle \alpha_{0}^{\bot}, \beta \rangle 
\xi_{\beta}^{2} \xi_{\gamma} 
\right)
\\
=& 
\frac{C_{\alpha_{0}}}{N} 
\sum_{\beta, \gamma \in \mathcal{H} \atop 
\beta \not= \gamma \not= \alpha_{0} \not= \beta} 
C_{\beta} C_{\gamma} d_{\alpha_{0}, \beta, \gamma}
\frac{\partial_{\xi, \alpha_{0}} 
\widetilde{Q}_{\alpha_{0}, \beta, \gamma}}
{\xi_{\alpha_{0}}} 
\left(
- 
\frac{2 \xi_{\beta}^{3} \xi_{\gamma}}
{\langle \alpha_{0}^{\bot}, \beta \rangle^{2}} 
+ 
\frac{3 \xi_{\beta}^{2} \xi_{\gamma}^{2}}
{\langle \alpha_{0}^{\bot}, \beta \rangle 
\langle \alpha_{0}^{\bot}, \gamma \rangle} 
- 
\frac{\xi_{\beta} \xi_{\gamma}^{3}}
{\langle \alpha_{0}^{\bot}, \gamma \rangle^{2}} 
\right)
\\
=& 
- 
\frac{C_{\alpha_{0}}}{N} 
\sum_{\beta, \gamma \in \mathcal{H} \atop 
\beta \not= \gamma \not= \alpha_{0} \not= \beta} 
\frac{C_{\beta} C_{\gamma} d_{\alpha_{0}, \beta, \gamma}}
{\langle \alpha_{0}^{\bot}, \beta \rangle^{2}}
\frac{\xi_{\beta}^{3} \xi_{\gamma}}
{\xi_{\alpha_{0}}} 
\partial_{\xi, \alpha_{0}} 
\widetilde{Q}_{\alpha_{0}, \beta, \gamma}
\\
=& 
- 
\frac{C_{\alpha_{0}}}{N} 
\sum_{\beta, \gamma \in \mathcal{H} \atop 
\beta \not= \gamma \not= \alpha_{0} \not= \beta} 
\frac{C_{\beta} C_{\gamma} d_{\alpha_{0}, \beta, \gamma}}
{\langle \alpha_{0}^{\bot}, \beta \rangle^{2}}
\frac{\xi_{\beta}^{3}}
{\xi_{\alpha_{0}}} 
\left\{
\partial_{\xi, \alpha_{0}} 
\left( 
\xi_{\gamma} 
\frac{\widetilde{P}_{2}^{\alpha_{0}} - \widetilde{P}_{2}^{\beta}}
{\langle \alpha_{0}^{\bot}, \beta \rangle \xi_{\gamma}} 
\right)
- \langle \gamma, \alpha_{0} \rangle 
\widetilde{Q}_{\alpha_{0}, \beta, \gamma}
\right\}
\\
=& 
- 
\frac{C_{\alpha_{0}}}{N} 
\sum_{\beta \in \mathcal{H} \atop 
\beta \not= \alpha_{0}} 
\frac{C_{\beta}}
{\langle \alpha_{0}^{\bot}, \beta \rangle^{3}}
\left(
\sum_{\gamma \in \mathcal{H} \atop 
\gamma \not= \alpha_{0}, \beta} 
C_{\gamma} d_{\alpha_{0}, \beta, \gamma}
\right) 
\frac{\xi_{\beta}^{3}}
{\xi_{\alpha_{0}}} 
\partial_{\xi, \alpha_{0}} 
(\widetilde{P}_{2}^{\alpha_{0}} - \widetilde{P}_{2}^{\beta}) 
\\
& \qquad 
+ 
\frac{C_{\alpha_{0}}}{N} 
\sum_{\beta, \gamma \in \mathcal{H} \atop 
\beta \not= \gamma \not= \alpha_{0} \not= \beta} 
\frac{\langle \gamma, \alpha_{0} \rangle 
C_{\beta} C_{\gamma} d_{\alpha_{0}, \beta, \gamma}}
{\langle \alpha_{0}^{\bot}, \beta \rangle^{2}}
\frac{\xi_{\beta}^{3}}
{\xi_{\alpha_{0}}} 
\widetilde{Q}_{\alpha_{0}, \beta, \gamma}
\\
=& 
- 
C_{\alpha_{0}} 
\sum_{\beta \in \mathcal{H} \atop 
\beta \not= \alpha_{0}} 
\frac{\langle \alpha_{0}, \beta \rangle C_{\beta}}
{\langle \alpha_{0}^{\bot}, \beta \rangle^{6}}
\frac{\xi_{\beta}^{3}}
{\xi_{\alpha_{0}}} 
\partial_{\xi, \alpha_{0}} 
(\widetilde{P}_{2}^{\alpha_{0}} - \widetilde{P}_{2}^{\beta}) 
+ 
\frac{C_{\alpha_{0}}}{N} 
\sum_{\beta, \gamma \in \mathcal{H} \atop 
\beta \not= \gamma \not= \alpha_{0} \not= \beta} 
\frac{\langle \gamma, \alpha_{0} \rangle 
C_{\beta} C_{\gamma} d_{\alpha_{0}, \beta, \gamma}}
{\langle \alpha_{0}^{\bot}, \beta \rangle^{2}}
\frac{\xi_{\beta}^{3}}
{\xi_{\alpha_{0}}} 
\widetilde{Q}_{\alpha_{0}, \beta, \gamma} 
\\
=& 
C_{\alpha_{0}} 
\sum_{\beta \in \mathcal{H} \atop \beta \not= \alpha_{0}} 
\frac{\langle \alpha_{0}, \beta \rangle C_{\beta}}
{\langle \alpha_{0}^{\bot}, \beta \rangle^{6}} 
\frac{\xi_{\beta}^{3}}{\xi_{\alpha_{0}}} 
\left(\partial_{\xi, \alpha_{0}} \widetilde{P}_{2}^{\beta} 
+ \frac{\langle \alpha_{0}^{\bot}, \beta \rangle^{3}}{N} 
\widetilde{Q}_{4}^{\alpha_{0}, \beta} 
- \partial_{\xi, \alpha_{0}} \widetilde{P}_{2}^{\alpha_{0}}
\right)
\\
& \qquad 
- 
\frac{C_{\alpha_{0}}}{N} 
\sum_{\beta, \gamma \in \mathcal{H} \atop 
\beta \not= \gamma \not= \alpha_{0} \not= \beta} 
\frac{C_{\beta} C_{\gamma} d_{\alpha_{0}, \beta, \gamma} 
}
{\langle \alpha_{0}^{\bot}, \beta \rangle^{3}} 
\frac{\xi_{\beta}^{3}}{\xi_{\alpha_{0}}} 
(\langle \alpha_{0}, \beta \rangle 
\langle \gamma^{\bot}, \alpha_{0} \rangle 
- \langle \gamma, \alpha_{0} \rangle 
\langle \alpha_{0}^{\bot}, \beta \rangle)
\widetilde{Q}_{\alpha_{0}, \beta, \gamma} 
\\
=& - S_{19} - S_{16}. 
\end{align*}
\end{proof}

\begin{lemma}\label{lemma:S12} 
\quad
$\displaystyle 
S_{17} 
= 
3 C_{\alpha_{0}} 
\sum_{\beta \in \mathcal{H} \atop \beta \not= \alpha_{0}} 
\frac{\langle \alpha_{0}, \beta \rangle 
|\beta|^{2} C_{\beta}}
{\langle \alpha_{0}^{\bot}, \beta \rangle^{5}}
\frac{D_{\theta} \widetilde{P}_{0}}{\xi_{\alpha_{0}}}$. 
\end{lemma} 
\begin{proof}
\begin{align*}
S_{17} =& 
\frac{3}{N} 
\sum_{\beta, \gamma \in \mathcal{H} \atop 
\beta \not= \gamma \not= \alpha_{0} \not= \beta} 
\frac{C_{\alpha_{0}} C_{\beta} C_{\gamma} 
d_{\alpha_{0}, \beta, \gamma}}
{\langle \alpha_{0}^{\bot}, \beta \rangle^{2} 
\langle \alpha_{0}^{\bot}, \gamma \rangle}
(-\langle \gamma^{\bot}, \alpha_{0} \rangle |\beta|^{2}
- \langle \alpha_{0}^{\bot}, \beta \rangle 
\langle \beta, \gamma \rangle) 
\xi_{\beta} \xi_{\gamma} 
\widetilde{Q}_{\alpha_{0}, \beta, \gamma} 
\\
=& 
\frac{3}{N} 
\sum_{\beta \in \mathcal{H} \atop 
\beta \not= \alpha_{0}} 
\frac{|\beta|^{2} C_{\alpha_{0}} C_{\beta}}
{\langle \alpha_{0}^{\bot}, \beta \rangle^{2}} 
\frac{D_{\theta} \widetilde{P}_{0}}{\xi_{\alpha_{0}}} 
\sum_{\gamma \in \mathcal{H} \atop 
\gamma \not= \alpha_{0}, \beta} 
C_{\gamma} 
d_{\alpha_{0}, \beta, \gamma}
- 
\frac{3}{N} 
\sum_{\beta, \gamma \in \mathcal{H} \atop 
\beta \not= \gamma \not= \alpha_{0} \not= \beta} 
\frac{\langle \beta, \gamma \rangle 
C_{\alpha_{0}} C_{\beta} C_{\gamma} 
d_{\alpha_{0}, \beta, \gamma}}
{\langle \alpha_{0}^{\bot}, \beta \rangle 
\langle \alpha_{0}^{\bot}, \gamma \rangle}
\xi_{\beta} \xi_{\gamma} 
\widetilde{Q}_{\alpha_{0}, \beta, \gamma}
\\
=&  
3 C_{\alpha_{0}} 
\sum_{\beta \in \mathcal{H} \atop \beta \not= \alpha_{0}} 
\frac{\langle \alpha_{0}, \beta \rangle 
|\beta|^{2} C_{\beta}}
{\langle \alpha_{0}^{\bot}, \beta \rangle^{5}}
\frac{D_{\theta} \widetilde{P}_{0}}{\xi_{\alpha_{0}}}. 
\end{align*}
\end{proof}
\begin{lemma}\label{lemma:D_theta^3} 
\quad 
$\displaystyle 
S_{18} = 
-6 |\alpha_{0}|^{2}
\sum_{\beta \in \mathcal{H} \atop \beta \not= \alpha_{0}} 
\frac{\langle \alpha_{0}, \beta \rangle |\beta|^{2} C_{\beta}}
{\langle \alpha_{0}^{\bot}, \beta \rangle^{5}}
\frac{D_{\theta} \widetilde{P}_{0}}{\xi_{\alpha_{0}}}$. 
\end{lemma} 
\begin{proof}
By using \eqref{eq:D_theta} and 
$|\alpha_{0}|^{2} \xi_{\beta} \partial_{\xi, \beta} 
+ 
|\beta|^{2} \xi_{\alpha_{0}} \partial_{\xi, \alpha_{0}} 
- \langle \alpha_{0}, \beta \rangle 
(\xi_{\beta} \partial_{\xi, \alpha_{0}} 
+ \xi_{\alpha_{0}} \partial_{\xi, \beta}) 
= 
\langle \alpha_{0}^{\bot}, \beta \rangle^{2} 
\langle \xi, \partial_{\xi} \rangle$,   
we can easily shown 
\begin{align*}
\xi_{\beta}^{3} \partial_{\xi, \alpha_{0}}^{3} \widetilde{P}_{0} 
&
- 3 \xi_{\alpha_{0}} \xi_{\beta}^{3} 
\partial_{\xi, \alpha_{0}}^{2} \widetilde{P}_{2}^{\beta} 
+ 3 \xi_{\alpha_{0}}^{3} \xi_{\beta} 
\partial_{\xi, \beta}^{2} \widetilde{P}_{2}^{\alpha_{0}} 
- \xi_{\alpha_{0}}^{3} 
\partial_{\xi, \beta}^{3} \widetilde{P}_{0} 
\\
=& 
\langle \alpha_{0}^{\bot}, \beta \rangle^{3} 
(D_{\theta}^{3} 
+ 3 \langle \xi, \partial_{\xi} \rangle D_{\theta} - 8 D_{\theta}) 
\widetilde{P}_{0} 
+ 
6 \langle \alpha_{0}^{\bot}, \beta \rangle |\alpha|^{2} |\beta|^{2} 
D_{\theta} 
\widetilde{P}_{0}. 
\end{align*}
Therefore, we have 
\begin{align*}
S_{18} =& - 
\left(
\sum_{\beta \in \mathcal{H} \atop \beta \not= \alpha_{0}} 
\frac{\langle \alpha_{0}, \beta \rangle C_{\beta}}
{\langle \alpha_{0}^{\bot}, \beta \rangle^{3}}
\right) 
\frac{
(D_{\theta}^{3} 
+ 3 \langle \xi, \partial_{\xi} \rangle D_{\theta} - 8 D_{\theta}) 
\widetilde{P}_{0}}
{\xi_{\alpha_{0}}}   
-6 
\sum_{\beta \in \mathcal{H} \atop \beta \not= \alpha_{0}} 
\frac{\langle \alpha_{0}, \beta \rangle C_{\beta}}
{\langle \alpha_{0}^{\bot}, \beta \rangle^{5}}
|\alpha_{0}|^{2} |\beta|^{2} 
\frac{D_{\theta} \widetilde{P}_{0}}{\xi_{\alpha_{0}}} 
\\
=& 
-6 |\alpha_{0}|^{2} 
\sum_{\beta \in \mathcal{H} \atop \beta \not= \alpha_{0}} 
\frac{\langle \alpha_{0}, \beta \rangle |\beta|^{2} C_{\beta}}
{\langle \alpha_{0}^{\bot}, \beta \rangle^{5}}
\frac{D_{\theta} \widetilde{P}_{0}}{\xi_{\alpha_{0}}}.
\end{align*}
\end{proof} 

\noindent\textsc{Proof} of Theorem~\ref{theorem:linear relation-2}. 

By the above long discussion, 
the equality $S_{1} + S_{2} + S_{3} = 0$ reduces to the equality
$S_{17} + S_{18} = 0$. 
By Lemma~\ref{lemma:S12} and Lemma~\ref{lemma:D_theta^3}, 
we have 
\[
0 
= S_{17} + S_{18} 
= 
3 (C_{\alpha_{0}} - 2 |\alpha_{0}|^{2}) 
\sum_{\beta \in \mathcal{H} \atop \beta \not= \alpha_{0}} 
\frac{\langle \alpha_{0}, \beta \rangle 
|\beta|^{2} C_{\beta}}
{\langle \alpha_{0}^{\bot}, \beta \rangle^{5}}
\frac{D_{\theta} \widetilde{P}_{0}}{\xi_{\alpha_{0}}}. 
\]
Since $\widetilde{P}_{0}$ is not a polynomial 
in $\xi_{1}^{2} + \xi_{2}^{2}$, 
$D_{\theta} \widetilde{P}_{0}$ is not zero, 
and the theorem is proved. 
\hspace*{\fill}$\square$\par\bigskip


\section{Possible deformation of root systems} 
\label{section:possibility} 

In this section, we investigate what kind of $\mathcal{H}$ and
$C_{\alpha}$ satisfy \eqref{eq:first relation} and 
\eqref{eq:second relation}. 

Let $\# \mathcal{H} = N$ and 
$\mathcal{H} = \{\alpha_{1}, \dots, \alpha_{N}\}$.  
For notational convenience, we define 
\begin{align*}
A_{ij} :=& 
\frac{\langle \alpha_{i}, \alpha_{j} \rangle}
{\langle \alpha_{i}^{\bot}, \alpha_{j} \rangle^{3}}
\qquad \mbox{(if $i \not= j$)},
& A_{ii} :=& 0 \qquad \mbox{(if $i = j$)}, 
& \mathcal{A} :=& (A_{ij})_{1 \leq i,j \leq N}, 
\\
B_{ij} :=& 
\frac{\langle \alpha_{i}, \alpha_{j} \rangle 
|\alpha_{i}|^{2} |\alpha_{j}|^{2}}
{\langle \alpha_{i}^{\bot}, \alpha_{j} \rangle^{5}}
\qquad \mbox{(if $i \not= j$)},
& 
B_{ii} :=& 
0 \qquad \mbox{(if $i = j$)}, 
& \mathcal{B} :=& (B_{ij})_{1 \leq i,j \leq N}, 
\\
C_{i} :=& C_{\alpha_{i}}, 
&
\boldsymbol{v} :=& 
{}^{t} (C_{1}, \dots, C_{N}). 
\end{align*}
Then, \eqref{eq:first relation} and \eqref{eq:second relation} are
equivalent to 
\begin{align}
& 
\mathcal{A} \boldsymbol{v} = \boldsymbol{0}, 
\label{eq:Av=0}
\\
& 
\mathrm{diag} 
(C_{1} - 2 |\alpha_{1}|^{2}, \dots, C_{N} - 2 |\alpha_{N}|^{2}) 
\mathcal{B} \boldsymbol{v} = \boldsymbol{0}, 
\label{eq:Bv=0}
\end{align}
respectively. 

If $N$ is odd, then $\det \mathcal{A} = 0$, since $\mathcal{A}$ is an
alternative matrix. 
Therefore, the solution space of \eqref{eq:Av=0} is at least one
dimensional. 
If the rank of $\mathcal{A}$ is $N-1$, 
the non-trivial solution of \eqref{eq:Av=0} is given by 
\begin{equation}
C_{i} = C (-1)^{i} \mathrm{Pf}_{i}(\mathcal{A})
\qquad 
(i = 1, 2, \dots, N), 
\end{equation}
where $C$ is a non-zero constant and 
$\mathrm{Pf}_{i}(\mathcal{A})$ is the Pfaffian of 
the $(N-1) \times (N-1)$ alternative matrix obtained by deleting the
$i$ th row and column of $\mathcal{A}$. 
Especially, if $N = 3$, then 
\begin{align}
& C_{1} = C A_{23}, 
& 
& C_{2} = C A_{31}, 
& 
& C_{3} = C A_{12}. 
\label{eq:C_i when H=3}
\end{align}
\begin{lemma}\label{lemma:C_1 not=2} 
If $N = 3$ and $C_{1} \not= 2 |\alpha_{1}|^{2}$, 
$\alpha_{2}$ and $\alpha_{3}$ are symmetric with
respect to the reflection $r_{\alpha_{1}}$. 
\end{lemma}
\begin{proof} 
Since $C_{\alpha_{1}} \not= 2 |\alpha_{1}|^{2}$, 
the equations \eqref{eq:Bv=0} and \eqref{eq:C_i when H=3}
imply 
\begin{align*}
B_{12} C_{2} + B_{13} C_{3} = 0 
& \quad \Leftrightarrow \quad 
- 
C 
\frac{
\langle \alpha_{1}, \alpha_{3} \rangle 
\langle \alpha_{1}, \alpha_{2} \rangle 
|\alpha_{2}|^{2}}
{\langle \alpha_{1}^{\bot}, \alpha_{3} \rangle^{3} 
\langle \alpha_{1}^{\bot}, \alpha_{2} \rangle^{5}}
+ 
C 
\frac{
\langle \alpha_{1}, \alpha_{2} \rangle 
\langle \alpha_{1}, \alpha_{3} \rangle 
|\alpha_{3}|^{2}}
{\langle \alpha_{1}^{\bot}, \alpha_{2} \rangle^{3} 
\langle \alpha_{1}^{\bot}, \alpha_{3} \rangle^{5}}
= 0  
\\
& \quad \Leftrightarrow \quad 
\langle \alpha_{1}^{\bot}, \alpha_{3} \rangle^{2} |\alpha_{2}|^{2} 
= 
\langle \alpha_{1}^{\bot}, \alpha_{2} \rangle^{3} |\alpha_{3}|^{2} 
\\
& \quad \Leftrightarrow \quad 
\left(
\frac{\langle \alpha_{1}^{\bot}, \alpha_{3} \rangle}
{|\alpha_{1}| |\alpha_{3}|}
\right)^{2} 
= 
\left(
\frac{\langle \alpha_{1}^{\bot}, \alpha_{2} \rangle}
{|\alpha_{1}| |\alpha_{2}|}
\right)^{2}
\\
& \quad \Leftrightarrow \quad 
\sin^{2} \theta_{2} = \sin^{2} \theta_{3}, 
\end{align*}
where $\theta_{i}$ are the angles from $\alpha_{1}$ to $\alpha_{i}$ 
($i = 2, 3$). 
Since $\alpha_{2}$ and $\alpha_{3}$ are not parallel, 
this implies the lemma. 
\end{proof} 
\begin{corollary}\label{corollary:A2, all not 2} 
If $\# \mathcal{H} = 3$ and (i) more than or equal to two of $C_{i}$'s
are not equal to $2 |\alpha_{i}|^{2}$ 
or 
(ii) all $C_{i}$'s are equal to $2 |\alpha_{i}|^{2}$, 
then $\mathcal{H}$ is a positive system of the $A_{2}$ type root
system and $L$ is the $A_{2}$ type CMS operator. 
\end{corollary}
\begin{proof}
The first assertion follows directly from the last lemma. 

If $C_{i} = 2 |\alpha_{i}|^{2}$ for $i = 1, 2, 3$, 
\eqref{eq:C_i when H=3} implies 
\begin{align*}
& 
\frac{\langle \alpha_{2}, \alpha_{3} \rangle}
{\langle \alpha_{2}^{\bot}, \alpha_{3} \rangle^{3} |\alpha_{1}|^{2}}  
= 
\frac{\langle \alpha_{3}, \alpha_{1} \rangle}
{\langle \alpha_{3}^{\bot}, \alpha_{1} \rangle^{3} |\alpha_{2}|^{2}} 
= 
\frac{\langle \alpha_{1}, \alpha_{2} \rangle}
{\langle \alpha_{1}^{\bot}, \alpha_{2} \rangle^{3} |\alpha_{3}|^{2}}
\\
& \Leftrightarrow \quad 
\cot(\theta_{3} - \theta_{2}) 
\{1 + \cot^{2} (\theta_{3} - \theta_{2})\} 
= 
- \cot \theta_{3} (1 + \cot^{2} \theta_{3}) 
= \cot \theta_{2} (1 + \cot^{2} \theta_{2}) 
\\
& \Leftrightarrow \quad 
\cot (\theta_{3} - \theta_{2}) 
= - \cot \theta_{3} 
= \cot \theta_{2} 
\\
& \Leftrightarrow \quad 
\cot \theta_{2} = - \cot \theta_{3} = \pm 1/\sqrt{3}. 
\end{align*}
By changing the norm of vectors if necessary, 
we may regard $\mathcal{H}$ to be a
positive system of $A_{2}$ type root system. 
\end{proof}
As a result of Lemma~\ref{lemma:C_1 not=2} and 
Corollary~\ref{corollary:A2, all not 2}, 
we obtain the following theorem. 
\begin{theorem}\label{theorem:A2} 
When $\# \mathcal{H} = 3$, the possible hypeplane arrangement
$\mathcal{H}$ is $\mathcal{H} = \{e_{1}, \pm a e_{1} + e_{2}\}$ 
($a \not= 0$). 
If $a \not= \pm 1/\sqrt{3}$, in other words if $\mathcal{H}$ is not a
positive system of $A_{2}$ type, 
the coupling constants for $\pm a e_{1} + e_{2}$ must be one. 
\end{theorem}

Next, let us consider the case $N=4$. 

\begin{proposition}\label{proposition:possibility, N=4}
If $\# \mathcal{H} = 4$, at least two vectors in $\mathcal{H}$ cross
at right angles. 
\end{proposition} 
\begin{proof} 
Put 
$\mathcal{H} 
= \{\alpha_{1}, \alpha_{2}, \alpha_{3}, \alpha_{4}\}$ 
and let $x_{i} = \cot \theta_{i}$ ($i = 2, 3, 4$), 
where $\theta_{i}$ is the angle from $\alpha_{1}$ to $\alpha_{i}$. 
Note that $x_{i}$ ($i = 2, 3, 4$) are all different since any two
vectors in $\mathcal{H}$ are not parallel. 
Since 
\begin{align*}
& A_{ij} = 
\frac{\langle \alpha_{i}, \alpha_{j} \rangle}
{\langle \alpha_{i}^{\bot}, \alpha_{j} \rangle^{3}} 
= \frac{1}{|\alpha_{i}|^{2} |\alpha_{j}|^{2}} 
\frac{\cos (\theta_{j} - \theta_{i})}
{\sin^{3} (\theta_{j} - \theta_{i})}
= 
\frac{1}{|\alpha_{i}|^{2} |\alpha_{j}|^{2}} 
\cot (\theta_{j} - \theta_{i}) 
(1 + \cot^{2} (\theta_{j} - \theta_{i})),
\\
& B_{ij} = 
\frac{\langle \alpha_{i}, \alpha_{j} \rangle 
|\alpha_{i}|^{2} |\alpha_{j}|^{2}}
{\langle \alpha_{i}^{\bot}, \alpha_{j} \rangle^{5}} 
= \frac{1}{|\alpha_{i}|^{2} |\alpha_{j}|^{2}} 
\frac{\cos (\theta_{j} - \theta_{i})}
{\sin^{5} (\theta_{j} - \theta_{i})}
= 
\frac{1}{|\alpha_{i}|^{2} |\alpha_{j}|^{2}} 
\cot (\theta_{j} - \theta_{i}) 
(1 + \cot^{2} (\theta_{j} - \theta_{i}))^{2}, 
\end{align*}
we have 
\begin{align*}
A_{1i} 
&= 
\frac{x_{i} (1 + x_{i}^{2})}{|\alpha_{1}|^{2} |\alpha_{i}|^{2}}, 
&
B_{1i} 
&= 
\frac{x_{i} (1 + x_{i}^{2})^{2}}{|\alpha_{1}|^{2} |\alpha_{i}|^{2}} 
& 
& (i = 2, 3, 4), 
\\
A_{ij} 
&= 
\frac{(1 + x_{i}^{2}) (1 + x_{j}^{2}) (1 + x_{i} x_{j})}
{|\alpha_{i}|^{2} |\alpha_{j}|^{2} (x_{i} - x_{j})^{3}}, 
& 
B_{ij} 
&= 
\frac{(1 + x_{i}^{2})^{2} (1 + x_{j}^{2})^{2} (1 + x_{i} x_{j})}
{|\alpha_{i}|^{2} |\alpha_{j}|^{2} (x_{i} - x_{j})^{5}} 
&
& (2 \leq i \not= j \leq 4). 
\end{align*}

Assume that any two vectors in $\mathcal{H}$ do not cross at right
angles and deduce contradiction. 
Note that  $x_{i} \not= 0$ for $i = 2, 3, 4$ and 
$1 + x_{i} x_{j} \not= 0$ for $2 \leq i \not= j \leq 4$ 
since $\langle \alpha_{i}, \alpha_{j} \rangle \not= 0$ for any $i, j$. 

We divide the proof of this theorem into three parts, since the method
of poof is different for the following cases: 
\begin{enumerate}
\item
More than or equal to three coupling constants are one. 
\item
More than or equal to three coupling constants are not one. 
\item
Just two coupling constants are one. 
\end{enumerate}

\noindent
(1) 
In this case, we may assume 
$C_{i} = 2 |\alpha_{i}|^{2}$ for $i = 2, 3, 4$. 
Let $p_{k} := \sum_{i = 2}^{4} x_{i}^{k}$ be the $k$-th power sum of
$x_{2}, x_{3}, x_{4}$. 
By \eqref{eq:Av=0} and \eqref{eq:Bv=0}, 
we have 
\[
\begin{cases} 
2(A_{12} |\alpha_{2}|^{2} 
+ A_{13} |\alpha_{3}|^{2} 
+ A_{14} |\alpha_{4}|^{2}) 
= 0 
\\
2( 
B_{12} |\alpha_{2}|^{2} 
+ B_{13} |\alpha_{3}|^{2} 
+ B_{14} |\alpha_{4}|^{2} 
)
= 0 
\end{cases}
\quad \Leftrightarrow \quad 
\begin{cases}
p_{1} + p_{3} = 0 
\\
p_{1} + 2 p_{3} + p_{5} = 0
\end{cases}
\quad \Leftrightarrow \quad 
p_{1} = - p_{3} = p_{5},  
\]
Since $6 p_{5} 
= p_{1}^{5} - 5 p_{1}^{3} p_{2} + 5 p_{1}^{2} p_{3} + 5 p_{2} p_{3}$, 
we have 
\begin{align*}
p_{1}
\{5(p_{1}^{2} + 1) p_{2} - (p_{1}^{4} - 5 p_{1}^{2} - 6)\} 
= 0 
& \quad \Leftrightarrow \quad 
p_{1} (p_{1}^{2} + 1) (5 p_{2} - p_{1}^{2} + 6) = 0 
\\
& \quad \Leftrightarrow \quad 
p_{1} (p_{1}^{2} + 1) 
\left(3 \sum_{i=2}^{4} x_{i}^{2} 
+ \sum_{2 \leq i < j \leq 4} (x_{i} - x_{j})^{2} + 6 
\right)
= 0
\\
& \quad \Rightarrow \quad 
p_{1} = 0. 
\end{align*}
If $p_{1} = 0$, 
then $p_{3} = x_{2}^{3} + x_{3}^{3} - (x_{2} + x_{3})^{3} 
= 3 x_{2} x_{3} x_{4} = 0$, 
which contradicts our assumption.

\noindent
(2) 
Assume that the coupling constants 
for $\alpha_{1}, \alpha_{2}, \alpha_{3}$ are not one. 
Then $C_{1}, \dots, C_{4}$ satisfy 
\begin{equation}\label{eq:B'v=0} 
\begin{pmatrix} 
0 & B_{12} & B_{13} & B_{14} 
\\
B_{21} & 0 & B_{23} & B_{24} 
\\
B_{31} & B_{32} & 0 & B_{34} 
\\
A_{41} & A_{42} & A_{43} & 0 
\end{pmatrix} 
\begin{pmatrix}
C_{1} \\ C_{2} \\ C_{3} \\ C_{4} 
\end{pmatrix} 
= 
\boldsymbol{0}. 
\end{equation}
Since all $C_{i}$'s are not zero, the determinant of the
coefficient matrix is zero; 
\begin{equation}\label{eq:detB'=0} 
(B_{12} B_{34} - B_{13} B_{24} + B_{14} B_{23}) 
(B_{12} A_{34} - B_{13} A_{24} + B_{23} A_{14}) = 0. 
\end{equation}

On the other hand, since $C_{1}, \dots, C_{4}$ satisfy
\eqref{eq:Av=0}, we have 
\[
\mathrm{Pf}(\mathcal{A}) 
=  A_{12} A_{34} - A_{13} A_{24} + A_{14} A_{23} 
= 0. 
\]
By our assumption that any two vectors in $\mathcal{H}$ do not cross
at right angles, each $A_{ij}$ is not zero. 
Therefore, the solution of \eqref{eq:Av=0} is expressed as 
\begin{align*}
& C_{1} = s A_{34}, 
& 
& C_{2} = t A_{34}, 
&
& C_{3} = -(s A_{14} + t A_{24}), 
& 
& C_{4} = s A_{13} + t A_{23}, 
&
& \quad (s, t \not= 0). 
\end{align*}
Since this satisfies \eqref{eq:B'v=0}, 
we have 
\[
s(A_{13} B_{14} - A_{14} B_{13}) 
+ t(B_{12} A_{34} - B_{13} A_{24} + B_{23} A_{14}) = 0. 
\]
Assume that $B_{12} A_{34} - B_{13} A_{24} + B_{23} A_{14} = 0$. 
In this case, 
\[
A_{13} B_{14} = A_{14} B_{13}
\quad \Leftrightarrow \quad 
x_{3} (1 + x_{3}^{2}) x_{4} (1 + x_{4}^{2})^{2} 
= x_{4} (1 + x_{4}^{2}) x_{3} (1 + x_{3}^{2})^{2} 
\]
and this implies $x_{3} = - x_{4}$, since $x_{3}, x_{4} \not= 0$ and 
$x_{3} \not= x_{4}$. 
Therefore, 
$A_{14} |\alpha_{4}|^{2} = - A_{13} |\alpha_{3}|^{2}$, 
$B_{14} |\alpha_{4}|^{2} = - B_{13} |\alpha_{3}|^{2}$, and we have 
\[
\begin{cases} 
A_{12} C_{2} + A_{13} C_{3} + A_{14} C_{4} = 0 
\\
B_{12} C_{2} + B_{13} C_{3} + B_{14} C_{4} = 0 
\end{cases} 
\quad \Leftrightarrow \quad 
\begin{cases} 
A_{12} C_{2} 
+ A_{13} (C_{3} -|\alpha_{3}|^{2} C_{4} / |\alpha_{4}|^{2}) 
= 0 
\\
B_{12} C_{2} 
+ B_{13} (C_{3} -|\alpha_{3}|^{2} C_{4} / |\alpha_{4}|^{2}) 
= 0. 
\end{cases} 
\]
By these equations, we have 
$(B_{13} A_{12} - B_{12} A_{13}) C_{2} = 0$,
which implies $x_{2}^{2} = x_{3}^{2}$. 
Since $x_{2} \not= x_{3}$, we have $x_{2} = - x_{3} = x_{4}$. 
But this contradicts the condition $x_{2} \not= x_{4}$. 

Therefore, $B_{12} A_{34} - B_{13} A_{24} + B_{23} A_{14} \not= 0$, 
and $A_{ij}$, $B_{ij}$ satisfy 
\[
\begin{cases} 
A_{12} A_{34} - A_{13} A_{24} + A_{14} A_{23} = 0 
\\
B_{12} B_{34} - B_{13} B_{24} + B_{14} B_{23} = 0 
\end{cases}  
\quad \Leftrightarrow \quad 
\begin{cases} 
\frac{x_{2} (1 + x_{3} x_{4})}{(x_{3} - x_{4})^{3}} 
+ 
\frac{x_{3} (1 + x_{4} x_{2})}{(x_{4} - x_{2})^{3}} 
+ 
\frac{x_{4} (1 + x_{2} x_{3})}{(x_{2} - x_{3})^{3}} 
= 0 
\\
\frac{x_{2} (1 + x_{3} x_{4})}{(x_{3} - x_{4})^{5}} 
+ 
\frac{x_{3} (1 + x_{4} x_{2})}{(x_{4} - x_{2})^{5}} 
+ 
\frac{x_{4} (1 + x_{2} x_{3})}{(x_{2} - x_{3})^{5}} 
= 0, 
\end{cases}
\]
by \eqref{eq:detB'=0}. 
From these equations, we have 
\begin{align}
& 
\frac{x_{2} (1 + x_{3} x_{4}) (2 x_{3} - x_{2} - x_{4})}
{(x_{3} - x_{4})^{6}} 
= 
\frac{x_{3} (1 + x_{2} x_{4}) (2 x_{2} - x_{3} - x_{4})}
{(x_{2} - x_{4})^{6}} 
\label{eq:x2 and x3} 
\\
& \Leftrightarrow \quad 
x_{2} (1 + x_{3} x_{4}) 
(2 x_{3} - x_{2} - x_{4}) (2 x_{4} - x_{2} - x_{3}) 
\notag 
\\
& \qquad \qquad \qquad 
\times 
\{(x_{2} - x_{4})^{4} + (x_{2} - x_{4})^{2} (x_{3} - x_{4})^{2} 
+ (x_{3} - x_{4})^{4}\} 
\notag 
\\
& \qquad \qquad 
+ (x_{2} + x_{3} + x_{4} + 3 x_{2} x_{3} x_{4}) (x_{3} - x_{4})^{6} 
= 0. 
\label{eq:x2 and x3, no.2} 
\end{align}
Here, we used $x_{2} \not= x_{3}$. 
Similarly, we have 
\begin{align}
& 
x_{2} (1 + x_{3} x_{4}) 
(2 x_{4} - x_{2} - x_{3}) (2 x_{3} - x_{2} - x_{4}) 
\notag 
\\
& \qquad \qquad 
\times 
\{(x_{2} - x_{3})^{4} + (x_{2} - x_{3})^{2} (x_{3} - x_{4})^{2} 
+ (x_{3} - x_{4})^{4}\} 
\notag 
\\
& \qquad 
+ (x_{2} + x_{3} + x_{4} + 3 x_{2} x_{3} x_{4}) (x_{3} - x_{4})^{6} 
= 0. 
\label{eq:x2 and x4, no.2} 
\end{align}
By \eqref{eq:x2 and x3, no.2} and \eqref{eq:x2 and x4, no.2}, 
we have 
\begin{align*}
& 
x_{2} (1 + x_{3} x_{4}) 
(2 x_{2} - x_{3} - x_{4}) (2 x_{3} - x_{4} - x_{2}) 
(2 x_{4} - x_{2} - x_{3}) 
\\
& \qquad \qquad 
\times
\{(x_{2} - x_{4})^{2} + (x_{2} - x_{3})^{2} + (x_{3} - x_{4})^{2}\} 
= 0.
\end{align*}
Since $x_{2} \not= 0$, $1 + x_{3} x_{4} \not= 0$ 
and $x_{2}, x_{3}, x_{4}$ are all different, 
$2x_{2} = x_{3} + x_{4}$, $2x_{3} = x_{4} + x_{2}$ or 
$2x_{4} = x_{2} + x_{3}$. 
But this is impossible since 
we obtain $x_{2} = x_{3}$ or $x_{4}(1 + x_{2} x_{3}) = 0$ 
from \eqref{eq:x2 and x3}. 

\noindent
(3) 
Let us assume the coupling constants for $\alpha_{3}$ and $\alpha_{4}$
are $1$ and those for $\alpha_{1}$ and $\alpha_{2}$ are not $1$. 
In this case, all the conditions for $A_{ij}$, $B_{ij}$ and $C_{i}$
are 
\begin{align*}
& 
A_{12} C_{2} + 2 |\alpha_{3}|^{2} A_{13} + 2 |\alpha_{4}|^{2} A_{14} 
= 0, 
& 
& 
B_{12} C_{2} + 2 |\alpha_{3}|^{2} B_{13} + 2 |\alpha_{4}|^{2} B_{14} 
= 0, 
\\
& 
A_{21} C_{1} + 2 |\alpha_{3}|^{2} A_{23} + 2 |\alpha_{4}|^{2} A_{24} 
= 0, 
& 
& 
B_{21} C_{1} + 2 |\alpha_{3}|^{2} B_{23} + 2 |\alpha_{4}|^{2} B_{24} 
= 0, 
\\
& 
\mathrm{Pf}(A) = A_{12} A_{34} - A_{13} A_{24} + A_{14} A_{23} = 0. 
& 
& 
\end{align*}
By eliminating $C_{1}$ and $C_{2}$, 
we have 
\begin{align}
& 
x_{2}^{2} = 
\frac{x_{3}^{2} - x_{3} x_{4} + x_{4}^{2} 
+ x_{3}^{4} - x_{3}^{3} x_{4} + x_{3}^{2} x_{4}^{2} 
- x_{3} x_{4}^{3} + x_{4}^{4}}
{1 + x_{3}^{2} - x_{3} x_{4} + x_{4}^{2}}, 
\label{eq:x2^2}
\\
& 
\frac{(1 + x_{3}^{2}) (1 + x_{2} x_{3}) 
(1 + 2 x_{2} x_{3} - x_{2}^{2})}
{(x_{2} - x_{3})^{5}} 
+ 
\frac{(1 + x_{4}^{2}) (1 + x_{2} x_{4}) 
(1 + 2 x_{2} x_{4} - x_{2}^{2})}
{(x_{2} - x_{4})^{5}} 
= 0, 
\label{eq:x2-x3x4}
\\
& 
\frac{x_{2} (1 + x_{3} x_{4})}{(x_{3} - x_{4})^{3}} 
+ 
\frac{x_{3} (1 + x_{4} x_{2})}{(x_{4} - x_{2})^{3}} 
+ 
\frac{x_{4} (1 + x_{2} x_{3})}{(x_{2} - x_{3})^{3}} 
= 0. 
\label{eq:PfA, explicit}
\end{align}
Let $F(x_{2})$ and $G(x_{2})$ be the left hand sides of
\eqref{eq:x2-x3x4} and \eqref{eq:PfA, explicit}, respectively. 
Since $\{F(x_{2}) G(-x_{2}) - F(-x_{2}) G(x_{2})\}/x_{2}$ is an even
polynomial in $x_{2}$, it is a polynomial in $x_{2}^{2}$. 
By substituting \eqref{eq:x2^2} for it 
and putting $x_{3} = s + t$, $x_{4} = s - t$, 
we have 
\begin{align*}
t^{6} & (s^{2} - t^{2}) 
\\
& \times 
(3 s^{2} + 24 s^{4} + 75 s^{6} + 120 s^{8} + 105 s^{10} + 48 s^{12} 
+ 9 s^{14} + 3 t^{2} + 92 s^{2} t^{2} + 559 s^{4} t^{2} 
+ 1408 s^{6} t^{2} 
\\
& \qquad 
+ 1745 s^{8} t^{2} + 1060 s^{10} t^{2} + 253 s^{12} t^{2} 
+ 44 t^{4} + 889 s^{2} t^{4} + 4368 s^{4} t^{4} + 8658 s^{6} t^{4} 
+ 7580 s^{8} t^{4} 
\\
& \qquad 
+ 2445 s^{10} t^{4} + 237 t^{6} + 3744 s^{2} t^{6} + 14538 s^{4} t^{6} 
+ 20616 s^{6} t^{6} + 9777 s^{8} t^{6} + 600 t^{8} + 7461 s^{2} t^{8} 
\\
& \qquad 
+ 20616 s^{4} t^{8} + 15675 s^{6} t^{8} + 773 t^{10} 
+ 6932 s^{2} t^{10} + 10263 s^{4} t^{10} + 492 t^{12} 
+ 2415 s^{2} t^{12} + 123 t^{14})
\\
& \times 
(5 s^{4} + 30 s^{6} + 75 s^{8} + 100 s^{10} + 75 s^{12} + 30 s^{14} 
+ 5 s^{16} + 50 s^{2} t^{2} + 478 s^{4} t^{2} + 1744 s^{6} t^{2} 
+ 3196 s^{8} t^{2} 
\\
& \qquad 
+ 3154 s^{10} t^{2} + 1606 s^{12} t^{2} + 332 s^{14} t^{2} 
+ 121 t^{4} + 1818 s^{2} t^{4} + 9358 s^{4} t^{4} + 22792 s^{6} t^{4} 
+ 28609 s^{8} t^{4} 
\\
& \qquad 
+ 17910 s^{10} t^{4} + 4432 s^{12} t^{4} + 1386 t^{6} 
+ 14280 s^{2} t^{6} + 52248 s^{4} t^{6} + 87916 s^{6} t^{6} 
+ 69246 s^{8} t^{6} 
\\
& \qquad 
+ 20684 s^{10} t^{6} + 6543 t^{8} + 48852 s^{2} t^{8} 
+ 123037 s^{4} t^{8} + 126170 s^{6} t^{8} + 45458 s^{8} t^{8} 
+ 16172 t^{10} 
\\
& \qquad 
+ 83490 s^{2} t^{10} + 126546 s^{4} t^{10} + 54532 s^{6} t^{10} 
+ 21879 t^{12} + 69266 s^{2} t^{12} + 45080 s^{4} t^{12} 
+ 15210 t^{14} 
\\
& \qquad 
+ 21860 s^{2} t^{14} + 4225 t^{16}) 
\\
& = 0. 
\end{align*}
This equation implies $t = 0$, $s = \pm t$ or $(s, t) = (0, 0)$, 
but these are impossible 
since $x_{3}, x_{4} \not= 0$ and $x_{3} \not= x_{4}$. 
\end{proof}

\begin{theorem}\label{theorem:4 lines arrangement}
When $\# \mathcal{H} = 4$, 
the possible singular locus $\mathcal{H}$ is 
$\mathcal{H} = \{e_{1}, e_{2}, \pm a e_{1} + e_{2}\}$ ($a \not =0$). 
\end{theorem} 
\begin{proof}
By Proposition~\ref{proposition:possibility, N=4}, at least two vectors in
$\mathcal{H}$, say $\alpha_{1}$ and $\alpha_{2}$, cross at right
angles. 
In this case, we have $A_{12} = 0$ and 
we may assume $\alpha_{2} = \alpha_{1}^{\bot}$. 
Note that 
$\alpha_{2}^{\bot} = \alpha_{1}^{\bot \bot} = -\alpha_{1}$. 
Then, 
\begin{align*}
\mathrm{Pf}(\mathcal{A}) = -A_{13} A_{24} + A_{14} A_{23} =0 
\quad & \Leftrightarrow \quad 
\frac{\langle \alpha_{1}, \alpha_{3} \rangle}
{\langle \alpha_{1}^{\bot}, \alpha_{3} \rangle^{3}} 
\frac{\langle \alpha_{2}, \alpha_{4} \rangle}
{\langle \alpha_{2}^{\bot}, \alpha_{4} \rangle^{3}} 
= 
\frac{\langle \alpha_{1}, \alpha_{4} \rangle}
{\langle \alpha_{1}^{\bot}, \alpha_{4} \rangle^{3}} 
\frac{\langle \alpha_{2}, \alpha_{3} \rangle}
{\langle \alpha_{2}^{\bot}, \alpha_{3} \rangle^{3}} 
\\
& \Leftrightarrow \quad 
\frac{\langle \alpha_{1}, \alpha_{3} \rangle}
{\langle \alpha_{1}^{\bot}, \alpha_{3} \rangle^{3}} 
(-1) 
\frac{\langle \alpha_{1}^{\bot}, \alpha_{4} \rangle}
{\langle \alpha_{1}, \alpha_{4} \rangle^{3}} 
= 
\frac{\langle \alpha_{1}, \alpha_{4} \rangle}
{\langle \alpha_{1}^{\bot}, \alpha_{4} \rangle^{3}} 
(-1) 
\frac{\langle \alpha_{1}^{\bot}, \alpha_{3} \rangle}
{\langle \alpha_{1}, \alpha_{3} \rangle^{3}} 
\\
& \Leftrightarrow \quad 
\left(
\frac{\langle \alpha_{1}, \alpha_{3} \rangle}
{\langle \alpha_{1}^{\bot}, \alpha_{3} \rangle} 
\right)^{4} 
= 
\left(
\frac{\langle \alpha_{1}, \alpha_{4} \rangle}
{\langle \alpha_{1}^{\bot}, \alpha_{4} \rangle} 
\right)^{4} 
\\
& \Leftrightarrow \quad 
x_{3}^{4} = x_{4}^{4} 
\\
& \Leftrightarrow \quad 
x_{3} = - x_{4},
\end{align*}
since $x_{3} \not= x_{4}$. 
By an appropriate coordinate change, 
we may assume $\alpha_{1} = e_{1}$. 
Then $\alpha_{2} = e_{1}^{\bot} = e_{2}$ and, by changing the norm of
$\alpha_{3}, \alpha_{4}$ if necessary, 
we have $\alpha_{3} = x_{3} e_{1} + e_{2}$ and 
$\alpha_{4} = - x_{3} e_{1} + e_{2}$. 
\end{proof}



\section{New deformation of $B_{2}$ type commutative pair}
\label{section:new deformation} 

In this section, we construct a pair of commuting differential
operators with the hyperplane 
arrangement 
$\mathcal{H} = \{\alpha_{1} := e_{1}, 
\alpha_{2} := e_{2}, 
\alpha_{\pm} := \pm a e_{1} + e_{2}\}$. 
If $a = \pm 1$, then $\mathcal{H}$ is the positive system of $B_{2}$
type root system and the commuting differential operators are known. 
Therefore, we assume $a \not= \pm 1$ in this section. 

If the coupling constants for $\alpha_{\pm}$ are one, 
the existence of such commuting operators is proved by
Veselov-Feigin-Chalykh for rational or trigonometric potential cases
(\cite{CFV}). 
Here, we consider the case 
where the coupling constants for $\alpha_{\pm}$ are two. 
In this case, there exists a pair of commuting differential
operators $L$ and $P$ for rational, trigonometric and even elliptic
cases. 
Remember the lowest order of the commutant $P$ for the original CMS
model is four. 
But in our case, we can not find a fourth order commutant $P$
because no fourth order operator satisfies
Proposition~\ref{proposition:review-2}. 
The lowest order of a commutant $P$ is six. 

Let the coupling constants for $\alpha_{\pm}$ be two and 
let $C_{\pm} : = C_{\alpha_{\pm}} = 2 \cdot (2 + 1) |\alpha_{\pm}|^{2} 
= 6 (a^{2} + 1)$. 
Since $C_{1}, C_{2}$ satisfy \eqref{eq:Av=0}, \eqref{eq:Bv=0}, 
we have 
\begin{align}
& a C_{1} + \frac{1}{a^{3}} C_{2} 
- \frac{1 - a^{2}}{8 a^{3}} 6(a^{2} + 1) 
= 0, 
\qquad 
a C_{1} + \frac{1}{a^{5}} C_{2} 
- \frac{(1 - a^{2})(a^{2} + 1)}{32 a^{5}} 6(a^{2} + 1) 
= 0 
\notag 
\\
& \Leftrightarrow \qquad 
C_{1} = 
\frac{3}{16 a^{2}} (a^{2} + 1) (3 a^{-2} - 1), 
\qquad 
C_{2} = \frac{3}{16} (a^{2} + 1) (3 a^{2} - 1). 
\notag
\end{align}
By these equations, $a \not = \pm \sqrt{3}, \pm 1/\sqrt{3}$, 
since $C_{1}, C_{2} \not= 0$.

For $\alpha \in \mathcal{H}$, 
let $u_{\alpha}$ be a function of the form 
\begin{align*}
& u_{\alpha} = u_{\alpha}(x_{\alpha}), 
& 
& u_{\alpha}(t) 
= \frac{C_{\alpha}}{t^{2}} 
+ \mbox{(real analytic at $t=0$)}, 
\end{align*}
and consider the equation $[L, P] = 0$ for 
\[
L = 
- (\partial_{x_{1}}^{2} + \partial_{x_{2}}^{2}) 
+ 
\sum_{\alpha \in \mathcal{H}} u_{\alpha}. 
\]

If $a$ is generic, $C_{1}, C_{2}$ are not 
of the form $k(k+1) |\alpha_{i}|^{2}$ ($k \in \boldsymbol{Z}$) 
for $i = 1, 2$. 
Therefore, the principal symbol $\widetilde{P}_{0}$ of $P$ is an even
polynomial in $\xi_{1}$ and $\xi_{2}$ because of 
Proposition~\ref{proposition:review-2} (1). 
Moreover, since the coupling constants for $\alpha_{\pm}$ are two, 
$\widetilde{P}_{0}$ must satisfy 
$\lim_{\xi_{\alpha_{\pm} \to 0}} 
\partial_{\xi, \alpha_{\pm}} \widetilde{P}_{0} 
= 
\lim_{\xi_{\alpha_{\pm} \to 0}} 
\partial_{\xi, \alpha_{\pm}}^{3} \widetilde{P}_{0} 
= 0$ because of Proposition~\ref{proposition:review-2} (2). 
Such a polynomial of degree six is unique up to constant multiple and
modulo $(\xi_{1}^{2} + \xi_{2}^{2})^{3}$. 
We choose 
\[
\widetilde{P}_{0} 
= 
a(4 - a^{2}) \xi_{1}^{6} 
+ 5a \xi_{1}^{4} \xi_{2}^{2} 
+ 5 a^{-1} \xi_{1}^{2} \xi_{2}^{4} 
+ a^{-1}(4 - a^{-2}) \xi_{2}^{6}. 
\]

\begin{proposition}\label{proposition:new potential functions} 
If $a^{2} \not= 7/3, 3/7, (13 \pm 4 \sqrt{10})/3$, 
then we have 
\begin{align*}
& 
u_{1}(t) 
= \frac{3(3 a^{-2} - 1) (a^{2} + 1)}{4} \wp(2 a t), 
&
& 
u_{2}(t) 
= \frac{3 (3 a^{2} - 1) (a^{2} + 1)}{4} \wp(2 t), 
&
& 
u_{\pm}(t) 
= 6 (a^{2} + 1) \wp(t), 
\end{align*}
modulo constant factors. 
\end{proposition}

\noindent
\textsc{Sketch of proof.} 
Let us consider the equation \eqref{eq:compatibility for P2}. 
Even if the potential functions $u_{\alpha}$ are not rational, 
we can show that $F(x, \xi)$ in \eqref{eq:compatibility for P2} is a
polynomial in $\xi$ of degree two. 
Since 
$[\partial_{x_{2}} \partial_{\xi_{1}} 
- \partial_{x_{1}} \partial_{\xi_{2}}, 
\langle \xi, \partial_{x} \rangle] 
= 0$, we have 
\begin{equation}\label{eq:compatibility condition for P4} 
(\partial_{x_{2}} \partial_{\xi_{1}} 
- \partial_{x_{1}} \partial_{\xi_{2}})^{3} 
\left(
\sum_{\alpha, \beta \in \mathcal{H} \atop \alpha \not= \beta} 
u_{\alpha}' u_{\beta} \partial_{\xi, \alpha} 
\widetilde{P}_{2}^{\beta} 
\right)
= 0. 
\end{equation}
Here, we used $\partial_{\xi, \alpha^{\bot}}^{3} \partial_{\xi, \alpha}
\widetilde{P}_{2}^{\alpha} = 0$ for any $\alpha \in \mathcal{H}$, 
which is a direct consequence of 
$\lim_{\xi_{\alpha} \to 0} \partial_{\xi, \alpha}^{3} 
\widetilde{P}_{0} 
= 0$. 

Choose $\alpha_{0} \in \mathcal{H}$. 
By taking the limit
$\lim_{x_{\alpha_{0}} \to 0} x_{\alpha_{0}}^{6} 
\times \eqref{eq:compatibility condition for P4}$, we obtain 
\[
\sum_{\beta \in \mathcal{H} \atop \beta \not= \alpha_{0}} 
(\partial_{\xi, \alpha_{0}^{\bot}}^{3} \partial_{\xi, \alpha_{0}} 
\widetilde{P}_{2}^{\beta}) 
u_{\beta} (\langle \alpha_{0}^{\bot}, \beta \rangle t) = 0, 
\]
Similarly, we obtain 
\[
\sum_{\beta \in \mathcal{H} \atop \beta \not= \alpha_{0}} 
\partial_{\xi, \alpha_{0}^{\bot}} 
(\partial_{\xi, \alpha_{0}}^{3} \widetilde{P}_{2}^{\beta} 
- 
C_{\alpha_{0}}
\partial_{\xi, \alpha_{0}} \widetilde{P}_{4}^{\alpha_{0}, \beta})
u_{\beta} (\langle \alpha_{0}^{\bot}, \beta \rangle t) = 0 
\]
from \eqref{eq:compatibility for P4}. 
These equations imply 
\begin{align*}
& (3 a^{-2} - 7) \{u_{+}(t) - u_{-}(t)\} = 0, 
\\
& 
(3 a^{2} - 7) \{u_{+}(-at) - u_{-}(at)\} = 0, 
\\
& 2 a^{2} u_{1}(-t) - 2 u_{2}(a t) + (a^{2} - 1) u_{-}(2 a t) = 0, 
\\
& 
2 a^{2} u_{1}(-t) - 2 u_{2}(-a t) + (a^{2} - 1) u_{+}(- 2 a t) = 0, 
\\
& 3 a (a^{2} - 1) (a^{2} + 1)^{2} 
(3 a^{4} - 26 a^{2} + 3) 
\{8 a^{2} u_{1}(t) + (a^{2} - 3) u_{+}(2 a t)\} 
= 0.  
\end{align*}
Here, we have abbreviated $u_{\alpha_{1}}$ to $u_{1}$ etc. 
If $a^{2} \not= 7/3, 3/7, (13 \pm 4 \sqrt{10})/3$, we obtain from
these that $u_{+}(t)$ is an even function, 
$u_{1}(t) = (3 a^{-2} - 1) u_{+}(2 a t) / 8$, 
$u_{2}(t) = (3 a^{2} - 1) u_{+}(2 t) / 8$ 
and $u_{-}(t) = u_{+}(t)$. 

Finally, we can show $u_{+}(t) = 6(a^{2} + 1) \wp(t)$ by 
the same method as in \S7 of \cite{OS}, 
namely, by studying the coefficients in the Laurent expansion of 
\eqref{eq:compatibility condition for P4} as a function of
$x_{\alpha_{+}}$. 
\hspace*{\fill}$\square$\par\bigskip

For such potential function, we can construct a commutant $P$ of $L$. 
Since we can check the commutativity by a direct method, 
we omit the proof and write the conclusion only. 

Before the statement of theorem, 
we introduce some notation. 
Let $g_{2}$, $g_{3}$ be the invariants of $\wp$ appearing in the
differential equation $(\wp')^{2} = 4 \wp^{3} - g_{2} \wp - g_{3}$. 
As above, we abbreviate $u_{\alpha_{1}}$ to $u_{1}$ etc. 
We put 
\begin{align*}
& L_{1} := \partial_{x_{1}}^{2} - u_{1}, 
& 
& L_{2} := \partial_{x_{2}}^{2} - u_{2}, 
& 
& L_{\pm} := \partial_{x, \alpha_{\pm}}^{2} - (a^{2} + 1) u_{\pm}  
\end{align*}
and
\[
A_{\pm}(5) 
:= \partial_{x, \alpha_{\pm}}^{5} 
- \frac{5}{2} u_{\pm} \partial_{x, \alpha_{\pm}}^{3} 
- \frac{15}{4} u_{\pm}' \partial_{x, \alpha_{\pm}}^{2} 
+ \frac{1}{8} (a^{2} + 1)^{2} 
\{15 u_{\pm}^{2} - 25 (a^{2} + 1) u_{\pm}''\} 
\partial_{x, \alpha_{\pm}}. 
\]
Note that $- (\partial_{x_{1}}^{2} + \partial_{x_{2}}^{2}) + u_{\pm}$ 
commutes with $L_{\pm}$, 
since $\partial_{x_{1}}^{2} + \partial_{x_{2}}^{2} 
= (a^{2} + 1)^{-1} 
(\partial_{x, \alpha_{\pm}}^{2} 
+ \partial_{x, \alpha_{\pm}^{\bot}}^{2})$. 
Moreover, since the coupling constants for $\alpha_{\pm}$ are two,
$L_{\pm}$ has a commutant of order five (\cite{BC}). 
The operator 
$A_{\pm}(5) - (21/8) (a^{2} + 1)^{4} g_{2} 
\partial_{x, \alpha_{\pm}}$ is such a commutant.

\begin{theorem}\label{theorem:new commutative pair} 
Let 
\[
L = 
- (\partial_{x_{1}}^{2} + \partial_{x_{2}}^{2}) 
+ u_{1} + u_{2} + u_{+} + u_{-}. 
\]
\end{theorem}
Then the following operator commutes with $L$. 
\begin{align*}
P =& 
a(4-a^{2}) L_{1}^{3} + 5 a L_{1}^{2} L_{2} + 5 a^{-1} L_{1} L_{2}^{2} 
+ a^{-1} (4 - a^{-2}) L_{2}^{3} + P_{2} 
\\
& \quad 
+ 
P_{4}
+ 
\frac{1}{2} \langle \partial_{x}, \partial_{\xi} \rangle 
\left. 
\widetilde{P}_{4}
\right|_{\xi \to \partial_{x}} 
+ \frac{1}{8} 
\langle \partial_{x}, \partial_{\xi} \rangle^{2} 
\widetilde{P}_{4} + Q_{4} 
+ P_{6}, 
\end{align*}
where 
\begin{align*}
P_{2} :=& \frac{1}{(a^{2} + 1)^{4}} 
[
- 20 a (a^{2} + 1) 
(u_{+} \partial_{x, \alpha_{+}^{\bot}}^{4} 
+ u_{-} \partial_{x, \alpha_{-}^{\bot}}^{4}) 
\\
& \hspace{17mm} - 10 a^{-1} (a^{-2} - 4 + a^{2}) 
\{
(L_{+}^{2} - \partial_{x, \alpha_{+}}^{4}) 
\partial_{x, \alpha_{+}^{\bot}}^{2} 
+ 
(L_{-}^{2} - \partial_{x, \alpha_{-}}^{4}) 
\partial_{x, \alpha_{-}^{\bot}}^{2} 
\}
\\
& \hspace{17mm}
- 2 (a^{2} - 1) (3 a^{-2} - 1) (3 a^{2} - 1) 
\{ 
(A_{+}(5) - \partial_{x, \alpha_{+}}^{5}) 
\partial_{x, \alpha_{+}^{\bot}} 
- (A_{-}(5) - \partial_{x, \alpha_{-}}^{5}) 
\partial_{x, \alpha_{-}^{\bot}} 
\} 
\\
& \hspace{17mm}
- a (a^{4} - 6 a^{2} + 6 - 6 a^{-2} + a^{-4}) 
\{
(L_{+}^{3} - \partial_{x, \alpha_{+}}^{6}) 
+ 
(L_{-}^{3} - \partial_{x, \alpha_{-}}^{6}) 
\}
],
\\
P_{4} 
= & 
u_{1} (u_{+} + u_{-}) 
\{ 
6a (4 - a^{2}) \partial_{x_{1}}^{2} 
+ 7 (a + a^{-1}) \partial_{x_{2}}^{2} 
\} 
+ u_{1} (u_{+} - u_{-}) (3 a^{2} - 7) 
\partial_{x_{1}} \partial_{x_{2}} 
\\
& + 
u_{2} (u_{+} + u_{-}) 
\{ 
7 (a + a^{-1}) \partial_{x_{1}}^{2} 
+ 6 a^{-1} (4 - a^{-2}) \partial_{x_{2}}^{2} 
\} 
+ u_{2} (u_{+} - u_{-}) (3 a^{-2} - 7) 
\partial_{x_{1}} \partial_{x_{2}} 
\\
& + 
\frac{1}{8} 
(u_{+} + u_{-}) 
\{ 
(35 a^{-1} + 156 a - 39 a^{3}) \partial_{x_{1}}^{2} 
+ 
(35 a + 156 a^{-1} - 39 a^{-3}) \partial_{x_{2}}^{2} 
\}, 
\\
Q_{4} 
= & 
- \frac{21}{16} 
(a^{2} + 1)^{2} (a - a^{-1}) (3 a^{2} - 1) (3 a^{-2} - 1) g_{2} 
\left(
L_{1} - L_{2} 
- \frac{a^{2} - 1}{a^{2} + 1} (u_{+} + u_{-}) 
\right) 
\end{align*}
and 
\begin{align*}
P_{6} = & 
3 a (4 - a^{2}) (u_{1}'' - u_{1}^{2}) (u_{+} + u_{-}) 
+ 
3 a^{-1} (4 - a^{-2}) (u_{2}'' - u_{2}^{2}) (u_{+} + u_{-}) 
\\
& + \frac{1}{8} 
\{
(15 a^{3} - 72 a - 7 a^{-1}) u_{1} 
+ 
(15 a^{-3} - 72 a^{-1} - 7 a) u_{2} 
\}(u_{+}^{2} + u_{-}^{2}) 
\\
& + 
\frac{a}{4} 
\{
(20 a^{4} - 83 a^{2} + 24 + 7 a^{-2}) u_{1} 
+ 
(20 a^{-4} - 83 a^{-2} + 24 + 7 a^{2}) u_{2} 
\}
(u_{+}'' + u_{-}'') 
\\
& + 
\frac{3}{16}
(a + a^{-1}) (7 a^{2} - 38 + 7 a^{-2}) 
(u_{+}^{2} u_{-} + u_{+} u_{-}^{2}) 
- \frac{5}{32} a (5 a^{4} - 202 + 5 a^{-4}) 
(u_{+}'' u_{-} + u_{+} u_{-}'') 
\\
& + \frac{3}{16} a (a^{2} - a^{-2}) 
(19 a^{2} - 122 + 19 a^{-2}) u_{+}' u_{-}' 
- 7 (a + a^{-1}) u_{1} u_{2} (u_{+} + u_{-}) 
\\
& + \frac{1}{8} 
\{
a (57 a^{2} - 216 + 7 a^{-2}) u_{1} 
+ 
a^{-1} (57 a^{-2} - 216 + 7 a^{2}) u_{2} 
\} u_{+} u_{-}. 
\end{align*}

\end{document}